\documentclass[prd,aps]{revtex4}
\usepackage{color}
\usepackage{mathtools}
\usepackage{array}
\usepackage{tabularx}
\usepackage{diagbox}
\usepackage{tikz}
\usetikzlibrary{decorations.pathmorphing}
\usepackage{bm}
\usepackage{amsmath,bm,amssymb,amsfonts,dcolumn,color,graphicx,latexsym,epsfig}
\usepackage{bbold}
\usepackage{rsfso}
\usepackage{microtype}
\usepackage{setspace} 
\usepackage{array, booktabs}
\usepackage[utf8]{inputenc}
\usepackage{hyperref}
\hypersetup{
colorlinks=true,
linkcolor=blue,
filecolor=magenta,      
urlcolor=magenta,
citecolor=magenta,
}
\usepackage{multirow}
\usepackage{natbib}
\usepackage{float}
\usepackage{booktabs}
\usepackage{pifont}
\usepackage{mathrsfs}
\usepackage{caption}
\usepackage{caption, threeparttable}
\begin{document}

\title{Displacement memory in regular black hole spacetimes}
\author{Ritwik Acharyya}
\email[Email address: ]{ritwikacharyya@kgpian.iitkgp.ac.in}
\affiliation{Department of Physics, Indian Institute of Technology, Kharagpur 721 302, India}

\author{Sayan Kar}
\email[Email address: ]{sayan@phy.iitkgp.ac.in}
\affiliation{Department of Physics, Indian Institute of Technology, Kharagpur 721 302, India}

\begin{abstract}
\noindent Displacement memory, induced by a wave pulse in a regular black hole spacetime, is studied using geodesic (timelike) separation and geodesic deviation.
The presence of the wave pulse in such a black hole is modeled via a function
$H(u)$ appearing in a restricted version of a generic Bondi-Sachs type line element. Choosing a sech-squared profile for $H(u)$, we first study 
(numerically) geodesic
separation and geodesic deviation in a flat background. Thereafter, similar 
investigations are carried out in the presence of the black hole, but in regions
far away from the vicinity of the horizon. Our results suggest the 
presence of a distinct displacement memory effect, which depends on the value of the regularisation parameter $g$
as well as the pulse height. Between different types of regular black holes, one
notices parameter-dependent changes in the net displacement memory. 
Further, a clear difference in the magnitude of displacement memory (at large $u$) in regular and singular black holes is also visible in our numerical results. 
% Finally, we perform the standard Bondi-Sachs analysis to
% check any dependencies on $g$ in the expression for the Bondi mass and other
% related quantities, up to the subleading order. 

\end{abstract}


\maketitle

%\newpage

\section{Introduction and summary} \label{introduction}

\noindent  The detection of gravitational waves (GW) has significantly advanced gravitational physics
over the past decade \cite{PhysRevLett.116.061102,theligoscientificcollaboration2025gw230814investigationloudgravitationalwave, akyüz2025potentialsciencegw250114}. Besides detection, such observations also provide indirect proof of binary black hole (or binary neutron star \cite{PhysRevLett.119.161101}, black hole-neutron star) merger events and the subsequent emergence of a perturbed black hole \cite{rezzolla2003gravitationalwavesperturbedblack}. The behaviour of gravity
in strong fields is, in a way, tested in these observations, which, in turn, 
test general relativity (GR) against other modified theories of gravity \cite{PhysRevD.102.044009, Shankaranarayanan_2022}.
In addition, questions on the nature of matter in compact objects characterised through their equations of state or the tidal response of a compact object \cite{Hinderer_2008,   Wang_2020, PhysRevD.103.124047} can be/are being addressed via such GW observations. 

\noindent Another remarkable, but as yet undetected consequence in gravitational wave physics and interferometric
detection is the possibility of detecting a new phenomenon known as 
gravitational wave memory \cite{PhysRevD.80.024002, Favata_2010, PhysRevD.84.124013}. GW  memory is a theoretically known effect in General Relativity (GR) 
(or even in other gravity theories \cite{Heisenberg_2023}). Till date, however, it has not
been confirmed observationally in astrophysical data \cite{PhysRevD.104.023004}. The concept was first proposed in the work by Zel’dovich and Polnarev \cite{Zeldovich:1974gvh} and developed further by Braginsky and Grishchuk \cite{Braginsky:1985vlg}. A fully non-linear treatment of GW memory within general relativity was discussed by Christodoulou \cite{PhysRevLett.67.1486} and later extended by Thorne and others \cite{PhysRevD.45.520, PhysRevD.44.R2945}. It was shown that gravitational waves propagating to null infinity generate an additional, permanent displacement in the detector response. This contribution, arising from the non-linear structure of Einstein’s equations, is now referred to as the \textit{nonlinear (Christodoulou) memory}. More recent developments by Bieri and Garfinkle \cite{PhysRevD.89.084039}, using gauge-invariant observables, revealed that the earlier linear–nonlinear classification is somewhat misleading, since both contributions can, in principle, arise within the framework of linearized gravity. Consequently, there has been a reclassification as ordinary and null memory effects, corresponding respectively to gravitational radiation sourced by massive and massless particles \cite{PhysRevD.90.044060}. 

\noindent Over the past few years, GW memory has been investigated in a wide range of theoretical and phenomenological contexts. It arises in
different incarnations which include displacement memory \cite{PhysRevD.107.064056, Bieri:2024ios, bhattacharya2025displacementmemorybmemorygeneralised}, velocity memory \cite{Grishchuk:1989qa, Zhang_2018}, spin memory \cite{Pasterski:2015tva, Nichols:2017rqr} and in electromagnetsim, as electromagnetic memory \cite{Bieri_2012, Winicour_2014, PhysRevD.104.084026}. Allied studies have formulated memory in the broader framework of persistent observables \cite{flanagan2019persistent, Flanagan:2019ezo} and examined them in detail  \cite{caldwell2025persistencenonlineargravitationalwave}. Extensions to post-Newtonian and self-force formalisms have provided further insight into its generation in compact binary systems \cite{Cunningham_2025_1}, while analogous effects have been explored in pp-wave \cite{ZHANG2017743, PhysRevD.99.024031, Chakraborty_2022} and cosmological spacetimes \cite{article, chakraborty2025prospectscosmologicalconstraintsusing}. Memory effects appear as an element in the so-called infrared triangle, which connects memory, asymptotic symmetries, and soft theorems \cite{Strominger:2014pwa, Blanchet_2023, Solanki:2023wmv}. 
Various other studies have addressed the influence of astrophysical environments on memory \cite{singh2025gravitationalmemorymeetsastrophysical, yyv5-3y1c}.
Numerical relativity simulations have confirmed that nonlinear memory grows significantly during the late inspiral and mergers \cite{Pollney:2010hs}, with its amplitude and detectability strongly influenced by the spin and mass ratios of the sources \cite{Mitman:2020pbt, Mitman:2021xkq, Mitman:2022kwt}. A comprehensive overview of theoretical and computational developments around GW memory may be found in \cite{Mitman_2024}. 

\noindent The essence of GW memory lies in a {\em permanent change caused by a
pulse.} It is found that once a gravitational wave pulse arrives and leaves the
detector, there is a lasting effective change in the interferometer arm-lengths through a change in the very definition of distance, caused by the pulse, in the vicinity
of the detector. The magnitude of this change is a couple of orders smaller than
the amplitudes observed in GW detection \cite{Favata_2009}. It may appear (theoretically) as a net d.c. effect \cite{PhysRevD.111.044045} on the observed GW strain, which adds to the time domain profile of a perturbed geometry. Hence, initial
and final geodesic separations will differ, leading to a measurable GW memory. Collectively, all the above-mentioned investigations 
project GW memory as a rapidly advancing area of research, providing deep insights into the symmetry structure of spacetime and offering promising prospects for future detection. The growing interest in gravitational wave (GW) memory is further driven by its potential observability with upcoming detectors such as Advanced LIGO and LISA \cite{PhysRevLett.117.061102, Talbot:2018sgr, Islo:2019qht}. Recent investigations have developed waveform models and data-analysis strategies aimed at identifying memory signatures in current and future observations \cite{PhysRevD.101.023011, Goncharov:2023woe, inchauspé2024measuringgravitationalwavememory, elhashash2025waveformmodelsgravitationalwavememory, agazie2025nanograv15yeardataset, PhysRevD.111.044044}. The wide range of studies around GW memory highlights its increasing relevance in the broader context of gravitational wave astronomy.

\noindent One can study GW memory in various ways, as mentioned above and in various articles and reviews. Here, we consider the approach where a restricted Bondi-Sachs type line element \cite{Shore_2001} is assumed, with a pulse in it signaling the brief presence of a wave. We study geodesic separation and geodesic deviation in typical 
background spacetimes by including the presence of the pulse in the line element.
In this way, we try to determine if there is a permanent change in 
future regions where the pulse is insignificant (almost absent), as an indicator of a
memory effect. This theoretical framework provides a direct means to connect the underlying spacetime geometry with signatures of GW memory and to probe nonlinear aspects of GR \cite{yyv5-3y1c}. Earlier work around this approach may be found in \cite{PhysRevD.106.104057, Datta_2024, Hadi_2024}.

\noindent Though initially we work in a flat background in Eddington-Finkelstein coordinates, later, we consider regular black holes \cite{1966JETP...22..241S, 1966JETP...22..378G}
as backgrounds. Let us briefly recall the idea of regular black holes. It is well-known that spacetime singularities remain an unresolved challenge in physics (see, for example, \cite{Joshi_2014} for an overview and \cite{NOVELLO_2008} for some cosmological implications). However, given the strict uniqueness theorems associated with the vacuum Einstein–Maxwell equations, any nonsingular (regular) black hole must incorporate some form of exotic matter, additional fields, or a modified internal structure. The
construction of such regular black holes has been a topic of active current research 
interest \cite{Kar:2023dko, huang2025regularblackholessingular, Kar:2025phe, bueno2025regularblackholesoppenheimersnyder}. Such regular spacetimes address the singularity problem in GR in a classical way and attempts to create and exploit scenarios circumventing the assumptions of the singularity theorems (e.g., violation of the Strong Energy Condition \cite{Zaslavskii_2010}). The early work of Bardeen, followed by later papers by Hayward, Neves--Saa, and others, provides examples of regular black hole geometries which we use in our analysis. There are other recent studies \cite{Walia_2022, PhysRevD.103.104047} where the parameters of the regular black holes are constrained using the Event Horizon Telescope (EHT) data. 
% Gravitational wave (GW) memory refers to a lasting displacement induced in test particle separation after a burst of gravitational wave has passed, a permanent imprint left in spacetime itself.
Signatures that distinguish regular black holes from the singular ones, offer an indirect test of alternative approaches to resolve the singularity problem in general relativity.
Likewise, through our studies here of parameter dependencies on displacement memory, we intend to uncover how such physical effects in regular black hole spacetimes may differ from similar effects in singular black holes. Thus, in addition to demonstrating the
presence of memory, we may also be able to recognize if the background spacetime has a singularity or is regular.

\noindent Throughout this work, we use natural units ($c=1$). Our article is organised as follows. In the next section, we first discuss regular black holes and thereafter introduce the restricted Bondi-Sachs line element used in our studies, in a generic way. 
We introduce a pulse in the geometry and write down the geodesic
equations for the modified line element. A brief analysis of the matter required to support such a geometry with a pulse is then presented. In Section III, we briefly 
recall the memory effect and state our approaches in studying memory in the
given context. Section IV contains the details of the numerical 
results on memory as obtained from our analysis. In Section V, we briefly present our results on geodesic deviation, while Section VI contains our conclusions.

\section{Regular black holes without and with a pulse}
\noindent The line element for a generic static, spherically symmetric spacetime, in the
Schwarzschild gauge may be written as 
\begin{equation}
    ds^2=-f(r)dt^2+\frac{dr^2}{f(r)}+r^2(d\theta^2+\sin^2 \theta d\phi^2)
    \label{general metric}
\end{equation}
where, \begin{equation}
    f(r)=1-\frac{2m(r)}{r}
\end{equation}
The following choice of $m(r)$ (as proposed in  \cite{Neves_2014}),  
\begin{equation}
    m(r)=\frac{M_0}{(1+(\frac{g}{r})^b)^{a/b}}
    \label{regular_mass_profile}
\end{equation}
gives a parametrized family of regular black holes.
Here $g$ is the regularisation parameter, related to the central density of a
de Sitter core, which replaces the central singularity, thereby yielding a regular spacetime. For $a=3$ and $b=2$, we have the well-known Bardeen class \cite{refId0} of regular black hole solutions, whereas, for $a=3$ and $b=3$, we have the Hayward class \cite{PhysRevLett.96.031103}. Numerous other solutions may be written down by choosing values of the parameters $a$ and $b$. We will use the above-mentioned $f(r)$ ($m(r)$) as well as other known forms of $f(r)$ representing regular black holes, while analysing displacement memory. 

\noindent Given a regular solution, we now need to include a pulse in the
spacetime metric in order to study a memory effect. To this end, we switch to outgoing Eddington--Finkelstein (EF) coordinates $(u,r,\theta,\phi)$, since the $u$ coordinate parametrizes outgoing null rays along which waves can propagate. We express the spacetime metric as the combination of the metric $g_{\mu \nu}$, which is given by \eqref{general metric}, and an additional piece $h_{\mu \nu}$, which encodes the pulse through a
chosen functional form. Therefore, in the EF coordinates, the
new geometry with the pulse may be chosen as 
\begin{equation}
     ds^2=-f(r)du^2 - 2 du dr+(r^2+ r H(u))d\theta^2+(r^2 - r H(u))\sin^2 \theta d\phi^2
    \label{general metric with pulse}
\end{equation}
where $H(u)$ encodes the details of the pulse (characterized by $H^{\prime \prime}(u)$). 
Note that the above specific form is a choice that we justify later.
The function $H(u)$ is further assumed as
\begin{equation}
    H(u) = A sech^2 (w (u-u_0))
    \label{pulse profile}
\end{equation}
which does represent a pulse (centre at $u=u_0$)
where $A$ and $w$ are its amplitude and width, respectively. Later, while writing down the geodesic deviation equation in such a geometry, we shall see that the Riemann tensor contributes terms on its R.H.S. which are proportional to $H^{\prime \prime} (u)$. Such terms act as a forcing \cite{Braginsky:1985vlg} over a finite interval (e.g., pulse) of the retarded time $u$. As we will see, this pulse induces a permanent displacement of test particles, thereby giving rise to the displacement memory effect.
\begin{figure}[H]
\centering
%\qquad
\includegraphics[width=0.45\textwidth]{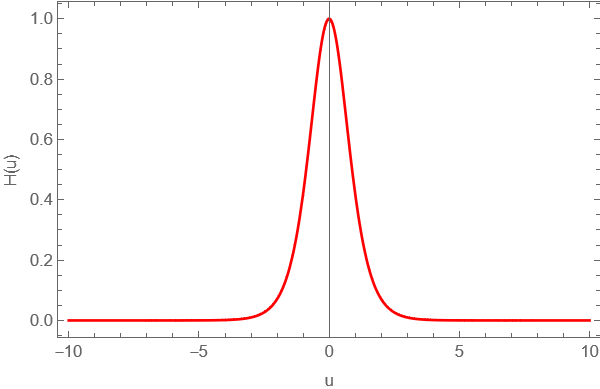}
\includegraphics[width=0.45\textwidth]{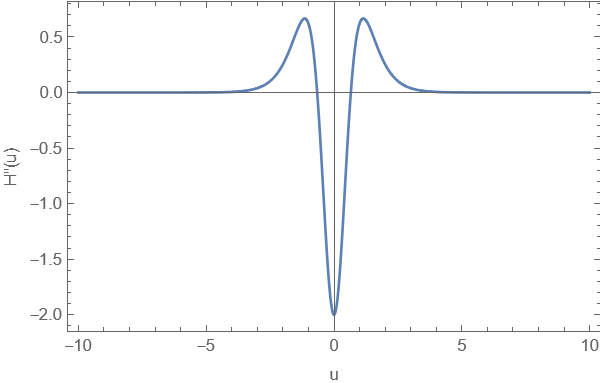}
\caption{Pulse profile ($H(u)$) with the parameters $A=1$, $u_0=0$, $w=1$ and its double derivative.} 
\label{Pulse profile image}
\end{figure}
\noindent We now attempt to interpret the nature of the metric in terms of the Bondi–Sachs framework. In 1960, Bondi introduced a novel approach \cite{Bondi:1960jsa} to study gravitational waves within Einstein’s theory of general relativity, formulated on the basis of null rays along which the waves propagate. Shortly thereafter, Sachs extended this formalism to non-axisymmetric spacetimes \cite{1961RSPSA.264..309S}, providing a systematic analysis of the asymptotic symmetries along outgoing null hypersurfaces approaching null infinity. The Bondi–Sachs formalism requires only six independent metric functions to fully characterize the asymptotic structure of a general spacetime. In this context, the line element given in Eq. \eqref{general metric with pulse} may be regarded as a restricted version of the generic Bondi–Sachs metric given by \cite{Shore_2001}, 
\begin{equation}
  ds^2 = -W du^2 -2e^{2 \beta} du dr + r^2 h_{ij} (dx^i- U^i du) (dx^j - U^j du)  
\label{bondisachs0}
\end{equation}
\noindent where
\begin{equation}
    h_{ij} dx^i dx^j= \frac{1}{2}\left(e^{2\gamma}+ e^{2 \delta}\right) d \theta^2 + 2 \sinh (\gamma- \delta) \sin \theta d \theta d \phi + \frac{1}{2} \left(e^{-2\gamma}+ e^{-2 \delta}\right) \sin^2 \theta d \phi^2.
    \label{hij bondi sachs}
\end{equation}
Here, $W$, $\beta$, $\gamma$, $\delta$, $U^i$ are functions of $u$, $r$ and angular coordinates $x^j$. From Eq. \eqref{bondisachs0}, we may obtain Eq. \eqref{general metric with pulse} with choices  $W= f(r)$, $\beta=0$, $U^i=0$ and $2\gamma (u, r) = 2\delta (u, r)= r H(u)$. It is easy to see that for
$\gamma$ small,
$e^{2\gamma} \sim 1 + 2\gamma$, $e^{-2\gamma} \sim  1-2\gamma$ and 
Eq. \eqref{bondisachs0} reduces to Eq. \eqref{general metric with pulse}.  
Considered as a vacuum solution of the Einstein equations as $r\rightarrow \infty$, the Bondi-Sachs geometry represents the gravitational field due to an isolated, radiating source, at 
${\cal I}^+$ \cite{1962RSPSA.269...21B, Shore_2001}. 
In our work here, we will only deal with a short-duration pulse 
(through the function $H(u)$), in a flat or a regular black hole
background.
\begin{itemize}
    \item In the flat background case $f(r)=1$ and the line element/metric in \eqref{general metric with pulse} 
    can be treated as a perturbation on Minkowski spacetime, which satisfies the transverse–traceless (TT) gauge
    condition. Thus, the interpretation of a gravitational wave propagating on a Minkowski background holds.
    \item When the background is not flat, i.e., $f(r)\neq 1$, however, the treatment of the deformation as a perturbation in the TT gauge
    is not quite valid -- the Lorentz gauge condition is violated globally.
    Thus, a residual gauge freedom remains. The deformation
    can still be wave--like and gravitational, but it is not similar to the
    `perturbative' sense in which we have a gravitational wave in the TT gauge in a flat background.
    
    \item A useful way to visualize this full geometry (including the pulse) is to consider a constant-$u$ hypersurface. In the absence of the pulse ($H(u)=0$), the angular part reduces to that of a round two-sphere of radius $r$. In contrast, whenever $H(u)$ is nonzero, the angular sector is no longer spherical: the geometry of the $u=constant$ slice is anisotropically deformed, resembling a squeezed or stretched sphere depending on the sign of $H(u)$.  For small perturbations $H(u) \ll r$, the distortion is slight, while for comparable $H(u)$ (w.r.t. $r$), the surface departs strongly from spherical symmetry. This illustrates explicitly how the wave pulse induces a transient deviation from the two-sphere, which disappears once the pulse has passed. The relic of this deformation persists in the relative separation between
    geodesics as well as the deviation vector, and gives rise to the displacement memory
    effect, as we elucidate below.
\end{itemize} 
To introduce this perturbation into our metric, we must incorporate an energy-momentum tensor in the matter sector
which serves as the source for the pulse. 
Considering $H/r$ to be very small, we can write the Einstein tensors up to the $\mathcal{O} (\frac{H}{r})$ and we can check the effect of the presence of the pulse in the Einstein tensors compared to the vacuum case: 
% (Complete Einstein tensors are given in Appendix I.)
\begin{align}
    G_{uu} &= \frac{r^4 f\left(1-f-rf^\prime \right) -r^3 fH}{r^6} & G_{ur}&= \frac{r^4 f\left(1-f-rf^\prime \right) -r^3 H}{r^6} \\
     G_{rr} &=0 & G_{u \theta} &= \frac{ \cot \theta H^\prime}{r} \\
    G_{r \theta} &= - \frac{H \cot \theta}{r^2} & G_{\theta \theta} &= \frac{4r^4 f^\prime + 2r^5 f^{\prime \prime} + 2r^3 H \left(2 f^\prime - r f^{\prime \prime} -3f^\prime\right)}{4r^3\left(1-\frac{H}{r}\right)^2} \\   G_{\phi \phi} &= \frac{\sin^2 \theta \left(4r^4 f^\prime + 2r^5 f^{\prime \prime} + H (2r^3 f^\prime + 4r^4 f^{\prime \prime} -2r^4 f^{\prime \prime} )\right) }{4r^3\left(1+\frac{H}{r}\right)^2}
\end{align}  
In the case of the Schwarzschild, where $f(r) = 1-2m/r$, we can write the Einstein tensors in the following way. The advantage of writing these tensors in this way is that we can see the effect of the pulse in the Einstein tensors more prominently up to the leading order of $\mathcal{O} (\frac{H}{r})$: 
\begin{align}
    G_{uu} &= \frac{- \left(1-\frac{2m}{r}\right)H}{r^3} & G_{ur}&= \frac{-H}{r^3} & G_{rr}&=0 & G_{u \theta}&= \frac{\cot \theta H^\prime}{r} \label{Einstein Tensors Schwarzschild 1}\\
    G_{r \theta}&= \frac{H \cot \theta}{r^2} &
    G_{\theta \theta}&= \frac{mH}{r^2} & G_{\phi \phi} &= \frac{-m H \sin^2 \theta}{r^2} \label{Einstein Tensors Schwarzschild 2}
\end{align}
The Einstein equations can equivalently be verified by decomposing the metric \eqref{general metric with pulse} as \begin{equation*}
    g_{\mu \nu} = \hat{g}_{\mu \nu}+ h_{\mu \nu}
\end{equation*} where $\hat{g}_{\mu \nu}$ denotes the background metric and $h_{\mu \nu}$ represents a small perturbation characterized by $H(u)$. Expanding the Einstein tensor to linear order in the perturbation amplitude yields 
\begin{equation*}
     G_{\mu \nu} = \hat{G}_{\mu \nu}+ \delta G_{\mu \nu}
\end{equation*}
with $\hat{G}_{\mu \nu}$ defined on the background spacetime and $\delta G_{\mu \nu}$ denoting the first-order correction $\mathcal{O}(H/r)$. For a vacuum background with $\hat{G}_{\mu \nu}=0$, the linear correction $\delta G_{\mu \nu}$ can be computed explicitly, yielding Eqs. \eqref{Einstein Tensors Schwarzschild 1} and \eqref{Einstein Tensors Schwarzschild 2}. This allows a consistent verification of the matter sector in the perturbed spacetime. We note that for the Schwarzschild metric with no pulse ($H(u) = 0$), the Ricci Tensors, Ricci Scalar, and Einstein tensors identically vanish, which is expected for a vacuum solution.

\section{Our approach in studying memory effects}
The gravitational wave memory effect has been studied in various ways in the literature. Among them are studies on the
change in (a) geodesic separation \cite{Datta_2024, Hadi_2024, Chakraborty_2022} of two nearby geodesics in the distant future as compared to chosen values in the past and (b) geodesic deviation analysis \cite{Zeldovich:1974gvh}.
The steps are outlined below.
\begin{itemize}
    \item To study the burst-like nature of gravitational radiation, we have chosen a specific profile given by Eq. \eqref{pulse profile}. The pulse is a function of the null coordinate $u$, which naturally characterizes the outgoing radiation. Such a choice is standard in studies of the gravitational-wave memory effect, as the net displacement memory is quantified by comparing the asymptotic behavior of geodesics at large $u$.
    \item The spacetime geometry is given by Eq. \eqref{general metric with pulse}. We integrate the geodesic equations to obtain the timelike geodesics parametrised by $u$. Analytical or numerical methods may be used to solve for the spatial coordinates as functions of $u$. In this work, we numerically solved the geodesics with the proper choice of initial conditions. 
    \item We have investigated the evolution of geodesic separation both in the presence and in the absence of a gravitational-wave pulse. The separation values before and after the passage of the pulse were compared. The resulting difference quantifies the \textit{displacement memory effect}. Once the pulse has passed, the geodesic separation does not return to its initial value. Using the same choice of parameters and numerically integrated geodesics, we further studied the geodesic deviation. 
    % The qualitative behavior of the deviation analysis at large $u$ is found to be consistent with the results obtained from the analysis of geodesic separation. 
    \item By differentiating the geodesic solutions and studying the evolution of their relative velocity, one can obtain a \textit{velocity memory effect}. In the context of pp-wave backgrounds, pulse profiles satisfying $\int_{-\infty}^{\infty} A(u) du \neq 0$ (where $A(u)$ is the function representing the pulse in the pp-wave geometry) show a nonzero velocity memory \cite{Chakraborty_2022, Divakarla_2021}.
    \end{itemize}

\noindent It is useful to note that while solving geodesics and finding geodesic separation, we are not making any assumption regarding the
geodesics being `neighboring', i.e., close to each other. On the other hand,
while finding the deviation using the geodesic deviation equation, the
equations themselves depend on the `neighboring geodesic' assumption. 
Further, the deviation equations form a set of linear, second-order,
coupled ordinary differential equations, whereas the geodesic equations
are always quasi-linear. 

\noindent In the next two sections, we will demonstrate our results for
displacement memory using analysis of geodesic separation and geodesic deviation.

\section{Geodesic separation and displacement memory}
Let us first write down the geodesic equations in the spacetime geometry given by Eq. \eqref{general metric with pulse}. The effective Lagrangian can be written as, 
\begin{equation}
    L= - f(r) \Dot{u}^2 - 2 \Dot{u} \Dot{r} + (r^2+ r H(u)) \Dot{\theta}^2 + (r^2 - r H(u)) \Dot{\phi}^2 
    \label{lagrangian}
\end{equation}
Hence, the geodesic equations are, 
\begin{align}
    \Ddot{u} - \frac{f^\prime}{2} \Dot{u}^2 + \left(\frac{2r + H}{2}\right)\Dot{\theta}^2 +  \left(\frac{2r - H}{2}\right) \sin^2 \theta \Dot{\phi}^2 & = 0 \label{u_geodesic_eqn} \\
    2\Ddot{r}+ f f^\prime  \Dot{u}^2 + 2 f^\prime  \Dot{r}\Dot{u} 
    + \left( r H^\prime - f(2r + H) \right)\Dot{\theta}^2+  \left( - r H^\prime - f(2r - H) \right) \sin^2 \theta \Dot{\phi}^2 & = 0 \label{r_geodesic_eqn} \\
    \Ddot{\theta} (r^2 + rH) + \Dot{r} \Dot{\theta} (2r+H) + \Dot{\theta} \Dot{u} r H^\prime  -\sin \theta \cos \theta (r^2-rH)\Dot{\phi}^2 &= 0 \label{theta_geodesic_eqn}\\
    \Ddot{\phi} \sin ^2 \theta (r^2 - r H) + 2\Dot{\theta} \Dot{\phi} \sin \theta \cos \theta (r^2 - r H) + \Dot{\phi} \Dot{r} \sin^2 \theta (2r - H) - r H^\prime  \sin^2 \theta \Dot{\phi} \Dot{u} & = 0 \label{phi_geodesic_eqn}
\end{align}
It can be checked that $\theta =\pi/2$ is a solution of the above set of geodesic equations. An overdot denotes the derivative with respect to the proper time ($\tau$), $f^\prime$ is $df(r)/dr$, and $H^\prime$ is $dH(u)/du$.

\noindent With $\theta=\frac{\pi}{2}$, the geodesic equations reduce to, 
\begin{align}
    \Ddot{u} - \frac{f^\prime (r)}{2} \Dot{u}^2 - \left( \frac{H(u) -2r}{2}\right) \Dot{\phi}^2 & = 0 \label{u geodesic theta 90} \\
    \Ddot{r} + \frac{f(r)}{2} f^\prime (r) \Dot{u}^2 + f^\prime (r) \Dot{r} \Dot{u} + \frac{f(r) H(u) - 2f(r) r - rH^\prime (u)}{2} \Dot{\phi}^2 & = 0 \label{r geodesic theta 90} \\
    \Ddot{\phi} - \frac{H^\prime (u)}{(r- H(u))} \Dot{\phi} \Dot{u} + \frac{2r -H(u)}{r^2 - r H(u)} \Dot{\phi} \Dot{r} & = 0 \label{phi geodesic theta 90}
\end{align}
We are looking for timelike geodesics that must satisfy the timelike condition: 
\begin{equation}
    -f(r) \Dot{u}^2 - 2 \Dot{u} \Dot{r} + (r^2- rH) \Dot{\phi}^2 = -1
    \label{timelike condition theta 90}
\end{equation}
\noindent The $\phi$ equation yields a first integral
\begin{equation}
\dot \phi = \frac{L_c}{r^2-r H}, 
\label{Conserved L}
\end{equation}
where $L_c$ is the conserved angular momentum corresponding to the $\phi$ coordinate. One can rewrite the above set of equations as a first-order dynamical system 
using the $u$  coordinate as an independent variable (note that the pulse is a function of $u$).  We rewrite the timelike condition in the following way using Eq. \eqref{Conserved L}: 
\begin{equation}
    -f (r) \Dot{u}^2 - 2 \Dot{u}^2 \frac{dr}{du} + \frac{L_c^2}{r^2-r H} = -1 \hspace{0.2 in} \implies \hspace{0.2 in} \Dot{u}^2 = \frac{1+\frac{L_c^2}{r^2-r H}}{f(r) + 2 \frac{dr}{du}} 
    \label{u dot square}
\end{equation}
Similarly, $\Ddot{r}$ and $\Ddot{\phi}$ can be written as
\begin{align}
    \Ddot{r} &= \frac{d^2 r}{d u^2} \Dot{u}^2 + \frac{dr}{du} \Dot{u}\frac{d}{du}(\Dot{u}) &  \Ddot{\phi} &= \frac{d^2 \phi}{d u^2} \Dot{u}^2 + \frac{d\phi}{du} \Dot{u}\frac{d}{du}(\Dot{u}) \label {r and phi double dot}
\end{align}
From \eqref{u geodesic theta 90} we can write: 
\begin{eqnarray}
    \Ddot{u} = \Dot{u} \frac{d}{du} (\Dot{u}) = \frac{f^\prime (r)}{2} \Dot{u}^2 + \left( \frac{H(u) -2r}{2}\right) \Dot{\phi}^2 
    \label{u double dot in u coordinate}
\end{eqnarray}
Hence using \eqref{Conserved L}, and substituting the expressions for quantities like, $\Dot{u}^2$ and $\Dot{u} \frac{d}{du} (\Dot{u})$ from \eqref{u dot square} and \eqref{u double dot in u coordinate} respectively, we can write the system of equations in $u$ coordinate, 
\begin{eqnarray}
    \frac{dp}{du}=-\frac{f f^\prime}{2} \left[3 p + f\right] - \frac{L_c^2}{(r^2-rH)^2} \frac{f(r) + 2 p}{1+\frac{L_c^2}{r^2-r H}} \left[p \frac{H -2r}{2} + \frac{fH-2rf-rH^\prime}{2} \right] \label{double derivative of r as function of u}\\
    \frac{dq}{du}=-q \left[\frac{f^{\prime}}{2}-\frac{H^\prime}{r-H} \right]- \frac{2r-H}{r^2-rH}pq- q\frac{H-2r}{2} \frac{L_c^2}{(r^2-rH)^2} \frac{f(r) + 2p}{1+\frac{L_c^2}{r^2-r H}} \label{double derivative of phi as function of u}\\
    \frac{dr}{du} = p \label{Equation of p}\\
    \frac{d\phi}{du} = q \label{Equation of q}
\end{eqnarray} 
One can thus find $p(u)$, $q(u)$,
$r(u)$ and $\phi(u)$ by numerical integration of the dynamical system. While solving, the timelike constraint has been added as a constraint on the variables in the dynamical system. Also, we have made careful choices of the initial conditions
so that they satisfy the timelike condition throughout the evolution. Furthermore, to guarantee that the geodesics remain outside the event horizon, 
the initial radial coordinate \(r\) is chosen to lie beyond the horizon, 
with \(\dot{r}\) set positive through an appropriate choice of initial conditions.
\noindent Velocity memory arises from \eqref{double derivative of r as function of u}, since for a pulse $dH/du$ is different in the asymptotic limits $u\rightarrow -\infty$ and
$u\rightarrow\infty$. From the above set of equations, it is important to note that the occurrence of the gravitational wave memory effect requires a non-zero $L_c$. This is because $H^{\prime} (u)$ enters the equations multiplied by $L_c$, and the asymptotic change in $H^{\prime}$ is precisely what gives rise to the memory effect. Such a memory will also be visible if one plots the $p(u)$ and $q(u)$ as a vector field in two dimensions at large positive and negative values of $u$.

\noindent We will now numerically solve Eqs \eqref{double derivative of r as function of u}--\eqref{Equation of q} and analyze various geometries, starting with the flat background.

\subsection{Flat background}
\begin{figure}[h]
\centering
%\qquad
\includegraphics[width=0.45\textwidth]{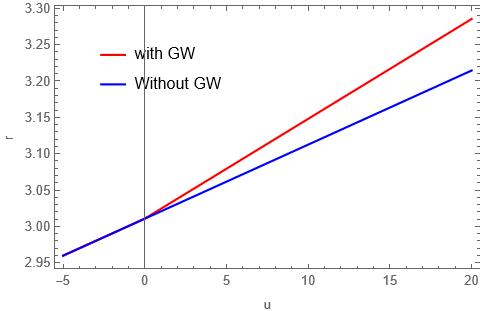}
\includegraphics[width=0.45\textwidth]{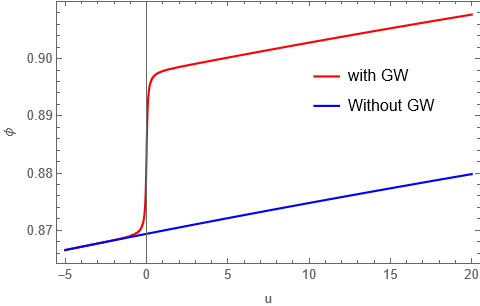}
\includegraphics[width=0.45\textwidth]{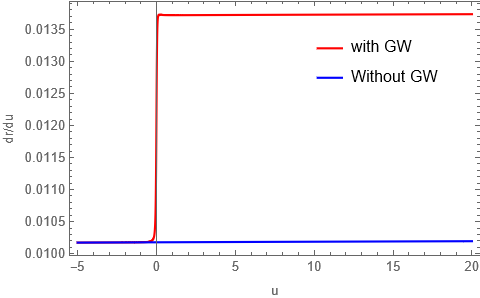}
\includegraphics[width=0.45\textwidth]{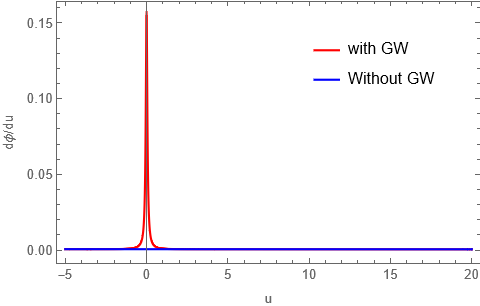}
\caption{\textbf{Top panel:} Geodesic components $r(u)$ and $\phi(u)$ with and without the GW pulse in flat spacetime. \textbf{ Bottom panel:} $dr/du$ and $ d\phi/du$ as functions of $u$, both in the presence and absence of GW pulse for flat metric. The values of the parameters are: $L_c=0.005$, and $A=3$. 
\label{geodesic plots as a function of u for flat metric}}
\end{figure}

\noindent For a  flat background we have $f(r)= 1$ in \eqref{general metric with pulse}. Using the pulse profile from Eq. \eqref{pulse profile}, we solve 
for geodesics and obtain the components of the geodesics $r(u)$, $\phi(u)$. The chosen parameters for the pulse profile are: $A=3$ and $w=1$
Thereafter, we investigate the separation of two nearby geodesics in the presence of the pulse and when it is absent. Numerical solutions of the geodesics are shown in Fig. \eqref{geodesic plots as a function of u for flat metric}. 
From the Fig. \eqref{geodesic plots as a function of u for flat metric}, it is evident that the geodesics evolve differently with respect to $u$ in the presence and absence of the GW pulse. We next calculate the geodesic separations and plot their evolution as a function of $u$. As evident from Eq. \eqref{pulse profile}, the pulse is significant around $u=0$ and subsequently decays rapidly. Below, we state the initial conditions used by us while
solving the geodesic equations. Large positive and negative $u$ values
correspond to the choices $u=100$ units and $u=-100$ units, respectively. The coordinates \(r (u)\) and \(\phi (u)\) are expressed in arbitrary units.
\begin{itemize}
    \item Geodesic 1: $r[-100] = 2$, $p[-100]= 0.01$, $\phi [-100]= \pi/4$. 
    \item Geodesic 2: $r[-100] = 4$, $p[-100]= 0.01$, $\phi [-100]= \pi/4$. The choice of $q[-100]$ for both the geodesics has been discussed below. 
    
\item Given the initial condition on $r$ and $p$, we  use them in Eq. \eqref{u dot square} to find $\Dot{u}$ at $u=-100$. We choose the positive root of $\Dot{u}$. Thereafter, we set the initial condition for $d\phi/ du$ using the positive root of $\Dot{u}$ and Eq. \eqref{Conserved L}.
    \item The parameter $L_c$ is chosen to be small, since for $L_c \gtrsim 0.005$ the separations in both $r$ and $\phi$ become negative at 
    large $u$, indicating a crossing of geodesics. We therefore restrict our choice of $L_c$ values appropriately.
\end{itemize}

\begin{figure}[H]
\centering
%\qquad
\includegraphics[width=0.45\textwidth]{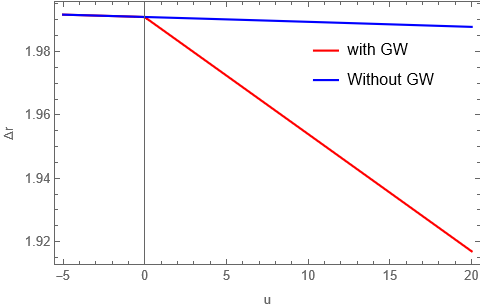}
\includegraphics[width=0.45\textwidth]{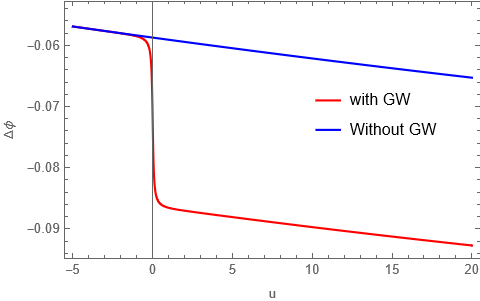}
\caption{ Variation of \(\Delta r\) and \(\Delta \phi\) with \(u\) for the flat metric, 
shown both in the presence and absence of a GW pulse. 
The chosen parameters are \(L_c = 0.005\) and \(A = 3\).
\label{displacement memory and velocity memory as function of u for flat metric}}
\end{figure}

\begin{figure}[H]
\centering
%\qquad
\includegraphics[width=0.45\textwidth]{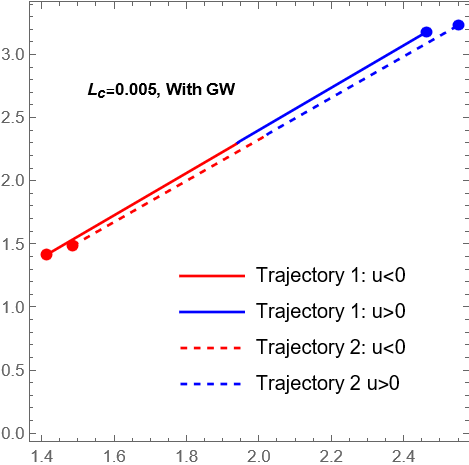}
\includegraphics[width=0.45\textwidth]{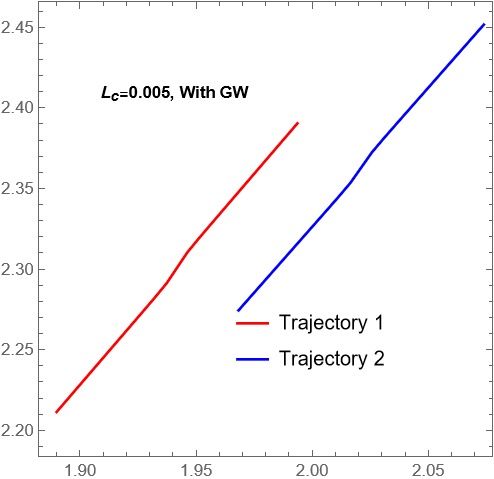}
\caption{\textbf{Left panel:} Trajectories of two geodesics have been shown in the presence of the GW with $L_c= 0.005$ and $A=1.5$. The red dot represents $u=-100$ and the blue dot marks $u=+100$. \textbf{Right panel:} Trajectories of the same pair of test particles shown in the region close to the GW pulse: $u \in [-10, 10]$.}
\label{Trajectory plots as function of u for flat metric}
\end{figure}
In Figure \eqref{geodesic plots as a function of u for flat metric} we have shown $r$, $\phi$, and their $u$ derivatives. 
Figure \eqref{displacement memory and velocity memory as function of u for flat metric} shows the differences $\Delta r$ and $\Delta \phi$. More precisely, we highlight the following key points.
\begin{itemize}
    \item The differences in the coordinates are defined as
    \begin{equation}
    \Delta r = r(\text{geodesic 2}) - r (\text{geodesic 1}) \hspace{1cm} \Delta \phi= \phi(\text{geodesic 2}) - \phi(\text{geodesic 1})
    \label{geodesic difference governing equation}
\end{equation}
\noindent  The geodesic separations evolve differently in the presence and absence of the wave pulse, and a residual difference persists at null infinity (which is different from the scenario with no pulse), indicating a displacement memory effect. However, the magnitude of the separation of two geodesics is very small at large $u$. 
    \item In a flat background, $\phi$ seems to be more sensitive to the GW pulse compared to $r$. In Fig. \eqref{displacement memory and velocity memory as function of u for flat metric}, this difference is clearly visible.
\end{itemize}
The picture becomes clearer if we plot the trajectories ($r (\phi)$) of two geodesics and examine their asymptotic differences.
\noindent Fig. \eqref{Trajectory plots as function of u for flat metric} shows the two trajectories in the interval $u \in [-100, 100]$, with the red colour indicating $u<0$ and blue indicating $u>0$. The point marked in red represents the starting point $u=-100$ and the blue point represents the end point $u=+100$. Having obtained $r(u)$ and $\phi(u)$, the trajectories are plotted using $x = r \cos \phi$ and $y= r \sin \phi$. For Fig. \eqref{Trajectory plots as function of u for flat metric}, we adopt the same set of boundary conditions as before, except that $r[-100]$ is chosen to be 2 and 2.1 for the two geodesics, respectively. The intersection of the red and blue trajectories marks the $u=0$ region, where the effect of the gravitational wave pulse is maximum. The right panel provides a zoomed-in view near the pulse, illustrating the nature of the transition in both trajectories around $u=0$. From the left panel of the Fig. \eqref{Trajectory plots as function of u for flat metric}, it is evident that the difference in trajectories between the two particles is smaller during the $u<0$ phase (red) compared to the $u>0$ phase (blue). This indicates that the GW pulse leaves an imprint on the relative trajectories, which remains even after the pulse has passed.

\noindent In the following subsections, we will carry out similar analyses
for geodesics in various regular black holes, with the aim of 
establishing displacement memory as a signature capable of distinguishing between them as well as the related singular spacetime.
% The right panel displays the difference between the trajectories in the presence and absence of the gravitational wave, with the $z$-axis representing the $u$ coordinate. The plot exhibits a finite difference between the two scenarios -- presence and absence of the gravitational waves at asymptotic null infinity. This demonstrates the displacement memory effect in flat spacetime.
\subsection{Analysis for Bardeen spacetime}
Let us begin by stating the $f(r)$ for the Bardeen spacetime. We have, 
\begin{equation}
    f(r)= 1- \frac{2}{r} \left[\frac{M}{\left(1+\left(\frac{g}{r}\right)^2\right)^{3/2}}\right]
    \label{Bardeen profile f(r)}
\end{equation}
Using this $f(r)$ along with the chosen pulse profile \eqref{pulse profile}, we solve for the geodesics. Our numerical solutions for the geodesics are presented below in Figure \eqref{geodesic plots as a function of u for Bardeen profile}. Throughout our analysis, we have considered $M=1$.
\begin{figure}[h]
\centering
%\qquad
\includegraphics[width=0.45\textwidth]{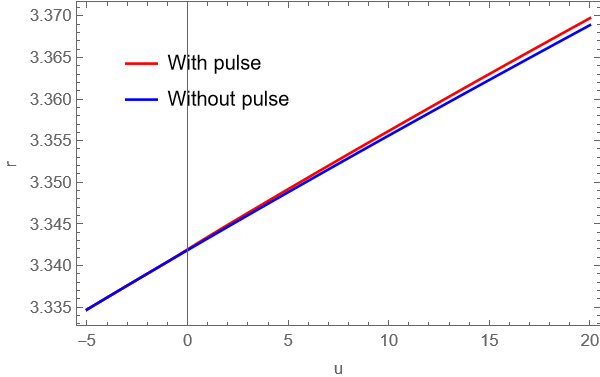}
\includegraphics[width=0.45\textwidth]{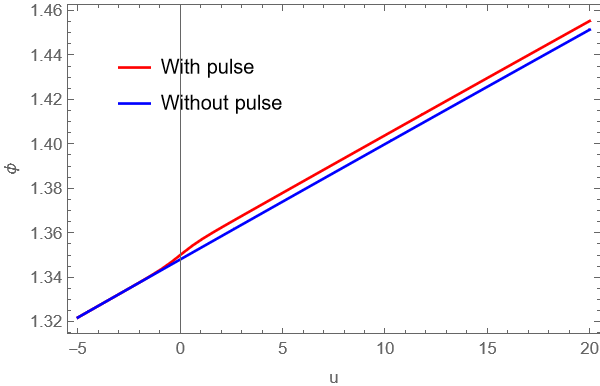}
\includegraphics[width=0.45\textwidth]{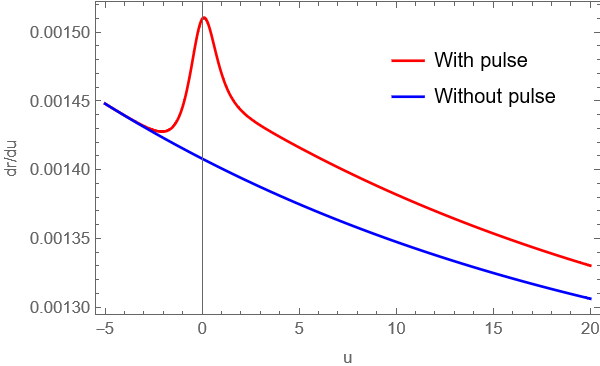}
\includegraphics[width=0.45\textwidth]{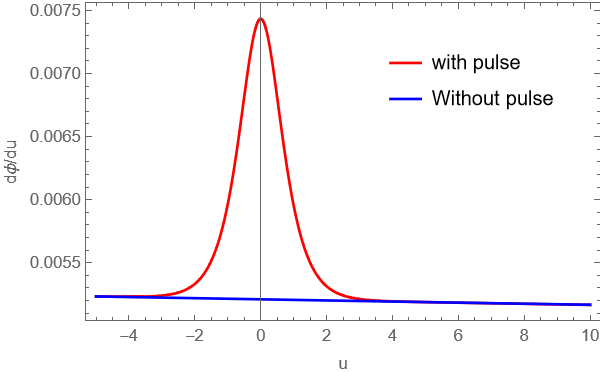}
\caption{\textbf{Top panel:} $r (u)$ and $\phi (u)$ both in the presence and absence of the pulse for Bardeen Profile. \textbf{ Bottom panel:} Variation of $dr/du$ and  $d\phi/ du$ with $u$ in the presence of pulse and without pulse for Bardeen Profile. The parameters are: $L_c= 0.1$, $g=0.1$ and $A=1$.} 
\label{geodesic plots as a function of u for Bardeen profile}
\end{figure}

\noindent Let us first state the initial conditions used while solving for the geodesics. 
\begin{itemize}
    \item Two geodesics are initialized at $u = - 100$, corresponding to $u \to -\infty$. The initial radial coordinates are $r[-100]= 3$ and $r[-100]= 3.1$.
    \item The initial condition for $\phi$, $dr/du$ are same as of the flat background case-- i.e $\phi [-100]= \pi/4$, $p [-100]= 0.01$ for both the geodesics. The initial condition for $q(u)$ can be fixed in the same way as done for the flat background case, thereby ensuring the timelike condition is satisfied. 
\end{itemize}
\begin{figure}[h]
\centering
%\qquad
\includegraphics[width=0.45\textwidth]{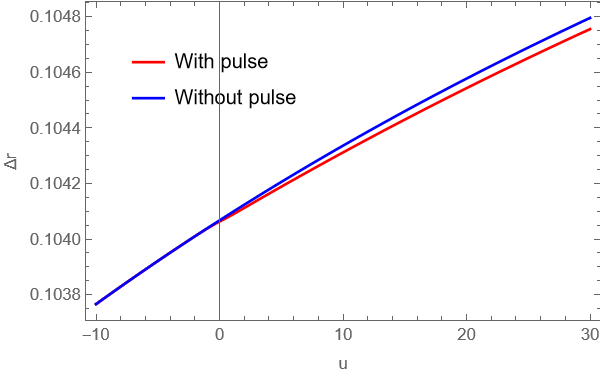}
\includegraphics[width=0.45\textwidth]{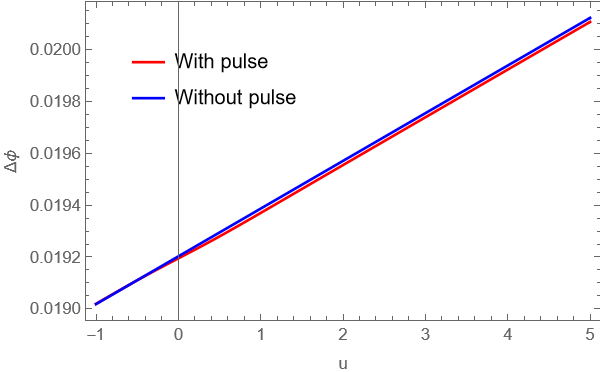}
\includegraphics[width=0.45\textwidth]{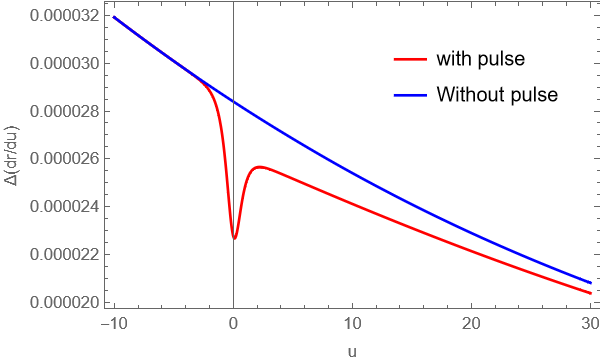}
\includegraphics[width=0.45\textwidth]{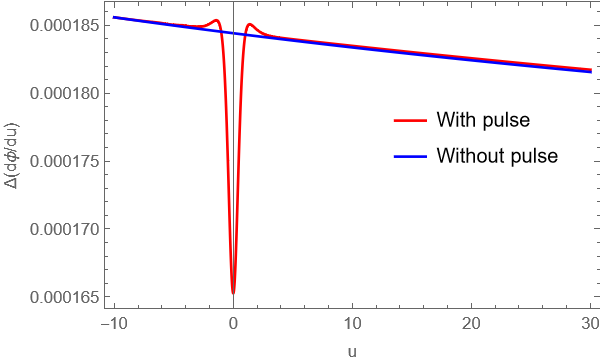 }
\caption{\textbf{Top panel:} Variation of $\Delta r$ and $\Delta \phi$ with $u$ in the presence of pulse and without pulse for Bardeen profile \textbf{ Bottom panel:} Variation of $\Delta$ $dr/du$ and $\Delta$ $d\phi/du$ with $u$ in the presence of pulse and without pulse for Bardeen profile. The parameters are: $L_c= 0.1$, $g=0.1$ and $A=1$. 
\label{displacement memory and velocity memory as function of u for Bardeen profile}}
\end{figure}

\noindent We now examine the evolution of the separation between two geodesics in Fig. \eqref{displacement memory and velocity memory as function of u for Bardeen profile} in the Bardeen spacetime. In this figure, the evolution of the separation between two nearby timelike geodesics (both $r$ and $\phi$ components) is plotted as a function of the $u$ coordinate, both with and without the presence of a pulse. A clear distinction between the two scenarios, with and without the pulse, emerges at large $u$. This plot indicates the displacement memory in the Bardeen spacetime with a fixed $g$ and $L_c$. The evolution of the tangent of the $r$ and $\phi$ coordinates (i.e., $dr/du$ and $d\phi/du$) differs between the cases with and without the wave pulse. However, the magnitude of these differences
% remain negligible, of order $10^{-5}$ and $10^{-4}$, respectively. They
are smaller than the displacement memory discussed above.\\

\noindent We considered \(M = 1\) and chose a wave pulse amplitude 
\(A = 1\), along with \(g = 0.1\), for Fig.~\eqref{geodesic plots as a function of u for Bardeen profile} 
and Fig.~\eqref{displacement memory and velocity memory as function of u for Bardeen profile}. We have taken the initial value of $dr/du$ as 0.01. In general, one can choose a higher initial value of  $dr/du$ and a higher $A$, which would also produce a noticeable displacement memory effect. To ensure that the effect of the wave pulse can be treated as a small perturbation, the amplitude $A$ has been deliberately chosen to be small. The solution described in Eq. \eqref{Bardeen profile f(r)} depends on two key parameters: $L_c$ and the regular parameter $g$. Changing $g$ alters the spacetime geometry, influencing the radius of the event horizon. There exists a critical value of $g/M$ (which we can find by setting $g_{tt} = 0$) beyond which the event horizon disappears, resulting in a naked singularity. If we make the discriminant of the $g_{tt}=0$ to be real, we get that for $g^2/M^2 \leq 48/81$ we will have horizons. For $g^2/M^2 > 48/81$, the spacetime doesn't have any event horizon, which we discarded in this discussion. Hence, we have allowed the maximum value of $g$ to be taken up to $\sqrt{48/81}$ (considering $M = 1$). \\
% The changes in the $dr/du$ component and $d\phi/du$ components turned out to be very small, which are quite negligible compared to the spatial components. Therefore, we do not present the separation of the corresponding components for the two timelike geodesics.\\
\begin{figure}[h]
\centering
%\qquad
\includegraphics[width=0.45\textwidth]{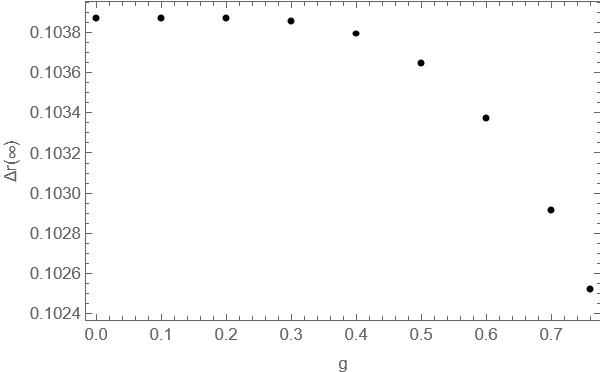}
\includegraphics[width=0.45\textwidth]{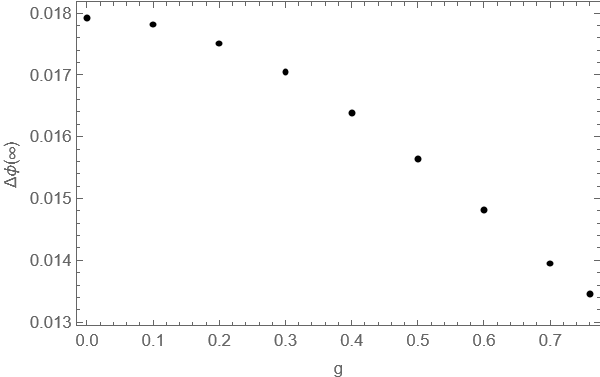}
\caption{ Variation of \(\Delta r\) and \(\Delta \phi\) at large \(u\) with \(g\) 
for the Bardeen profile (\(L_c = 0.05\), \(A = 1\), \(M = 1\)).
\label{memory dependence on the g and l for Bardeen profile (1)}}
\end{figure}
\begin{figure}[h]
\centering
%\qquad
\includegraphics[width=0.45\textwidth]{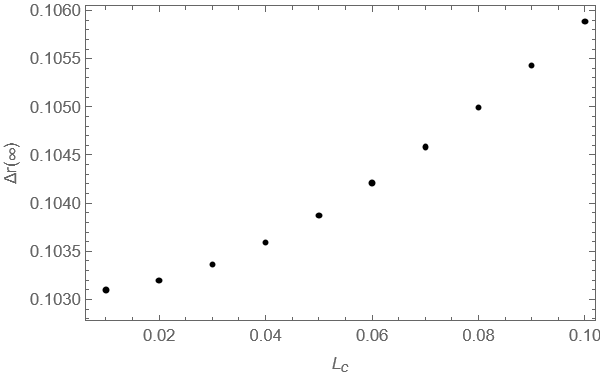}
\includegraphics[width=0.45\textwidth]{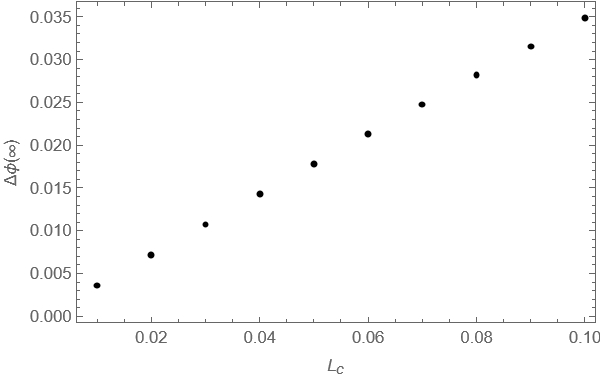}
\caption{ Variation of \(\Delta r\) and \(\Delta \phi\) at large \(u\) with \(L_c\) 
for the Bardeen profile ($g= 0.1$, $A=1$, $M=1$). \label{memory dependence on the g and l for Bardeen profile (2)}}
\end{figure}
\noindent Finally, we plot the difference between the two geodesics at positive large $u$ ($ u\rightarrow\infty$) and how that changes with the variation of the $g$ and $L_c$ to explore how the memory depends on those parameters. That has been shown in Fig. \eqref{memory dependence on the g and l for Bardeen profile (1)} and \eqref{memory dependence on the g and l for Bardeen profile (2)}. A few points corresponding to the geodesic separation analysis have been listed below :
 \begin{itemize}
     \item The differences in both $r$ and $\phi$ components of the two geodesics at large $u$ decrease with increasing $g$, for a fixed value of $L_c$.
     \item In the chosen spacetime with the pulse, the conserved angular momentum ($L_c$) is another parameter that is a constant motion; thus, we have studied the variation of the asymptotic $\Delta r$ and $\Delta \phi$ to check how the asymptotic differences depend on the values of $L_c$ in the figure \eqref{memory dependence on the g and l for Bardeen profile (2)}, and it turned out that as $L_c$ is increased, the separation increases. This same trend for both $L_c$ and $g$ has been seen for other values [$0.1-10$] of the pulse amplitudes as well.
     \item We adopt the generalized Neves–Saa mass profile from Eq. \eqref{regular_mass_profile}, which, in the limit $g=0$, reduces to the standard Schwarzschild geometry. From Fig.~\eqref{memory dependence on the g and l for Bardeen profile (1)}, we can note that the difference between the two geodesics is most prominent for the Schwarzschild geometry (\(g = 0\)) for a constant $L_c$.
     % Specifically, for $L_c =0.05$ and pulse amplitude, $A=1$, the final asymptotic values are: 
     % \begin{align*}
     %     \Delta r (\infty) &= 0.103872 & \Delta \phi (\infty) &= 0.0179165
     % \end{align*} 
 \end{itemize}

\begin{figure}[h]
\centering
%\qquad
\includegraphics[width=0.30\textwidth]{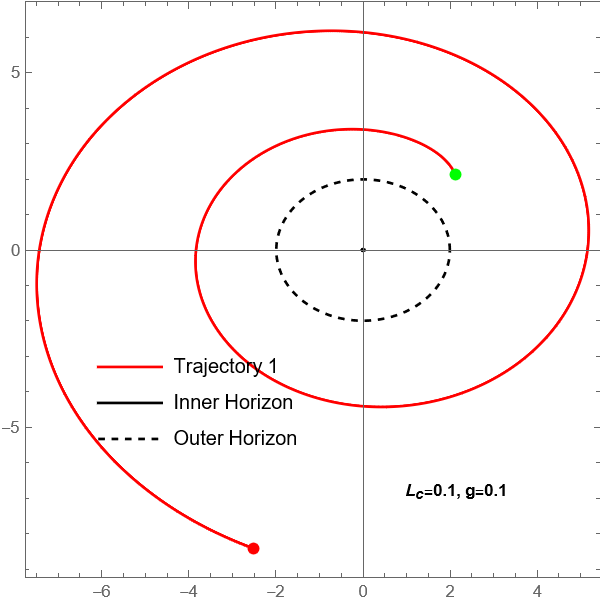}
\includegraphics[width=0.30\textwidth]{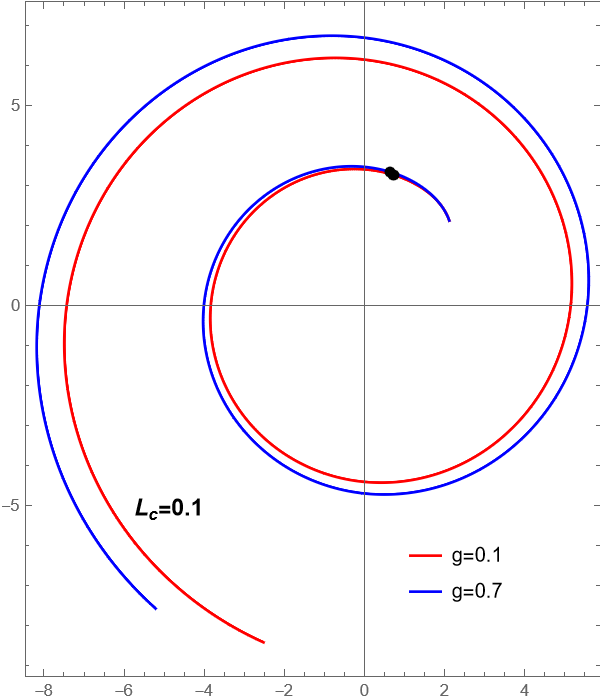}
\includegraphics[width=0.30\textwidth]{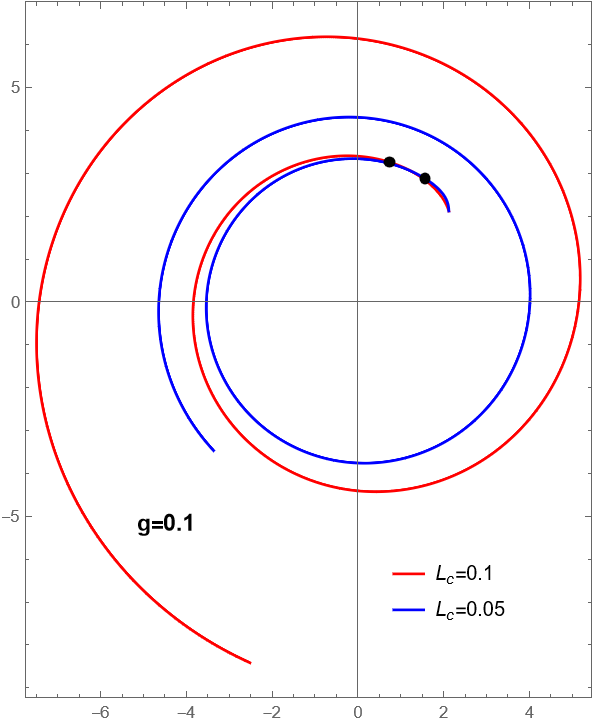}
\caption{Trajectories in the Bardeen spacetime for different $L_c$ and $g$ with $u \in [-100, 5000]$ and pulse amplitude $A=1$. \textbf{Left Panel}: Trajectory of a particle in the Bradeen spacetime. The black dotted curve denotes the outer horizon. Green and red dots are the endpoints corresponding to u=-100 and u=5000, respectively. \textbf{Middle Panel}: Trajectories for different $g$ values for a fixed $L_c$. The black dots represent $u=0$ for both the trajectories. \textbf{Right panel}: Trajectories for different values of $L_c$ at fixed $g$. The black dots mark $u=0$.
\label{Bardeen polar plot}}
\end{figure}
\begin{figure}[H]
\centering
%\qquad
\includegraphics[width=0.45\textwidth]{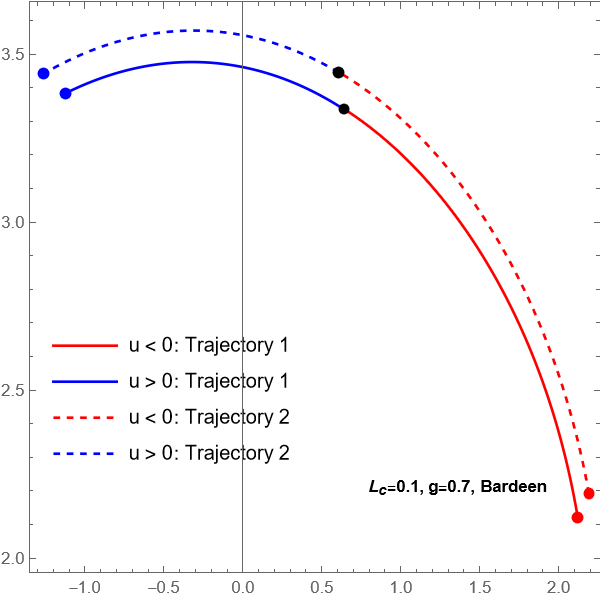}
\includegraphics[width=0.45\textwidth]{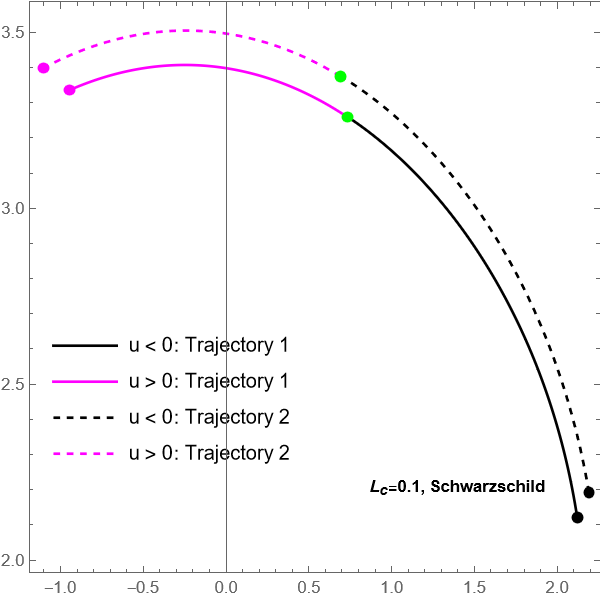}
\caption{\textbf{Left Panel} Two trajectories (obtained from the geodesic equations using the initial conditions discussed earlier) for the Bardeen case ($g=0.7$) in the presence of the pulse. \textbf{Right Panel}: Two trajectories for the Schwarzschild case ($g=0$) in the presence of the pulse. For both cases, $L_c=0.1$.
\label{Bardeen trajectory plot}}
\end{figure}

Having solved the geodesics, we now plot their trajectories. Figure \eqref{Bardeen polar plot} illustrates the trajectories under consideration.
\begin{itemize}
    \item The left panel of Fig. \eqref{Bardeen polar plot} shows the trajectory of a particle in the $x$–$y$ plane in Bardeen spacetime for a fixed $g$ and $L_c$ value. The black curves represent the horizon, and the initial conditions are chosen such that the trajectories remain outside it, as illustrated in the figure. Each trajectory is plotted from an initial \(u = -100\) (green dot) to a final \(u = 5000\) (red dot).

    \item From Fig. \eqref{Bardeen polar plot}, it is evident that higher values of the conserved angular momentum $L_c$ result in trajectories that spiral outward more prominently. This behavior is consistent with Eq. \eqref{Conserved L}, where an increase in $L_c$ leads to a larger $\dot{\phi}$, producing faster outward spiraling, while a smaller $L_c$ yields slower spirals closer to the central mass. Furthermore, for a fixed $L_c$, increasing the regular parameter widens the spiral. 
    \item Since the spacetime contains a time-dependent pulse, it does not have a timelike Killing vector. As a result, energy is not conserved along the geodesics, and the orbits are therefore not closed.
    \item Our primary aim is to study the difference between trajectories as $u \to \infty$, which was taken to be $u=100$ in the previous geodesic separation analysis. At $u=5000$, the $\mathrm{sech}^2$ function attains very small values, potentially introducing numerical errors. Therefore, we focus on the portion of the trajectory from $u=-100$  to $u=100$  to analyze the memory effect in the Fig. \eqref{Bardeen trajectory plot}.
\end{itemize}
\noindent Following the flat metric analysis, we examine the trajectories in the presence of the pulse. For $g=0$, the spacetime reduces to Schwarzschild geometry. Hence, we will compare the differences in trajectories both in Schwarzschild and Bardeen spacetimes. In the Fig. \eqref{Bardeen trajectory plot}, we have shown the two trajectories for both the Bardeen and the Schwarzschild geometry in the presence of the pulse. From the figure, it is evident that the difference between the trajectories at $u \to \infty$ is larger than at $u \to -\infty$. The $u<0$ phase is shown in red and the $u>0$ phase in blue, with the trajectory starting at $u=-100$ (red dot) and ending at $u=+100$ (blue dot). The pulse becomes significant near $u=0$ (black dot). This figure also demonstrates that the separation between trajectories is larger in the Schwarzschild geometry compared to the Bardeen geometry.

\begin{figure}[H]
\centering
%\qquad
\includegraphics[width=0.45\textwidth]{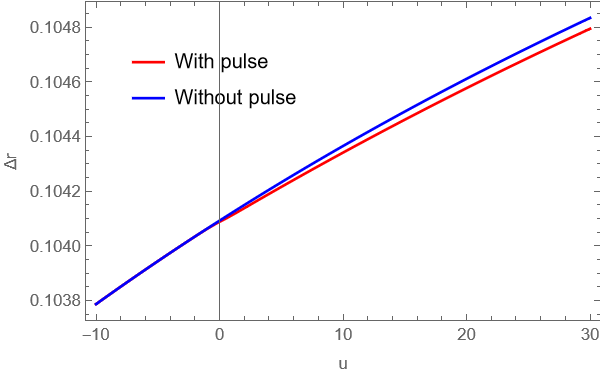}
\includegraphics[width=0.45\textwidth]{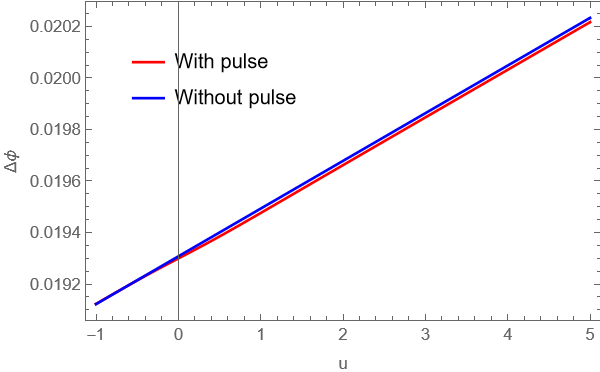}
\includegraphics[width=0.45\textwidth]{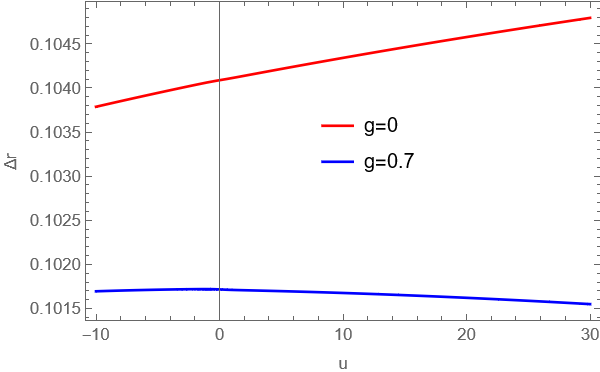}
\includegraphics[width=0.45\textwidth]{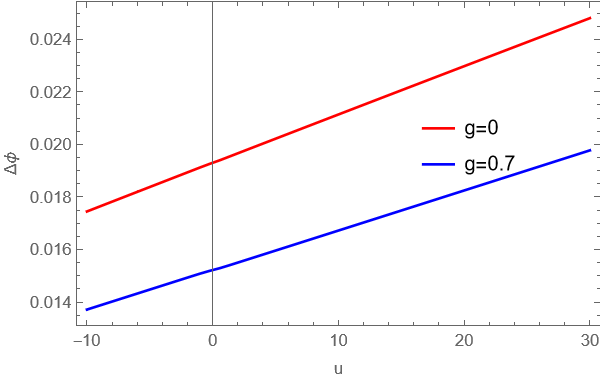}
\caption{\textbf{Top Panel}:  Variation of $\Delta$r and $\Delta$ $\phi$ with $u$ in the presence of pulse and without pulse for the Schwarzschild case. The parameters are $L_c=0.1$ and $A=1$. ($g=0$).\textbf{Bottom Panel}: Evolution of $\Delta r$ and $\Delta \phi$ for the Schwarzschild geometry and the Bardeen geometry for a fixed $L_c=0.1$, and $A=1$.
\label{Schwarzschild memory plot}}
\end{figure}

\noindent To further illustrate the above, we show the differences in the evolution of $\Delta r$ and $\Delta \phi$ for the Schwarzschild geometry in both the presence and absence of the pulse in the upper panel of the figure \eqref{Schwarzschild memory plot}. 
The lower panel of the same figure shows that \(\Delta r\) and \(\Delta \phi\) attain larger magnitudes 
for the Schwarzschild geometry ($g=0$) in the presence of the pulse compared to the Bardeen geometry ($g=0.7$), for a fixed \(L_c\). This behavior is consistent with our earlier conclusion that, for $g=0$, both $\Delta r$ and $\Delta \phi$ are larger compared to other $g$ values at $u \to \infty$. The trajectory differences at large positive \(u\) show a distinguishable measure between regular and singular black hole spacetimes. This demonstrates that the displacement memory effect might be used as an independent observational probe of black hole geometry.
\
\subsection{Analysis for the Hayward spacetime}

We now move on to the Hayward regular black hole, which, like the Bardeen spacetime, reduces to Schwarzschild geometry for $g=0$, and is flat for $M=0$.
The function $f(r)$ for the Hayward spacetime is \cite{PhysRevLett.96.031103}
\begin{equation}
    f (r) = 1- \frac{2}{r} \left[\frac{M}{\left(1+\left(\frac{g}{r}\right)^3\right)}\right].
    \label{Hayward profile}
\end{equation}
Ae before, we solve Eqs. \eqref{double derivative of r as function of u}–\eqref{Equation of q} using \eqref{Hayward profile}, with and without the pulse. The resulting geodesics as functions of $u$ are obtained numerically and shown below in Fig. \eqref{geodesic plots as a function of u for Hayward profile}.

\begin{figure}[H]
\centering
%\qquad
\includegraphics[width=0.45\textwidth]{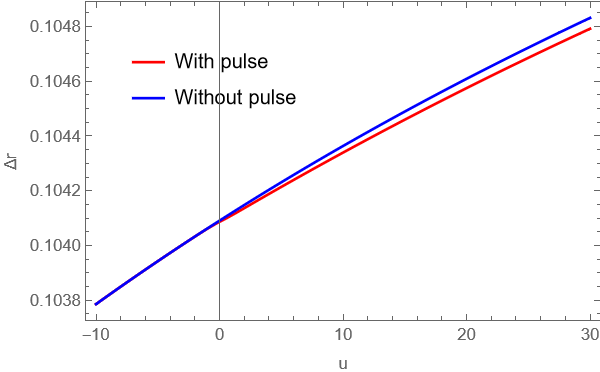}
\includegraphics[width=0.45\textwidth]{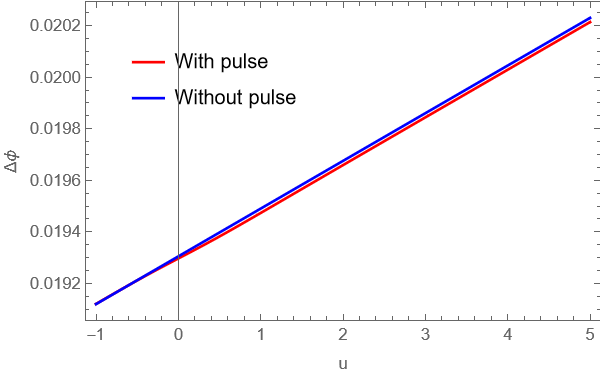}
\caption{\textbf{ Left panel:} Variation of $\Delta r$ and $\Delta \phi$ with $u$ in the presence of pulse and without pulse for the Hayward profile. The parameters are: $L_c=0.1$, $g=0.1$, and $A=1$
\label{displacement memory and velocity memory for Hayward with pulse}}
\end{figure}

To study displacement memory in the Hayward spacetime with a pulse, we solve the geodesic equations for two different timelike geodesics and use Eq. \eqref{geodesic difference governing equation} to capture their evolution as functions of $u$. The initial conditions, implemented at $u=-100$, are the same as in the Bardeen case. Figure \eqref{displacement memory and velocity memory for Hayward with pulse} shows the radial ($\Delta r$) and angular ($\Delta \phi$) separations of two geodesics, with and without the pulse, as functions of $u$. 
% The deviation from the no-pulse case becomes significant near $u=0$, where the pulse is significantly present. Although the pulse amplitude decays rapidly beyond $u=0$, the difference in separation between the two scenarios persists at large $u$, indicating displacement memory.
The Hayward profile exhibits qualitative features similar to the Bardeen case. However, the memory contribution from the wave pulse is more in magnitude in $\Delta r$ and $\Delta \phi$, a feature we discuss more later on. The critical value of $g^2/M^2$ remains $48/81$ for both types of spacetimes,
Bardeen and Hayward.
\begin{figure}[H]
\centering
%\qquad
\includegraphics[width=0.45\textwidth]{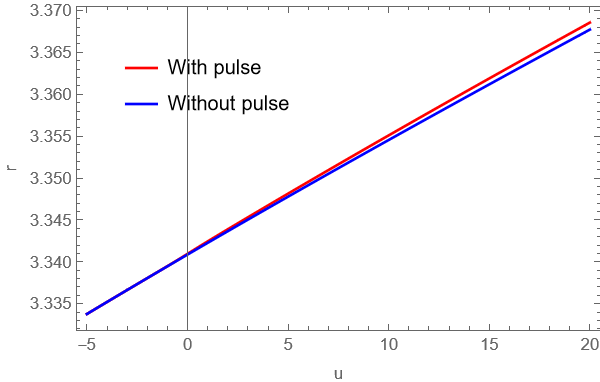}
\includegraphics[width=0.45\textwidth]{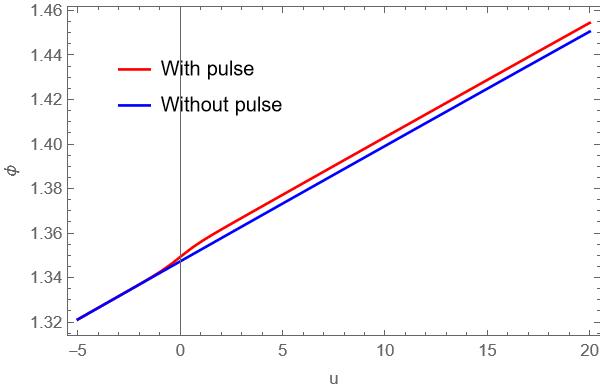}
\includegraphics[width=0.45\textwidth]{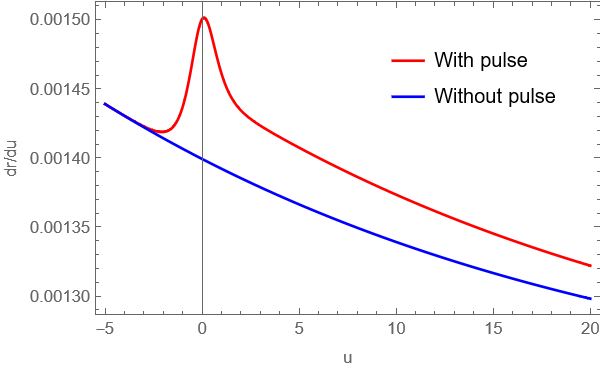}
\includegraphics[width=0.45\textwidth]{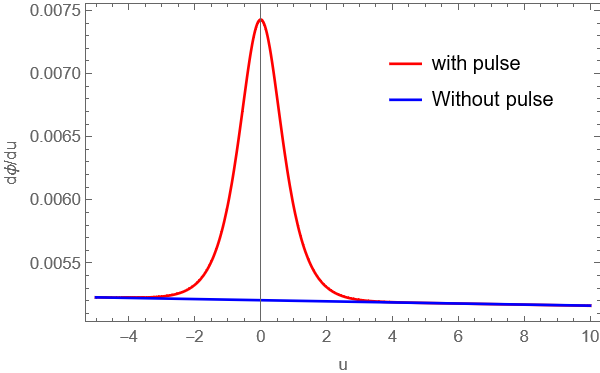}
\caption{\textbf{Top panel:} $r$ and $\phi$ as functions of $u$ in the presence 
and without pulse, for the Hayward Profile \textbf{ Bottom panel:} Variation of $dr/du$ and $d\phi/du$ with $u$ for the Hayward profile, with and without the pulse. The parameters are: $L_c= 0.1$, $g=0.1$ and $A=1$. 
\label{geodesic plots as a function of u for Hayward profile}}
\end{figure}

\begin{figure}[H]
\centering
%\qquad
\includegraphics[width=0.45\textwidth]{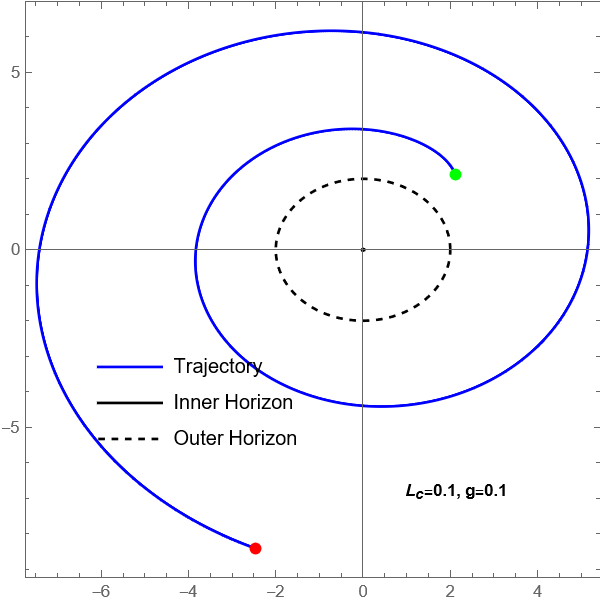}
\includegraphics[width=0.45\textwidth]{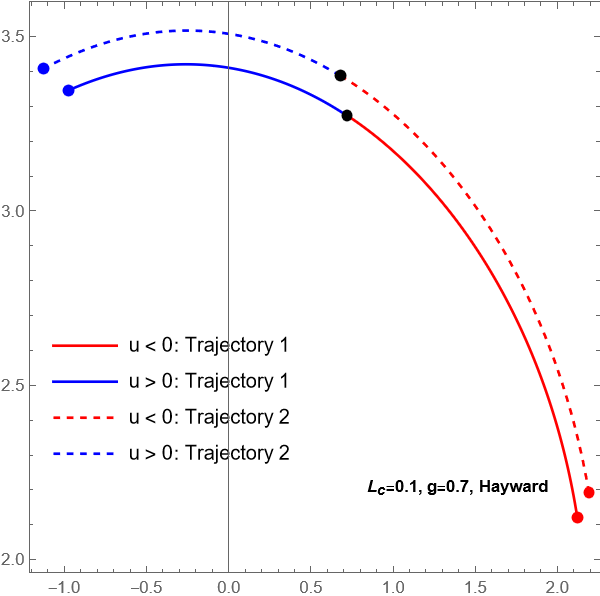}
\caption{\textbf{Left Panel}: Trajectory in Hayward spacetime in the presence of the pulse. The green and red dots denote $u=-100$ and $u=5000$, respectively, while the black dotted curve indicates the outer horizon. \textbf{Right panel}: Trajectories in the presence of the pulse, for $u<0$ (red) and $u>0$ (blue). The red and blue dots correspond to $u=-100$ and $u=100$, respectively, and the black dot marks $u=0$, where the pulse is strongest.
\label{Hayward trajectory plot}}
\end{figure}

Analogous to the Bardeen case, we now examine the trajectories in the Hayward spacetime. The left panel of Fig. \eqref{Hayward trajectory plot} shows the trajectory of the particle with the horizon of the spacetime. The trajectories remain outside the horizon and do not form closed orbits. Their dependence on $g$ and $L_c$ follows the same trend as in the Bardeen case. The right panel shows two trajectories in the presence of the pulse; their difference is small at $u=-100$ but increases at $u=100$ due to the pulse acting near $u=0$ (black dot), illustrating the memory effect. The pulse acts significantly only near $u=0$, while its influence is negligible in other regions.

\begin{figure}[h]
\centering
%\qquad
\includegraphics[width=0.45\textwidth]{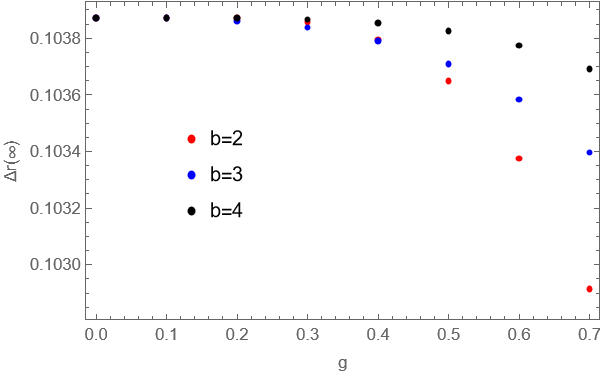}
\includegraphics[width=0.45\textwidth]{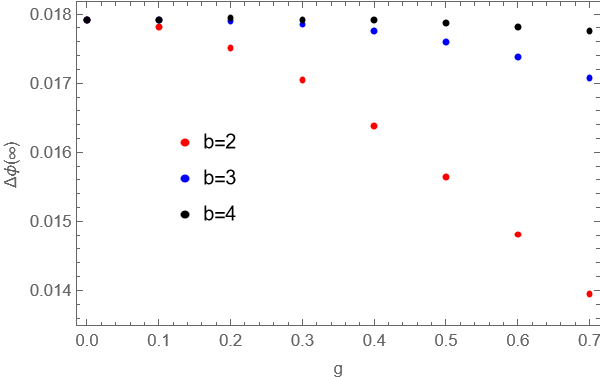}
\includegraphics[width=0.45\textwidth]{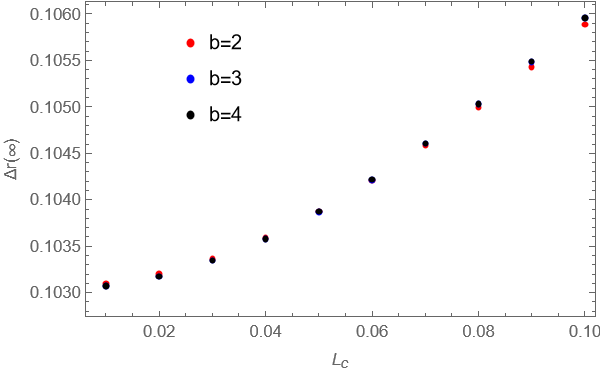}
\includegraphics[width=0.45\textwidth]{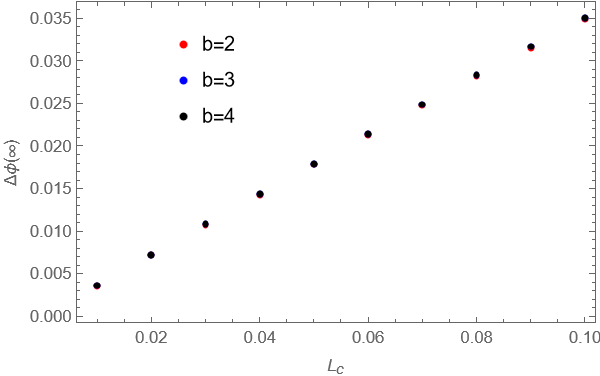}
\caption{\textbf{Top Panel:} Variation of the $\Delta r$ and $\Delta \phi$ calculated at large $u$ with $g$ for different values of $b$. The chosen parameters are: $L_c =0.05$ and $A=1$
\textbf{Bottom Panel:} Variation of the $\Delta r$ and $\Delta \phi$ calculated at large $u$ with $L_c$ for different values of $b$. The chosen parameters are: $g=0.1$ and $A=1$
 \label{memory dependence on b}}
\end{figure}
\noindent Both the Bardeen ($b=2$) and Hayward ($b=3$) profiles follow from the general mass profile proposed by Neves and Saa, given in Eq. \eqref{regular_mass_profile}. Setting $g=0$ in Eq. \eqref{regular_mass_profile} reduces all these geometries to the Schwarzschild spacetime. As a consistency check, we compute the 
coordinate differences at large $u$ ($u=100$) for the Hayward case with $g=0$. We then compare these results with the corresponding Schwarzschild values. The match between them confirms the validity of our numerical analysis for the memory effect analysis. Next, we examine the dependence of the displacement memory on the parameter $b$ of \eqref{regular_mass_profile} in the figure \eqref{memory dependence on b}.
Our inferences from the figure \eqref{memory dependence on b} are mentioned below.
\begin{itemize}
    \item At large $u$, the differences between geodesics vary with $g$ and $L_c$ in the same qualitative way. This behavior is the same for all spacetime families derived from the Neves--Saa profile.
    \item The analysis of the figure \eqref{memory dependence on b} indicates a clear trend: as the parameter $b$ decreases, the difference between geodesics ($r$ and $\phi$ components) at large $u$ decreases. Hence, the memory effect is expected to be more prominent for the Hayward spacetime compared to the Bardeen spacetime for the same set of parameters. For higher values of $g$, this difference is much more prominent. The difference is less at low $g$, whereas at higher $g$ it is visually distinctive from the plot.
    \item For $g=0$, $\Delta r(\infty)$ coincides exactly for different geometries (different $b$ values). This is expected, since for $g=0$ the family of regular black hole spacetimes described by Eq. \eqref{regular_mass_profile} reduces to the Schwarzschild geometry.
    \item The change in geodesic separation with respect to the regularization parameter $g$ is more pronounced for the Bardeen profile than for other regular black holes in this family.
    \item Thus, for probing regular black holes with GW memory, the effect is expected to be prominent at larger $b$ and $L_c$ and smaller $g$.
\end{itemize} 
\subsection{Analysis for a different class of regular black holes} 
We now choose a different metric function $f(r)$ \cite{Kar:2023dko},
\begin{equation}
    f(r) = 1 - \frac{b_0^2 r^2}{(r^2 + g^2)^2}
    \label{Kar-Kar Profile}
\end{equation}
Where $b_0$ is a constant and $g$ is the regular parameter.

\begin{figure}[H]
\centering
%\qquad
\includegraphics[width=0.45\textwidth]{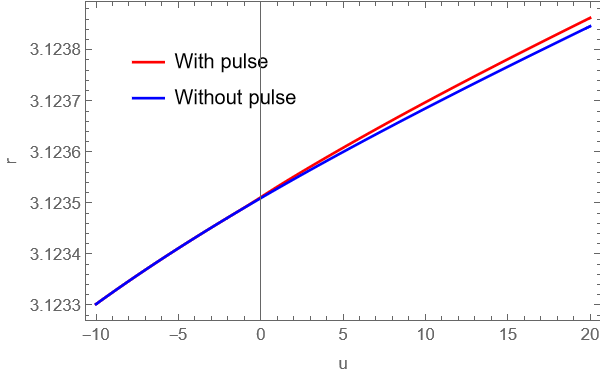}
\includegraphics[width=0.45\textwidth]{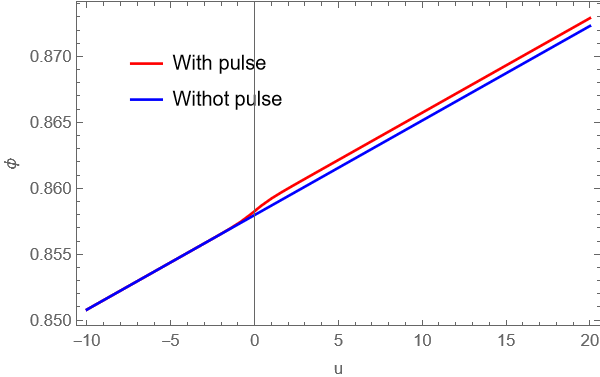}
\caption{$r$ and $\phi$ components as functions of $u$ in the presence and absence of the pulse for profile \eqref{Kar-Kar Profile}. The parameters are: $L_c= 0.01$, $g=0.1$, $b_0=1$ and $A=1$.}
\label{geodesic plots as a function of u for kar-kar profile}
\end{figure}
In this $f(r)$ too, the parameter $g$ influences the behavior of the spacetime geometry. However, the difference to note is that this metric does not fall within the same class as the regular black hole spacetimes developed by Bardeen, Hayward, or Simpson-Visser. Such classic regular black hole solutions typically reduce to the well-known Schwarzschild geometry in the absence of the regularizing parameter— i.e., when $g^2=0$. In contrast, the $f(r)$ above shows a different limiting behavior. When $g^2=0$, instead of recovering the singular Schwarzschild solution, the spacetime reduces to a singular and deformed version of the Reissner–Nordstrom (RN) geometry \cite{Kar:2023dko}. It can be shown that for $g^2/b_0^2 > 0.25$, there are no horizons in the geometry. 
Since we are not interested in geometries with no horizon, we have constrained the values of $g$ to $g=0.50$. We also choose $b_0=1$. 
\noindent First, we solve the coupled equations of motion using \eqref{double derivative of r as function of u}--\eqref{Equation of q}, using the $f(r)$ from Eq. \eqref{Kar-Kar Profile} and $H(u)$ given in Eq. \eqref{pulse profile}. Our numerical solutions are presented in the figure \eqref{geodesic plots as a function of u for kar-kar profile}

\begin{figure}[H]
\centering
%\qquad
\includegraphics[width=0.45\textwidth]{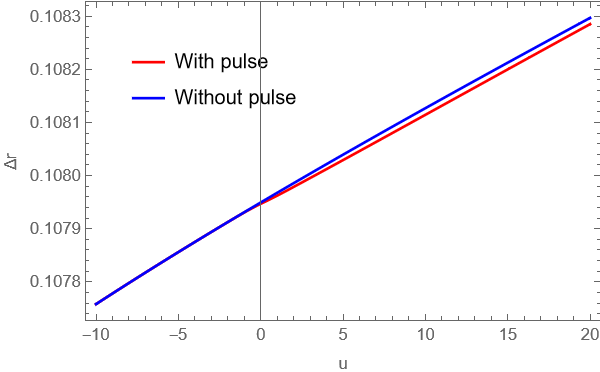}
\includegraphics[width=0.45\textwidth]{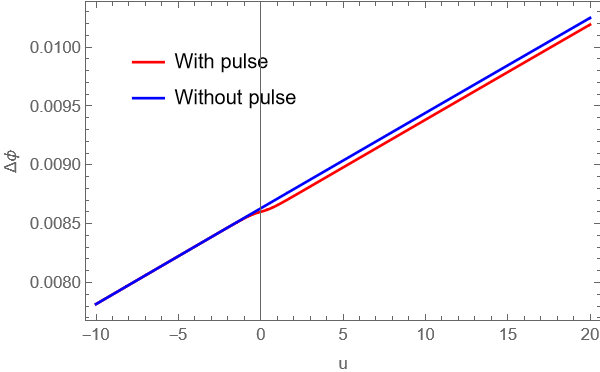}
\caption{\textbf{left panel:} Variation of $\Delta$r with $u$ in the presence of pulse and without pulse \textbf{right panel:} Variation of $\Delta$ $\phi$ with $u$ in the presence of pulse and without pulse. The chosen parameters are: $L_c=0.01$, $g=0.1$, and $A=1$.
\label{displacement and velocity memory for Kar-Kar profile}}
\end{figure} 

\begin{figure}[h]
\centering
%\qquad
\includegraphics[width=0.45\textwidth]{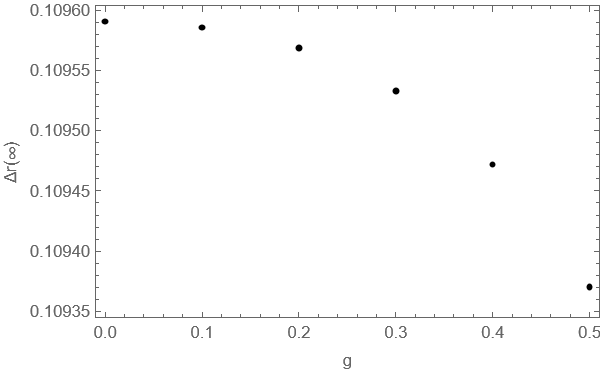}
\includegraphics[width=0.45\textwidth]{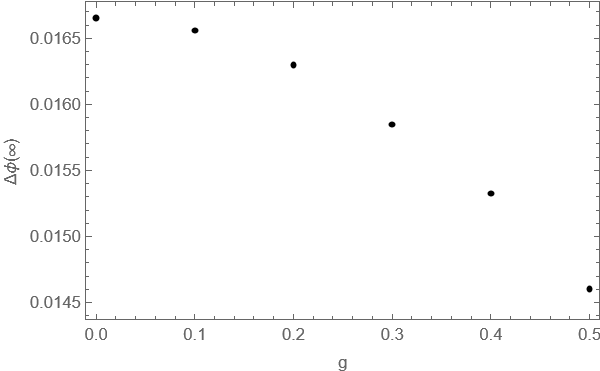}
\includegraphics[width=0.45\textwidth]{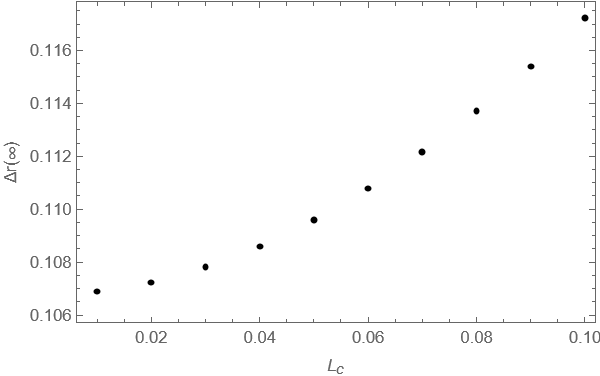}
\includegraphics[width=0.45\textwidth]{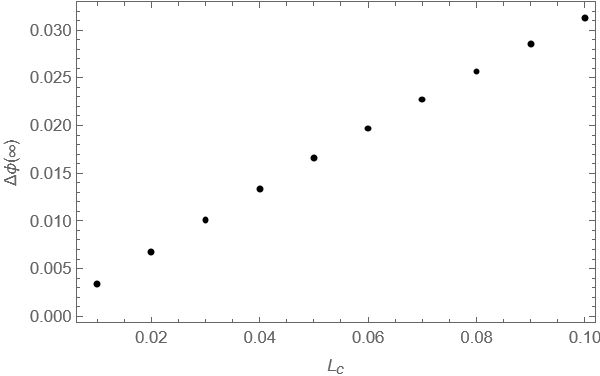}
\caption{\textbf{Top Panel :} Variation of $\Delta r$ and $\Delta \phi$ at large $u$ with the regular parameter (g) for the profile \eqref{Kar-Kar Profile}. The chosen parameters are: $L_c=0.05$, $b_0=1$, and $A=1$.
\textbf{Bottom Panel :} Variation of $\Delta r$ and $\Delta \phi$ at large $u$ with $L_c$. The chosen parameters are: $g=0.1$, $b_0=1$, and $A=1$.}
\label{memory dependence on the g and l for Kar-Kar profile}
\end{figure}

For the analysis of the displacement memory effect, the same methodology as described in Section 3 is employed. Figure \eqref{displacement and velocity memory for Kar-Kar profile} illustrates the separation between two neighboring geodesics as a function of $u$, comparing cases both with and without the presence of a wave pulse. Here too, the displacement memory effect is prominent. In contrast to the profiles discussed earlier, the geodesic separation in the $\phi$ component is notably more prominent (in magnitude) when the GW pulse is present. Examining the magnitudes of the geodesic separation, it is evident that the memory effect 
is larger for this profile compared to that in the previously discussed class of solutions. 
This might suggest that the curvature properties of this geometry tend to enhance the permanent 
displacement induced by the GW pulse, resulting in an increased displacement memory. To further illustrate the above feature, we explore how $\Delta r$ and $\Delta \phi$ at large $u$ change with the regularization parameter $g$ and $L_c$ in the figure \eqref{memory dependence on the g and l for Kar-Kar profile}.

A few points corresponding to the geodesic separation analysis for this profile 
are listed below:
 \begin{itemize}
     \item The differences between two geodesics in \(r\) and \(\phi\) at large \(u\) decrease 
     with increasing \(g\), whereas they increase with larger values of \(L_c\). The same trend has been observed for other regular black hole profiles as well, which we discussed earlier. Similar behaviour is noted for both $L_c$ and $g$
     for other values ($0.1-10$) of the pulse amplitude. 
     \item  From the figure \eqref{memory dependence on the g and l for Kar-Kar profile} it is evident that the asymptotic values of the corresponding memories are most prominent for the $g=0$. We note that for $L_c =0.05$ and pulse amplitude $A=1$, the magnitudes of the separation at large $u$ are: 
     \begin{align*}
           \Delta r (\infty) &= 0.1096 & \Delta \phi (\infty) &= 0.0167
     \end{align*}
    This is different from the results for Schwarzschild geometry.  It is expected because the geometry does not reduce to Schwarzschild for $g=0$. For the Neves--Saa cases, the magnitudes of the separation (for $g=0$) of the corresponding components at large $u$ are
    \begin{align*}
    \Delta r (\infty) &= 0.1039 & \Delta \phi (\infty) &= 0.0179.
    \end{align*} 
    \item Thus, the memory effect carries distinct signatures in different regular black hole models, thereby providing a way to distinguish between them.
 \end{itemize} 
\section{Geodesic deviation and displacement memory}

\noindent Recall the pulsed metric (in the coordinate basis) given as \eqref{general metric with pulse},
\begin{equation}
     ds^2=-f(r)du^2 - 2 du dr+(r^2+ r H(u))d\theta^2+(r^2 - r H(u))\sin^2 \theta d\phi^2
    \label{general metric with pulse1}
\end{equation}
To carry out a deviation analysis, we write the deviation equation in a chosen tetrad basis as follows. The tetrad is given as: 
\begin{align}
    e^0 &= \sqrt{f(r)} du + \frac{dr}{\sqrt{f(r)}} & e^1 &= \frac{dr}{\sqrt{f(r)}} & e^2 &= \sqrt{r^2 + r H(u)} d\theta & e^3 &= \sqrt{r^2 - r H(u)} \sin \theta d\phi,
    \label{tetrad basis}
\end{align}
and the coordinate basis is given as,
\begin{align}
    e^{0^{\prime}} &= du & e^{1^{\prime}} &= dr & e^{2^{\prime}} &= d\theta & e^{3^{\prime}} &= d\phi
    \label{coordinate basis}
\end{align}
Furthermore, we restrict our attention to geodesics lying in the equatorial plane, i.e., $\theta = \pi/2$. The deviation vector components are computed with respect to the numerically obtained geodesics, as discussed in the earlier sections of this article. The explicit expressions for the non-vanishing components of the Riemann tensor are provided in Appendix A. For the geodesic deviation equations, the indices 
${0,1,2,3}$ correspond to the coordinates in tetrad basis, which we adopt for our analysis.

In the previous section, we derived and solved the geodesic equations in the $u$ coordinate. Following the same approach, we now express the geodesic deviation equations in terms of $u$, making use of \eqref{u dot square} and \eqref{u double dot in u coordinate}. All relevant Riemann tensor components must be evaluated along the numerically obtained geodesics. Accordingly, we rewrite the deviation equations in the $ u$ coordinate and solve them numerically for different parameter choices corresponding to the different geometries. The deviation equation reduces to the following set of equations,
\begin{align}
   \frac{d^2 \eta^0}{d u^2}
    + \frac{d \eta^0}{ du} \left( \frac{f^{\prime} (r)}{2}
    + \frac{f(r)+ 2 p(u)}{1+ \tfrac{L_c^2}{r^2 - rH}}
      \frac{H-2r}{2} \frac{L_c^2}{(r^2-rH)^2}\right)
    &= -R_{003}^{0} \eta^0 q(u)^2 
       -R_{013}^{0} \eta^1 q(u)^2 
       + R_{003}^{0} \eta^3 q(u) 
       + R_{013}^{0} \eta^3 p(u) q(u) 
       \label{deviation u} \\[1ex]
    \frac{d^2 \eta^1}{d u^2}
    + \frac{d \eta^1}{ du} \left( \frac{f^{\prime} (r)}{2}
    + \frac{f(r)+ 2 p(u)}{1+ \tfrac{L_c^2}{r^2 - rH}}
      \frac{H-2r}{2} \frac{L_c^2}{(r^2-rH)^2}\right)
    &= -R_{001}^{1} \eta^0 p(u) -R_{303}^{1} \eta^0 q(u)^2 \notag \\
    &\quad -R_{010}^{1} \eta^1 -R_{313}^{1} \eta^1 q(u)^2
           -R_{330}^{1} \eta^3 q(u) -R_{331}^{1} \eta^3 p(u)q(u) \label{deviation r}
   \end{align}
% \begin{align}
%     \frac{d^2 \eta^0}{d u^2}+ \frac{d \eta^0}{ du} \left( \frac{f^{\prime} (r)}{2}+ \frac{f(r)+ 2 p(u)}{1+ \frac{L_c^2}{r^2 - rH}} \frac{H-2r}{2} \frac{L_c^2}{(r^2-rH)^2}\right)&=-R_{003}^{3} \eta^0 q(u)^2 -R_{013}^{3} \eta^1 q(u)^2 + R_{003}^{3} \eta^3 q(u) + R_{013}^{3} \eta^3 p(u) q(u) \label{deviation u} \\
%     \frac{d^2 \eta^1}{d u^2}+ \frac{d \eta^1}{ du} \left( \frac{f^{\prime} (r)}{2}+ \frac{f(r)+ 2 p(u)}{1+ \frac{L_c^2}{r^2 - rH}} \frac{H-2r}{2} \frac{L_c^2}{(r^2-rH)^2}\right)&= -R_{001}^{1} \eta^0 p(u) -R_{303}^{1} \eta^0 q(u)^2 -R_{010}^{1} \eta^1 -R_{313}^{1} \eta^1 q(u)^2-R_{330}^{1} \eta^3 q(u) -R_{331}^{1} \eta^3 p(u)q(u) 
% \end{align}
\begin{align}
    \frac{d^2 \eta^2}{d u^2}
    + \frac{d \eta^2}{ du} \left( \frac{f^{\prime} (r)}{2}
    + \frac{f(r)+ 2 p(u)}{1+ \tfrac{L_c^2}{r^2 - rH}}
      \frac{H-2r}{2} \frac{L_c^2}{(r^2-rH)^2}\right)
    &= -R_{020}^{2} \eta^2 -R_{021}^{2} \eta^2 p(u) 
       -R_{120}^{2} \eta^2 p(u) \notag \\
    &\quad -R_{121}^{2} \eta^2 p(u)^2 -R_{323}^{2} \eta^2 q(u)^2 \label{deviation theta}\\
    \frac{d^2 \eta^3}{d u^2}
    + \frac{d \eta^3}{ du} \left( \frac{f^{\prime} (r)}{2}
    + \frac{f(r)+ 2 p(u)}{1+ \tfrac{L_c^2}{r^2 - rH}}
      \frac{H-2r}{2} \frac{L_c^2}{(r^2-rH)^2}\right)
    &= -R_{003}^{3} \eta^0 q (u)-R_{103}^{3} \eta^0 p(u) q(u)  -R_{013}^{3} \eta^1 q(u) -R_{113}^{3} \eta^1 p(u) q(u)  \notag \\
    &\quad -R_{030}^{3} \eta^3 -R_{031}^{3} \eta^3 p(u) - R_{130}^{3} \eta^3 p(u) - R_{131}^{3} \eta^3 p(u)^2.\label{deviation phi}
\end{align}
Here, $p(u) = \tfrac{dr}{du}$ and $q(u) = \tfrac{d\phi}{du}$. 
The functions $r(u)$, $\phi(u)$, $p(u)$, and $q(u)$ have already been obtained 
in the geodesic analysis for different geometries. 
We now use those numerical solutions to integrate 
Eqs.\eqref{deviation u}–\eqref{deviation phi}, using appropriate initial conditions 
and parameter choices for different geometries. The solutions obtained in the tetrad basis are then transformed back into the coordinate basis in order to facilitate a direct qualitative comparison with the earlier geodesic analysis, which was performed in the coordinate basis. The numerical results for the $r$ and $\phi$ components of the deviation vector are presented in the following sections, where we emphasize the comparative features of the deviation behavior across different spacetime geometries.
 
\subsection{Flat Metric}
We solve Eqs.~\eqref{deviation u}–\eqref{deviation phi} for the flat metric, where 
$f(r)=1$. The geodesics for this case have already been computed numerically in the 
earlier section. We now use those solutions to first write down and then
numerically integrate the deviation equation. The initial conditions are listed in Table~\ref{tab:init_flat}. These conditions are 
chosen to enable a qualitative comparison between the $r$ and $\phi$ components of 
the geodesic separation and the geodesic deviation vector. 
% In Section 4, the same difference between two geodesics was used to study the memory 
% effect.
\begin{figure}[H]
\centering
%\qquad
\includegraphics[width=0.45\textwidth]{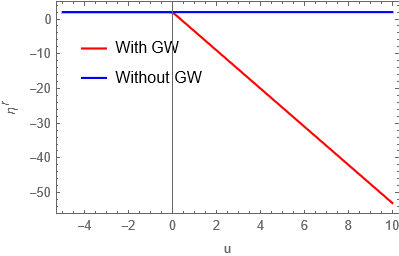}
\includegraphics[width=0.45\textwidth]{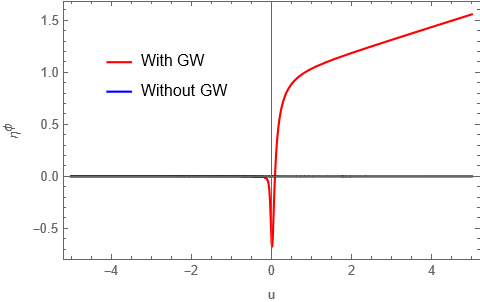}
\caption{$r$ and $\phi$ components of the deviation vector in the flat metric, both in the presence and in the absence of the pulse. The parameters are: $L_c=0.005, A=3$.}
 \label{deviation for flat}
\end{figure}
\begin{table}[h!]
\centering

\begin{tabular}{c|c}
\hline
Quantity & Value at $u=-100$ \\
\hline
$\eta^u$, $\eta^{\theta}$, $\eta^{\phi}$ & $0$ \\
$\dfrac{d\eta^u}{du}$, $\dfrac{d\eta^r}{du}$, $\dfrac{d\eta^{\theta}}{du}$, $\dfrac{d\eta^{\phi}}{du}$ & $0$ \\
$\eta^r$ & $2$ \\
\hline
\end{tabular}
\caption{Initial conditions for solving the deviation equations in the flat metric.}
\label{tab:init_flat}
\end{table}
 Figure~\eqref{deviation for flat} shows the numerical results for the $r$ and $\phi$ 
components of the deviation vector as functions of $u$. The qualitative behaviour of 
the $r$-component is consistent with that obtained from the geodesic separation analysis, 
whereas the $\phi$-component exhibits distinct features in the transient region. Here, we observe that the corresponding components of the deviation vector begin 
to evolve differently in the two scenarios—with and without the GW pulse—from $u=0$, 
the region where the effect of the pulse is introduced. At large values of $u$, the deviation vector exhibits a lasting difference between the two cases, which corresponds to the displacement memory effect. 

\subsection{Bardeen Profile}

\begin{figure}[h]
\centering
%\qquad
\includegraphics[width=0.45\textwidth]{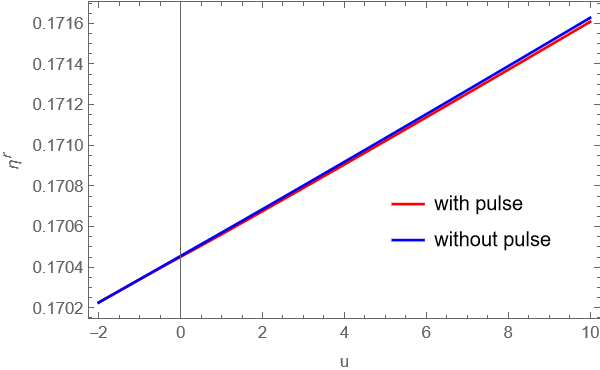}
\includegraphics[width=0.45\textwidth]{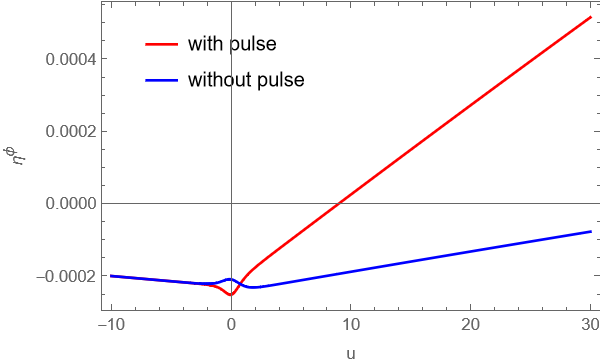}
\caption{$r$ and $\phi$ component of the deviation vector in case of Bardeen metric, both in the presence and in the absence of the pulse. The parameters are: $g=0.1, L_c = 0.1, A=1, m=1$}
\label{memory for Bardeen metric via geodesic deviation analysis}
\end{figure}
\noindent For the Bardeen profile, $f(r)$ is given by \eqref{Bardeen profile f(r)}. We plot the $r$ and $\phi$ components of the deviation vectors in the figure \eqref{memory for Bardeen metric via geodesic deviation analysis}. The initial conditions are the same as in the flat metric case 
(Table \ref{tab:init_flat}), except for $\eta^r[-100] = 0.1$. Here, also, the $r$-component shows qualitative similarity with the separation vector $\Delta r$, while the $\phi$-component exhibits a different behaviour, Fig. \eqref{displacement memory and velocity memory as function of u for Bardeen profile}.

\begin{figure}[h]
\centering
%\qquad
\includegraphics[width=0.45\textwidth]{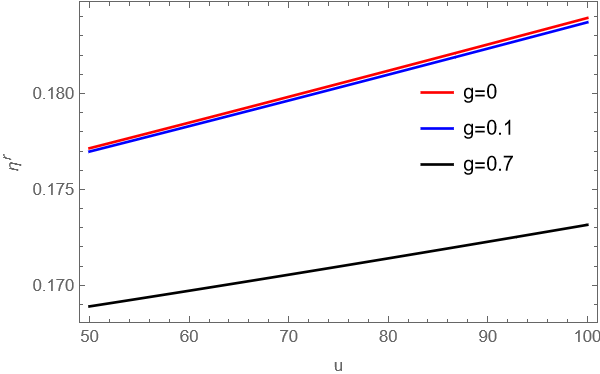}
\includegraphics[width=0.45\textwidth]{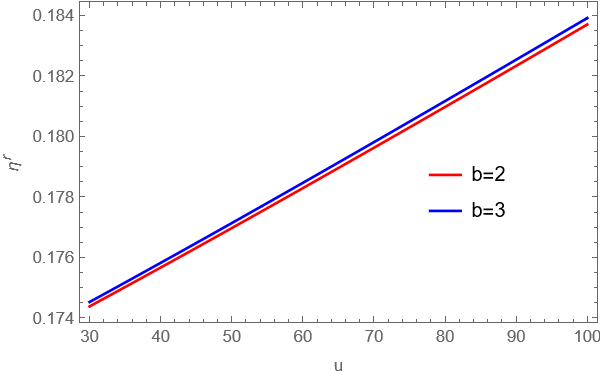}
\caption{\textbf{Left panel:} $r$ component of deviation vector for different values of $g$ with $L_c=0.1$ and $A=1$. \textbf{Right panel} $r$ component of deviation vector for different values of $b$ with $L_c=0.1$, $g=0.1$, and $A=1$.}
\label{memory dependence on the parameters via deviation analysis}
\end{figure}
In Fig.~\eqref{memory dependence on the parameters via deviation analysis}, we show the 
$r$-component of the deviation vector for different values of $g$ and $b$. In the 
geodesic separation analysis, we found that $\Delta r$ at large $u$ for the Schwarzschild case 
($g=0$), is larger compared to that for the regular black hole 
cases. The left panel of Fig.~\eqref{memory dependence on the parameters via deviation analysis} 
confirms this behaviour, showing that the $r$-component of the deviation vector is indeed larger
in magnitude for $g=0$ than for nonzero $g$. Furthermore, the memory effect is smaller 
for nonzero lower values of $g$. The dependence on $b$ is also shown in the right panel, where 
larger values of $b$ correspond to larger values of $\eta^r$. Overall, the dependence 
of the memory on these parameters confirms that the deviation analysis and the geodesic 
analysis are qualitatively consistent, as they are expected to be.
We note that while the radial (\(r\)) component of the separation and the deviation vector exhibit nearly identical qualitative behavior, the $\phi$ component displays a transient that differs between the two approaches. This discrepancy may be understood as follows. The geodesic separation method involves two geodesics that are not necessarily infinitesimally close whereas the deviation--vector analysis always assumes infinitesimal deviations about a central geodesic. Thus, while both the deviation--vector and geodesic--separation methods yield the same qualitative behaviour at large $u$, they differ at small $u$, as manifest in the $\eta^\phi$ component.

\section{Conclusion}

In this work, we have investigated the displacement memory effect in regular black hole 
spacetimes using geodesic separation and geodesic deviation. Our main findings may be summarized 
as follows:  

\begin{enumerate}
    \item \textbf{Studies on the memory effect:}  
    \begin{itemize}
        \item The separation between two geodesics after the pulse passes differs between the cases with and without the pulse. This demonstrates that the wave pulse leaves a permanent imprint, known as displacement memory. Both geodesic separation and geodesic deviation analyses consistently show that a wave pulse induces a permanent displacement at large $u$.
        \item The qualitative behavior of the $r$ coordinate at large $u$ is in good agreement between the two approaches, whereas the transient behavior of the $\phi$ component differs.
    \end{itemize}

    \item \textbf{Parameter dependence:}  
    \begin{itemize}
        \item  Displacement memory depends on the regularization parameter \(g\), 
        the angular momentum \(L_c\), and the pulse profile parameter \(b\).  
        \item The magnitude of displacement memory decreases with increasing \(g\), while it increases with larger \(L_c\) and \(b\).  
        \item The Schwarzschild limit (\(g=0\)) yields the largest memory, marking a clear distinction between singular and regular black hole geometries. This suggests that the memory effect can serve as a tool to distinguish regular black hole models from the singular Schwarzschild spacetime. It may also be used to differentiate between various regular black hole models.
    \end{itemize}

    \item \textbf{Comparison across regular black hole geometries:}  
    \begin{itemize}
        \item The Bardeen, Hayward, and related Neves--Saa profiles exhibit distinct signatures which are manifest through the magnitude of displacement memory.  
        \item A different class of regular black holes \eqref{Kar-Kar Profile} shows even stronger memory effects, highlighting how curvature variations
        also influence permanent displacements.  
        \item Such characteristic variations indicate that memory could be exploited to distinguish between different families of regular black holes.  
    \end{itemize}
\end{enumerate}
From both geodesic separation and geodesic deviation analyses, we find that the memory effect depends on the regularization parameter $g$. Consequently, following the Bondi-Sachs analysis, the Bondi functions are expected to reflect this dependence. To observe the influence of $g$ on the Bondi mass and parameters, the asymptotic expansion of the mass function \eqref{regular_mass_profile} must include $g$ in the $1/r$ or $1/r^2$ terms:
\begin{align}
m(r) &\simeq M \Bigg[1 - \left(\frac{a}{b}\right)\left(\frac{g}{r}\right)^b + \frac{a/b (a/b+1)}{2} \left(\frac{g}{r}\right)^{2b} - \dots \Bigg], &
f(r) &= 1 - \frac{2m(r)}{r}.
\end{align}
It follows that $g$ affects the Bondi functions only if $b \le 1$. Choosing $a = b = 1$, the large-$r$ metric becomes
\begin{equation}
ds^2 = - \left(1 - \frac{2M}{r} + \frac{2 M g^2}{r^2}\right) du^2 - 2 du dr + r^2 d\Omega^2.
\label{asymptotic expansion of regular black hole}
\end{equation}
Within the Bondi--Sachs framework, $g$ explicitly enters the subleading asymptotic terms, directly linking the geodesic memory effect to asymptotic flux-balance laws at large $u$. We have performed a complete analysis of the $g$-dependence in the Bondi functions (for $b=1$), but do not report it here, as $g$ contributes only at subleading order for $b=1$. For the regular black holes considered in our work—for example, Bardeen ($b=2$) and Hayward ($b=3$)—the $g$-dependence appears in even higher-order Bondi terms and is therefore negligible. \\
\noindent The present study establishes that gravitational wave memory carries distinct imprints related to the 
background black hole geometry and its regular or singular nature. Looking ahead, several avenues remain open:  
\begin{itemize}
    \item Extending the analysis to \textit{nonlinear (Christodoulou) memory} and 
    \textit{tail effects} could provide a more complete picture of memory in regular spacetimes. A perturbative analysis could reveal higher-order effects, enhancing both theoretical insight and potential observational relevance.
    \item Incorporating \textit{realistic astrophysical sources} of gravitational radiation, such as binary mergers in regular black hole backgrounds, would enhance the observational relevance of these results.   
\end{itemize}
\section*{ACKNOWLEDGEMENTS}

\noindent RA thanks the Indian Institute of Technology Kharagpur, India, for supporting
him through a fellowship allowing access to available computational facilities.

\bibliographystyle{apsrev} %abbrvnat %apsrev
\bibliography{bibfile}

@inbook{Joshi_2014,
   title={Spacetime Singularities},
   ISBN={9783642419928},
   ISSN={2522-8706},
   url={http://dx.doi.org/10.1007/978-3-642-41992-8_20},
   DOI={10.1007/978-3-642-41992-8_20},
   booktitle={Springer Handbook of Spacetime},
   publisher={Springer Berlin Heidelberg},
   author={Joshi, Pankaj S.},
   year={2014},
   pages={409–436} }

@article{NOVELLO_2008,
   title={Bouncing cosmologies},
   volume={463},
   ISSN={0370-1573},
   url={http://dx.doi.org/10.1016/j.physrep.2008.04.006},
   DOI={10.1016/j.physrep.2008.04.006},
   number={4},
   journal={Physics Reports},
   publisher={Elsevier BV},
   author={Novello, M and Bergliaffa, S},
   year={2008},
   month=jul, pages={127–213} }

@article{Neves_2014,
   title={Regular rotating black holes and the weak energy condition},
   volume={734},
   ISSN={0370-2693},
   url={http://dx.doi.org/10.1016/j.physletb.2014.05.026},
   DOI={10.1016/j.physletb.2014.05.026},
   journal={Physics Letters B},
   publisher={Elsevier BV},
   author={Neves, J.C.S. and Saa, Alberto},
   year={2014},
   month=jun, pages={44–48} }

@article{ refId0,
	author = {{Frolov, Valeri P.}},
	title = {Remarks on non-singular black holes},
	DOI= "10.1051/epjconf/201816801001",
	url= "https://doi.org/10.1051/epjconf/201816801001",
	journal = {EPJ Web Conf.},
	year = 2018,
	volume = 168,
	pages = "01001",
}

@article{Kar:2023dko,
    author = "Kar, Anjan and Kar, Sayan",
    title = "{Novel regular black holes: geometry, source and shadow}",
    eprint = "2308.12155",
    archivePrefix = "arXiv",
    primaryClass = "gr-qc",
    doi = "10.1007/s10714-024-03238-4",
    journal = "Gen. Rel. Grav.",
    volume = "56",
    number = "5",
    pages = "52",
    year = "2024"
}

@article{Bondi:1960jsa,
    author = "Bondi, H.",
    title = "{Gravitational Waves in General Relativity}",
    doi = "10.1038/186535a0",
    journal = "Nature",
    volume = "186",
    number = "4724",
    pages = "535--535",
    year = "1960"
}

@ARTICLE{1961RSPSA.264..309S,
       author = {{Sachs}, R.},
        title = "{Gravitational Waves in General Relativity. VI. The Outgoing Radiation Condition}",
      journal = {Proceedings of the Royal Society of London Series A},
         year = 1961,
        month = nov,
       volume = {264},
       number = {1318},
        pages = {309-338},
          doi = {10.1098/rspa.1961.0202},
       adsurl = {https://ui.adsabs.harvard.edu/abs/1961RSPSA.264..309S}
}

@article{Hadi_2024,
   title={Gravitational wave pulse and memory effects for hairy Kiselev black hole and its analogy with Bondi–Sachs formalism},
   volume={41},
   ISSN={1361-6382},
   url={http://dx.doi.org/10.1088/1361-6382/ad3caf},
   DOI={10.1088/1361-6382/ad3caf},
   number={10},
   journal={Classical and Quantum Gravity},
   publisher={IOP Publishing},
   author={Hadi, H and Rezaei Akbarieh, Amin and Mota, David F},
   year={2024},
   month=apr, pages={105005} }

@article{Chakraborty_2022,
   title={A simple analytic example of the gravitational wave memory effect},
   volume={137},
   ISSN={2190-5444},
   url={http://dx.doi.org/10.1140/epjp/s13360-022-02593-y},
   DOI={10.1140/epjp/s13360-022-02593-y},
   number={4},
   journal={The European Physical Journal Plus},
   publisher={Springer Science and Business Media LLC},
   author={Chakraborty, Indranil and Kar, Sayan},
   year={2022},
   month=apr }

@article{Divakarla_2021,
   title={First-order velocity memory effect from compact binary coalescing sources},
   volume={104},
   ISSN={2470-0029},
   url={http://dx.doi.org/10.1103/PhysRevD.104.064001},
   DOI={10.1103/physrevd.104.064001},
   number={6},
   journal={Physical Review D},
   publisher={American Physical Society (APS)},
   author={Divakarla, Atul K. and Whiting, Bernard F.},
   year={2021},
   month=sep }

@article{Favata_2010,
   title={The gravitational-wave memory effect},
   volume={27},
   ISSN={1361-6382},
   url={http://dx.doi.org/10.1088/0264-9381/27/8/084036},
   DOI={10.1088/0264-9381/27/8/084036},
   number={8},
   journal={Classical and Quantum Gravity},
   publisher={IOP Publishing},
   author={Favata, Marc},
   year={2010},
   month=apr, pages={084036} }

@article{Braginsky:1985vlg,
    author = "Braginsky, V. B. and Grishchuk, L. P.",
    title = "{Kinematic Resonance and Memory Effect in Free Mass Gravitational Antennas}",
    journal = "Sov. Phys. JETP",
    volume = "62",
    pages = "427--430",
    year = "1985"
}

@article{Mitman_2024,
   title={A review of gravitational memory and BMS frame fixing in numerical relativity},
   volume={41},
   ISSN={1361-6382},
   url={http://dx.doi.org/10.1088/1361-6382/ad83c2},
   DOI={10.1088/1361-6382/ad83c2},
   number={22},
   journal={Classical and Quantum Gravity},
   publisher={IOP Publishing},
   author={Mitman, Keefe and Boyle, Michael and Stein, Leo C and Deppe, Nils and Kidder, Lawrence E and Moxon, Jordan and Pfeiffer, Harald P and Scheel, Mark A and Teukolsky, Saul A and Throwe, William and Vu, Nils L},
   year={2024},
   month=oct, pages={223001} }

@article{PhysRevD.106.104057,
  title = {Gravitational wave memory in wormhole spacetimes},
  author = {Chakraborty, Indranil and Bhattacharya, Soumya and Chakraborty, Sumanta},
  journal = {Phys. Rev. D},
  volume = {106},
  issue = {10},
  pages = {104057},
  numpages = {12},
  year = {2022},
  month = {Nov},
  publisher = {American Physical Society},
  doi = {10.1103/PhysRevD.106.104057},
  url = {https://link.aps.org/doi/10.1103/PhysRevD.106.104057}
}

@article{PhysRevLett.116.061102,
  title = {Observation of Gravitational Waves from a Binary Black Hole Merger},
  author = {Abbott, B. P. and Abbott, R. and Abbott, T. D. and Abernathy, M. R. and Acernese, F. and Ackley, K. and Adams, C. and Adams, T. and Addesso, P. and Adhikari, R. X. and Adya, V. B. and Affeldt, C. and Agathos, M. and Agatsuma, K. and Aggarwal, N. and Aguiar, O. D. and Aiello, L. and Ain, A. and Ajith, P. and Allen, B. and Allocca, A. and Altin, P. A. and Anderson, S. B. and Anderson, W. G. and Arai, K. and Arain, M. A. and Araya, M. C. and Arceneaux, C. C. and Areeda, J. S. and Arnaud, N. and Arun, K. G. and Ascenzi, S. and Ashton, G. and Ast, M. and Aston, S. M. and Astone, P. and Aufmuth, P. and Aulbert, C. and Babak, S. and Bacon, P. and Bader, M. K. M. and Baker, P. T. and Baldaccini, F. and Ballardin, G. and Ballmer, S. W. and Barayoga, J. C. and Barclay, S. E. and Barish, B. C. and Barker, D. and Barone, F. and Barr, B. and Barsotti, L. and Barsuglia, M. and Barta, D. and Bartlett, J. and Barton, M. A. and Bartos, I. and Bassiri, R. and Basti, A. and Batch, J. C. and Baune, C. and Bavigadda, V. and Bazzan, M. and Behnke, B. and Bejger, M. and Belczynski, C. and Bell, A. S. and Bell, C. J. and Berger, B. K. and Bergman, J. and Bergmann, G. and Berry, C. P. L. and Bersanetti, D. and Bertolini, A. and Betzwieser, J. and Bhagwat, S. and Bhandare, R. and Bilenko, I. A. and Billingsley, G. and Birch, J. and Birney, I. A. and Birnholtz, O. and Biscans, S. and Bisht, A. and Bitossi, M. and Biwer, C. and Bizouard, M. A. and Blackburn, J. K. and Blair, C. D. and Blair, D. G. and Blair, R. M. and Bloemen, S. and Bock, O. and Bodiya, T. P. and Boer, M. and Bogaert, G. and Bogan, C. and Bohe, A. and Bojtos, P. and Bond, C. and Bondu, F. and Bonnand, R. and Boom, B. A. and Bork, R. and Boschi, V. and Bose, S. and Bouffanais, Y. and Bozzi, A. and Bradaschia, C. and Brady, P. R. and Braginsky, V. B. and Branchesi, M. and Brau, J. E. and Briant, T. and Brillet, A. and Brinkmann, M. and Brisson, V. and Brockill, P. and Brooks, A. F. and Brown, D. A. and Brown, D. D. and Brown, N. M. and Buchanan, C. C. and Buikema, A. and Bulik, T. and Bulten, H. J. and Buonanno, A. and Buskulic, D. and Buy, C. and Byer, R. L. and Cabero, M. and Cadonati, L. and Cagnoli, G. and Cahillane, C. and Bustillo, J. Calder\'on and Callister, T. and Calloni, E. and Camp, J. B. and Cannon, K. C. and Cao, J. and Capano, C. D. and Capocasa, E. and Carbognani, F. and Caride, S. and Diaz, J. Casanueva and Casentini, C. and Caudill, S. and Cavagli\`a, M. and Cavalier, F. and Cavalieri, R. and Cella, G. and Cepeda, C. B. and Baiardi, L. Cerboni and Cerretani, G. and Cesarini, E. and Chakraborty, R. and Chalermsongsak, T. and Chamberlin, S. J. and Chan, M. and Chao, S. and Charlton, P. and Chassande-Mottin, E. and Chen, H. Y. and Chen, Y. and Cheng, C. and Chincarini, A. and Chiummo, A. and Cho, H. S. and Cho, M. and Chow, J. H. and Christensen, N. and Chu, Q. and Chua, S. and Chung, S. and Ciani, G. and Clara, F. and Clark, J. A. and Cleva, F. and Coccia, E. and Cohadon, P.-F. and Colla, A. and Collette, C. G. and Cominsky, L. and Constancio, M. and Conte, A. and Conti, L. and Cook, D. and Corbitt, T. R. and Cornish, N. and Corsi, A. and Cortese, S. and Costa, C. A. and Coughlin, M. W. and Coughlin, S. B. and Coulon, J.-P. and Countryman, S. T. and Couvares, P. and Cowan, E. E. and Coward, D. M. and Cowart, M. J. and Coyne, D. C. and Coyne, R. and Craig, K. and Creighton, J. D. E. and Creighton, T. D. and Cripe, J. and Crowder, S. G. and Cruise, A. M. and Cumming, A. and Cunningham, L. and Cuoco, E. and Canton, T. Dal and Danilishin, S. L. and D'Antonio, S. and Danzmann, K. and Darman, N. S. and Da Silva Costa, C. F. and Dattilo, V. and Dave, I. and Daveloza, H. P. and Davier, M. and Davies, G. S. and Daw, E. J. and Day, R. and De, S. and DeBra, D. and Debreczeni, G. and Degallaix, J. and De Laurentis, M. and Del\'eglise, S. and Del Pozzo, W. and Denker, T. and Dent, T. and Dereli, H. and Dergachev, V. and DeRosa, R. T. and De Rosa, R. and DeSalvo, R. and Dhurandhar, S. and D\'{\i}az, M. C. and Di Fiore, L. and Di Giovanni, M. and Di Lieto, A. and Di Pace, S. and Di Palma, I. and Di Virgilio, A. and Dojcinoski, G. and Dolique, V. and Donovan, F. and Dooley, K. L. and Doravari, S. and Douglas, R. and Downes, T. P. and Drago, M. and Drever, R. W. P. and Driggers, J. C. and Du, Z. and Ducrot, M. and Dwyer, S. E. and Edo, T. B. and Edwards, M. C. and Effler, A. and Eggenstein, H.-B. and Ehrens, P. and Eichholz, J. and Eikenberry, S. S. and Engels, W. and Essick, R. C. and Etzel, T. and Evans, M. and Evans, T. M. and Everett, R. and Factourovich, M. and Fafone, V. and Fair, H. and Fairhurst, S. and Fan, X. and Fang, Q. and Farinon, S. and Farr, B. and Farr, W. M. and Favata, M. and Fays, M. and Fehrmann, H. and Fejer, M. M. and Feldbaum, D. and Ferrante, I. and Ferreira, E. C. and Ferrini, F. and Fidecaro, F. and Finn, L. S. and Fiori, I. and Fiorucci, D. and Fisher, R. P. and Flaminio, R. and Fletcher, M. and Fong, H. and Fournier, J.-D. and Franco, S. and Frasca, S. and Frasconi, F. and Frede, M. and Frei, Z. and Freise, A. and Frey, R. and Frey, V. and Fricke, T. T. and Fritschel, P. and Frolov, V. V. and Fulda, P. and Fyffe, M. and Gabbard, H. A. G. and Gair, J. R. and Gammaitoni, L. and Gaonkar, S. G. and Garufi, F. and Gatto, A. and Gaur, G. and Gehrels, N. and Gemme, G. and Gendre, B. and Genin, E. and Gennai, A. and George, J. and Gergely, L. and Germain, V. and Ghosh, Abhirup and Ghosh, Archisman and Ghosh, S. and Giaime, J. A. and Giardina, K. D. and Giazotto, A. and Gill, K. and Glaefke, A. and Gleason, J. R. and Goetz, E. and Goetz, R. and Gondan, L. and Gonz\'alez, G. and Castro, J. M. Gonzalez and Gopakumar, A. and Gordon, N. A. and Gorodetsky, M. L. and Gossan, S. E. and Gosselin, M. and Gouaty, R. and Graef, C. and Graff, P. B. and Granata, M. and Grant, A. and Gras, S. and Gray, C. and Greco, G. and Green, A. C. and Greenhalgh, R. J. S. and Groot, P. and Grote, H. and Grunewald, S. and Guidi, G. M. and Guo, X. and Gupta, A. and Gupta, M. K. and Gushwa, K. E. and Gustafson, E. K. and Gustafson, R. and Hacker, J. J. and Hall, B. R. and Hall, E. D. and Hammond, G. and Haney, M. and Hanke, M. M. and Hanks, J. and Hanna, C. and Hannam, M. D. and Hanson, J. and Hardwick, T. and Harms, J. and Harry, G. M. and Harry, I. W. and Hart, M. J. and Hartman, M. T. and Haster, C.-J. and Haughian, K. and Healy, J. and Heefner, J. and Heidmann, A. and Heintze, M. C. and Heinzel, G. and Heitmann, H. and Hello, P. and Hemming, G. and Hendry, M. and Heng, I. S. and Hennig, J. and Heptonstall, A. W. and Heurs, M. and Hild, S. and Hoak, D. and Hodge, K. A. and Hofman, D. and Hollitt, S. E. and Holt, K. and Holz, D. E. and Hopkins, P. and Hosken, D. J. and Hough, J. and Houston, E. A. and Howell, E. J. and Hu, Y. M. and Huang, S. and Huerta, E. A. and Huet, D. and Hughey, B. and Husa, S. and Huttner, S. H. and Huynh-Dinh, T. and Idrisy, A. and Indik, N. and Ingram, D. R. and Inta, R. and Isa, H. N. and Isac, J.-M. and Isi, M. and Islas, G. and Isogai, T. and Iyer, B. R. and Izumi, K. and Jacobson, M. B. and Jacqmin, T. and Jang, H. and Jani, K. and Jaranowski, P. and Jawahar, S. and Jim\'enez-Forteza, F. and Johnson, W. W. and Johnson-McDaniel, N. K. and Jones, D. I. and Jones, R. and Jonker, R. J. G. and Ju, L. and Haris, K. and Kalaghatgi, C. V. and Kalogera, V. and Kandhasamy, S. and Kang, G. and Kanner, J. B. and Karki, S. and Kasprzack, M. and Katsavounidis, E. and Katzman, W. and Kaufer, S. and Kaur, T. and Kawabe, K. and Kawazoe, F. and K\'ef\'elian, F. and Kehl, M. S. and Keitel, D. and Kelley, D. B. and Kells, W. and Kennedy, R. and Keppel, D. G. and Key, J. S. and Khalaidovski, A. and Khalili, F. Y. and Khan, I. and Khan, S. and Khan, Z. and Khazanov, E. A. and Kijbunchoo, N. and Kim, C. and Kim, J. and Kim, K. and Kim, Nam-Gyu and Kim, Namjun and Kim, Y.-M. and King, E. J. and King, P. J. and Kinzel, D. L. and Kissel, J. S. and Kleybolte, L. and Klimenko, S. and Koehlenbeck, S. M. and Kokeyama, K. and Koley, S. and Kondrashov, V. and Kontos, A. and Koranda, S. and Korobko, M. and Korth, W. Z. and Kowalska, I. and Kozak, D. B. and Kringel, V. and Krishnan, B. and Kr\'olak, A. and Krueger, C. and Kuehn, G. and Kumar, P. and Kumar, R. and Kuo, L. and Kutynia, A. and Kwee, P. and Lackey, B. D. and Landry, M. and Lange, J. and Lantz, B. and Lasky, P. D. and Lazzarini, A. and Lazzaro, C. and Leaci, P. and Leavey, S. and Lebigot, E. O. and Lee, C. H. and Lee, H. K. and Lee, H. M. and Lee, K. and Lenon, A. and Leonardi, M. and Leong, J. R. and Leroy, N. and Letendre, N. and Levin, Y. and Levine, B. M. and Li, T. G. F. and Libson, A. and Littenberg, T. B. and Lockerbie, N. A. and Logue, J. and Lombardi, A. L. and London, L. T. and Lord, J. E. and Lorenzini, M. and Loriette, V. and Lormand, M. and Losurdo, G. and Lough, J. D. and Lousto, C. O. and Lovelace, G. and L\"uck, H. and Lundgren, A. P. and Luo, J. and Lynch, R. and Ma, Y. and MacDonald, T. and Machenschalk, B. and MacInnis, M. and Macleod, D. M. and Maga\~na-Sandoval, F. and Magee, R. M. and Mageswaran, M. and Majorana, E. and Maksimovic, I. and Malvezzi, V. and Man, N. and Mandel, I. and Mandic, V. and Mangano, V. and Mansell, G. L. and Manske, M. and Mantovani, M. and Marchesoni, F. and Marion, F. and M\'arka, S. and M\'arka, Z. and Markosyan, A. S. and Maros, E. and Martelli, F. and Martellini, L. and Martin, I. W. and Martin, R. M. and Martynov, D. V. and Marx, J. N. and Mason, K. and Masserot, A. and Massinger, T. J. and Masso-Reid, M. and Matichard, F. and Matone, L. and Mavalvala, N. and Mazumder, N. and Mazzolo, G. and McCarthy, R. and McClelland, D. E. and McCormick, S. and McGuire, S. C. and McIntyre, G. and McIver, J. and McManus, D. J. and McWilliams, S. T. and Meacher, D. and Meadors, G. D. and Meidam, J. and Melatos, A. and Mendell, G. and Mendoza-Gandara, D. and Mercer, R. A. and Merilh, E. and Merzougui, M. and Meshkov, S. and Messenger, C. and Messick, C. and Meyers, P. M. and Mezzani, F. and Miao, H. and Michel, C. and Middleton, H. and Mikhailov, E. E. and Milano, L. and Miller, J. and Millhouse, M. and Minenkov, Y. and Ming, J. and Mirshekari, S. and Mishra, C. and Mitra, S. and Mitrofanov, V. P. and Mitselmakher, G. and Mittleman, R. and Moggi, A. and Mohan, M. and Mohapatra, S. R. P. and Montani, M. and Moore, B. C. and Moore, C. J. and Moraru, D. and Moreno, G. and Morriss, S. R. and Mossavi, K. and Mours, B. and Mow-Lowry, C. M. and Mueller, C. L. and Mueller, G. and Muir, A. W. and Mukherjee, Arunava and Mukherjee, D. and Mukherjee, S. and Mukund, N. and Mullavey, A. and Munch, J. and Murphy, D. J. and Murray, P. G. and Mytidis, A. and Nardecchia, I. and Naticchioni, L. and Nayak, R. K. and Necula, V. and Nedkova, K. and Nelemans, G. and Neri, M. and Neunzert, A. and Newton, G. and Nguyen, T. T. and Nielsen, A. B. and Nissanke, S. and Nitz, A. and Nocera, F. and Nolting, D. and Normandin, M. E. N. and Nuttall, L. K. and Oberling, J. and Ochsner, E. and O'Dell, J. and Oelker, E. and Ogin, G. H. and Oh, J. J. and Oh, S. H. and Ohme, F. and Oliver, M. and Oppermann, P. and Oram, Richard J. and O'Reilly, B. and O'Shaughnessy, R. and Ott, C. D. and Ottaway, D. J. and Ottens, R. S. and Overmier, H. and Owen, B. J. and Pai, A. and Pai, S. A. and Palamos, J. R. and Palashov, O. and Palomba, C. and Pal-Singh, A. and Pan, H. and Pan, Y. and Pankow, C. and Pannarale, F. and Pant, B. C. and Paoletti, F. and Paoli, A. and Papa, M. A. and Paris, H. R. and Parker, W. and Pascucci, D. and Pasqualetti, A. and Passaquieti, R. and Passuello, D. and Patricelli, B. and Patrick, Z. and Pearlstone, B. L. and Pedraza, M. and Pedurand, R. and Pekowsky, L. and Pele, A. and Penn, S. and Perreca, A. and Pfeiffer, H. P. and Phelps, M. and Piccinni, O. and Pichot, M. and Pickenpack, M. and Piergiovanni, F. and Pierro, V. and Pillant, G. and Pinard, L. and Pinto, I. M. and Pitkin, M. and Poeld, J. H. and Poggiani, R. and Popolizio, P. and Post, A. and Powell, J. and Prasad, J. and Predoi, V. and Premachandra, S. S. and Prestegard, T. and Price, L. R. and Prijatelj, M. and Principe, M. and Privitera, S. and Prix, R. and Prodi, G. A. and Prokhorov, L. and Puncken, O. and Punturo, M. and Puppo, P. and P\"urrer, M. and Qi, H. and Qin, J. and Quetschke, V. and Quintero, E. A. and Quitzow-James, R. and Raab, F. J. and Rabeling, D. S. and Radkins, H. and Raffai, P. and Raja, S. and Rakhmanov, M. and Ramet, C. R. and Rapagnani, P. and Raymond, V. and Razzano, M. and Re, V. and Read, J. and Reed, C. M. and Regimbau, T. and Rei, L. and Reid, S. and Reitze, D. H. and Rew, H. and Reyes, S. D. and Ricci, F. and Riles, K. and Robertson, N. A. and Robie, R. and Robinet, F. and Rocchi, A. and Rolland, L. and Rollins, J. G. and Roma, V. J. and Romano, J. D. and Romano, R. and Romanov, G. and Romie, J. H. and Rosi\ifmmode \acute{n}\else \'{n}\fi{}ska, D. and Rowan, S. and R\"udiger, A. and Ruggi, P. and Ryan, K. and Sachdev, S. and Sadecki, T. and Sadeghian, L. and Salconi, L. and Saleem, M. and Salemi, F. and Samajdar, A. and Sammut, L. and Sampson, L. M. and Sanchez, E. J. and Sandberg, V. and Sandeen, B. and Sanders, G. H. and Sanders, J. R. and Sassolas, B. and Sathyaprakash, B. S. and Saulson, P. R. and Sauter, O. and Savage, R. L. and Sawadsky, A. and Schale, P. and Schilling, R. and Schmidt, J. and Schmidt, P. and Schnabel, R. and Schofield, R. M. S. and Sch\"onbeck, A. and Schreiber, E. and Schuette, D. and Schutz, B. F. and Scott, J. and Scott, S. M. and Sellers, D. and Sengupta, A. S. and Sentenac, D. and Sequino, V. and Sergeev, A. and Serna, G. and Setyawati, Y. and Sevigny, A. and Shaddock, D. A. and Shaffer, T. and Shah, S. and Shahriar, M. S. and Shaltev, M. and Shao, Z. and Shapiro, B. and Shawhan, P. and Sheperd, A. and Shoemaker, D. H. and Shoemaker, D. M. and Siellez, K. and Siemens, X. and Sigg, D. and Silva, A. D. and Simakov, D. and Singer, A. and Singer, L. P. and Singh, A. and Singh, R. and Singhal, A. and Sintes, A. M. and Slagmolen, B. J. J. and Smith, J. R. and Smith, M. R. and Smith, N. D. and Smith, R. J. E. and Son, E. J. and Sorazu, B. and Sorrentino, F. and Souradeep, T. and Srivastava, A. K. and Staley, A. and Steinke, M. and Steinlechner, J. and Steinlechner, S. and Steinmeyer, D. and Stephens, B. C. and Stevenson, S. P. and Stone, R. and Strain, K. A. and Straniero, N. and Stratta, G. and Strauss, N. A. and Strigin, S. and Sturani, R. and Stuver, A. L. and Summerscales, T. Z. and Sun, L. and Sutton, P. J. and Swinkels, B. L. and Szczepa\ifmmode \acute{n}\else \'{n}\fi{}czyk, M. J. and Tacca, M. and Talukder, D. and Tanner, D. B. and T\'apai, M. and Tarabrin, S. P. and Taracchini, A. and Taylor, R. and Theeg, T. and Thirugnanasambandam, M. P. and Thomas, E. G. and Thomas, M. and Thomas, P. and Thorne, K. A. and Thorne, K. S. and Thrane, E. and Tiwari, S. and Tiwari, V. and Tokmakov, K. V. and Tomlinson, C. and Tonelli, M. and Torres, C. V. and Torrie, C. I. and T\"oyr\"a, D. and Travasso, F. and Traylor, G. and Trifir\`o, D. and Tringali, M. C. and Trozzo, L. and Tse, M. and Turconi, M. and Tuyenbayev, D. and Ugolini, D. and Unnikrishnan, C. S. and Urban, A. L. and Usman, S. A. and Vahlbruch, H. and Vajente, G. and Valdes, G. and Vallisneri, M. and van Bakel, N. and van Beuzekom, M. and van den Brand, J. F. J. and Van Den Broeck, C. and Vander-Hyde, D. C. and van der Schaaf, L. and van Heijningen, J. V. and van Veggel, A. A. and Vardaro, M. and Vass, S. and Vas\'uth, M. and Vaulin, R. and Vecchio, A. and Vedovato, G. and Veitch, J. and Veitch, P. J. and Venkateswara, K. and Verkindt, D. and Vetrano, F. and Vicer\'e, A. and Vinciguerra, S. and Vine, D. J. and Vinet, J.-Y. and Vitale, S. and Vo, T. and Vocca, H. and Vorvick, C. and Voss, D. and Vousden, W. D. and Vyatchanin, S. P. and Wade, A. R. and Wade, L. E. and Wade, M. and Waldman, S. J. and Walker, M. and Wallace, L. and Walsh, S. and Wang, G. and Wang, H. and Wang, M. and Wang, X. and Wang, Y. and Ward, H. and Ward, R. L. and Warner, J. and Was, M. and Weaver, B. and Wei, L.-W. and Weinert, M. and Weinstein, A. J. and Weiss, R. and Welborn, T. and Wen, L. and We\ss{}els, P. and Westphal, T. and Wette, K. and Whelan, J. T. and Whitcomb, S. E. and White, D. J. and Whiting, B. F. and Wiesner, K. and Wilkinson, C. and Willems, P. A. and Williams, L. and Williams, R. D. and Williamson, A. R. and Willis, J. L. and Willke, B. and Wimmer, M. H. and Winkelmann, L. and Winkler, W. and Wipf, C. C. and Wiseman, A. G. and Wittel, H. and Woan, G. and Worden, J. and Wright, J. L. and Wu, G. and Yablon, J. and Yakushin, I. and Yam, W. and Yamamoto, H. and Yancey, C. C. and Yap, M. J. and Yu, H. and Yvert, M. and Zadro\ifmmode \dot{z}\else \.{z}\fi{}ny, A. and Zangrando, L. and Zanolin, M. and Zendri, J.-P. and Zevin, M. and Zhang, F. and Zhang, L. and Zhang, M. and Zhang, Y. and Zhao, C. and Zhou, M. and Zhou, Z. and Zhu, X. J. and Zucker, M. E. and Zuraw, S. E. and Zweizig, J.},
  collaboration = {LIGO Scientific Collaboration and Virgo Collaboration},
  journal = {Phys. Rev. Lett.},
  volume = {116},
  issue = {6},
  pages = {061102},
  numpages = {16},
  year = {2016},
  month = {Feb},
  publisher = {American Physical Society},
  doi = {10.1103/PhysRevLett.116.061102},
  url = {https://link.aps.org/doi/10.1103/PhysRevLett.116.061102}
}

@misc{rezzolla2003gravitationalwavesperturbedblack,
      title={Gravitational Waves from Perturbed Black Holes and Relativistic Stars}, 
      author={Luciano Rezzolla},
      year={2003},
      eprint={gr-qc/0302025},
      archivePrefix={arXiv},
      primaryClass={gr-qc},
      url={https://arxiv.org/abs/gr-qc/0302025}, 
}

@article{PhysRevD.104.023004,
  title = {Memory remains undetected: Updates from the second LIGO/Virgo gravitational-wave transient catalog},
  author = {H\"ubner, Moritz and Lasky, Paul and Thrane, Eric},
  journal = {Phys. Rev. D},
  volume = {104},
  issue = {2},
  pages = {023004},
  numpages = {7},
  year = {2021},
  month = {Jul},
  publisher = {American Physical Society},
  doi = {10.1103/PhysRevD.104.023004},
  url = {https://link.aps.org/doi/10.1103/PhysRevD.104.023004}
}

@ARTICLE{1966JETP...22..241S,
       author = {{Sakharov}, A.~D.},
        title = "{The Initial Stage of an Expanding Universe and the Appearance of a Nonuniform Distribution of Matter}",
      journal = {Soviet Journal of Experimental and Theoretical Physics},
         year = 1966,
        month = jan,
       volume = {22},
        pages = {241},
       adsurl = {https://ui.adsabs.harvard.edu/abs/1966JETP...22..241S}
}

@ARTICLE{1966JETP...22..378G,
       author = {{Gliner}, E.~B.},
        title = "{Algebraic Properties of the Energy-momentum Tensor and Vacuum-like States o$^{+}$ Matter}",
      journal = {Soviet Journal of Experimental and Theoretical Physics},
         year = 1966,
        month = feb,
       volume = {22},
        pages = {378},
       adsurl = {https://ui.adsabs.harvard.edu/abs/1966JETP...22..378G}
}

@misc{huang2025regularblackholessingular,
      title={Regular black holes and their singular families}, 
      author={Hyat Huang and Xiao-Pin Rao},
      year={2025},
      eprint={2503.13133},
      archivePrefix={arXiv},
      primaryClass={gr-qc},
      url={https://arxiv.org/abs/2503.13133}, 
}

@article{Kar:2025phe,
    author = "Kar, Anjan and Kar, Sayan",
    title = "{Diverse regular spacetimes using a parametrised density profile}",
    eprint = "2504.12042",
    archivePrefix = "arXiv",
    primaryClass = "gr-qc",
    doi = "10.1140/epjc/s10052-025-14483-5",
    journal = "Eur. Phys. J. C",
    volume = "85",
    number = "7",
    pages = "773",
    year = "2025"
}

@misc{bueno2025regularblackholesoppenheimersnyder,
      title={Regular black holes from Oppenheimer-Snyder collapse}, 
      author={Pablo Bueno and Pablo A. Cano and Robie A. Hennigar and Ángel J. Murcia and Aitor Vicente-Cano},
      year={2025},
      eprint={2505.09680},
      archivePrefix={arXiv},
      primaryClass={gr-qc},
      url={https://arxiv.org/abs/2505.09680}, 
}

@article{Shore_2001,
   title={Accelerating photons with gravitational radiation},
   volume={605},
   ISSN={0550-3213},
   url={http://dx.doi.org/10.1016/S0550-3213(01)00137-7},
   DOI={10.1016/s0550-3213(01)00137-7},
   number={1–3},
   journal={Nuclear Physics B},
   publisher={Elsevier BV},
   author={Shore, G.M.},
   year={2001},
   month=jul, pages={455–466} }

@article{PhysRevLett.96.031103,
  title = {Formation and Evaporation of Nonsingular Black Holes},
  author = {Hayward, Sean A.},
  journal = {Phys. Rev. Lett.},
  volume = {96},
  issue = {3},
  pages = {031103},
  numpages = {4},
  year = {2006},
  month = {Jan},
  publisher = {American Physical Society},
  doi = {10.1103/PhysRevLett.96.031103},
  url = {https://link.aps.org/doi/10.1103/PhysRevLett.96.031103}
}

@article{Walia_2022,
doi = {10.3847/1538-4357/ac9623},
url = {https://dx.doi.org/10.3847/1538-4357/ac9623},
year = {2022},
month = {nov},
publisher = {The American Astronomical Society},
volume = {939},
number = {2},
pages = {77},
author = {Walia, Rahul Kumar and Ghosh, Sushant G. and Maharaj, Sunil D.},
title = {Testing Rotating Regular Metrics with EHT Results of Sgr A*},
journal = {The Astrophysical Journal}
}

@article{PhysRevD.103.104047,
  title = {Constraints on black-hole charges with the 2017 EHT observations of M87*},
  author = {Kocherlakota, Prashant and Rezzolla, Luciano and Falcke, Heino and Fromm, Christian M. and Kramer, Michael and Mizuno, Yosuke and Nathanail, Antonios and Olivares, H\'ector and Younsi, Ziri and Akiyama, Kazunori and Alberdi, Antxon and Alef, Walter and Algaba, Juan Carlos and Anantua, Richard and Asada, Keiichi and Azulay, Rebecca and Baczko, Anne-Kathrin and Ball, David and Balokovi\ifmmode \acute{c}\else \'{c}\fi{}, Mislav and Barrett, John and Benson, Bradford A. and Bintley, Dan and Blackburn, Lindy and Blundell, Raymond and Boland, Wilfred and Bouman, Katherine L. and Bower, Geoffrey C. and Boyce, Hope and Bremer, Michael and Brinkerink, Christiaan D. and Brissenden, Roger and Britzen, Silke and Broderick, Avery E. and Broguiere, Dominique and Bronzwaer, Thomas and Byun, Do-Young and Carlstrom, John E. and Chael, Andrew and Chan, Chi-kwan and Chatterjee, Shami and Chatterjee, Koushik and Chen, Ming-Tang and Chen, Yongjun and Chesler, Paul M. and Cho, Ilje and Christian, Pierre and Conway, John E. and Cordes, James M. and Crawford, Thomas M. and Crew, Geoffrey B. and Cruz-Osorio, Alejandro and Cui, Yuzhu and Davelaar, Jordy and De Laurentis, Mariafelicia and Deane, Roger and Dempsey, Jessica and Desvignes, Gregory and Doeleman, Sheperd S. and Eatough, Ralph P. and Farah, Joseph and Fish, Vincent L. and Fomalont, Ed and Fraga-Encinas, Raquel and Friberg, Per and Ford, H. Alyson and Fuentes, Antonio and Galison, Peter and Gammie, Charles F. and Garc\'{\i}a, Roberto and Gentaz, Olivier and Georgiev, Boris and Goddi, Ciriaco and Gold, Roman and G\'omez, Jos\'e L. and G\'omez-Ruiz, Arturo I. and Gu, Minfeng and Gurwell, Mark and Hada, Kazuhiro and Haggard, Daryl and Hecht, Michael H. and Hesper, Ronald and Ho, Luis C. and Ho, Paul and Honma, Mareki and Huang, Chih-Wei L. and Huang, Lei and Hughes, David H. and Ikeda, Shiro and Inoue, Makoto and Issaoun, Sara and James, David J. and Jannuzi, Buell T. and Janssen, Michael and Jeter, Britton and Jiang, Wu and Jimenez-Rosales, Alejandra and Johnson, Michael D. and Jorstad, Svetlana and Jung, Taehyun and Karami, Mansour and Karuppusamy, Ramesh and Kawashima, Tomohisa and Keating, Garrett K. and Kettenis, Mark and Kim, Dong-Jin and Kim, Jae-Young and Kim, Jongsoo and Kim, Junhan and Kino, Motoki and Koay, Jun Yi and Kofuji, Yutaro and Koch, Patrick M. and Koyama, Shoko and Kramer, Carsten and Krichbaum, Thomas P. and Kuo, Cheng-Yu and Lauer, Tod R. and Lee, Sang-Sung and Levis, Aviad and Li, Yan-Rong and Li, Zhiyuan and Lindqvist, Michael and Lico, Rocco and Lindahl, Greg and Liu, Jun and Liu, Kuo and Liuzzo, Elisabetta and Lo, Wen-Ping and Lobanov, Andrei P. and Loinard, Laurent and Lonsdale, Colin and Lu, Ru-Sen and MacDonald, Nicholas R. and Mao, Jirong and Marchili, Nicola and Markoff, Sera and Marrone, Daniel P. and Marscher, Alan P. and Mart\'{\i}-Vidal, Iv\'an and Matsushita, Satoki and Matthews, Lynn D. and Medeiros, Lia and Menten, Karl M. and Mizuno, Izumi and Moran, James M. and Moriyama, Kotaro and Moscibrodzka, Monika and M\"uller, Cornelia and Musoke, Gibwa and Mej\'{\i}as, Alejandro Mus and Nagai, Hiroshi and Nagar, Neil M. and Nakamura, Masanori and Narayan, Ramesh and Narayanan, Gopal and Natarajan, Iniyan and Neilsen, Joseph and Neri, Roberto and Ni, Chunchong and Noutsos, Aristeidis and Nowak, Michael A. and Okino, Hiroki and Ortiz-Le\'on, Gisela N. and Oyama, Tomoaki and \"Ozel, Feryal and Palumbo, Daniel C. M. and Park, Jongho and Patel, Nimesh and Pen, Ue-Li and Pesce, Dominic W. and Pi\'etu, Vincent and Plambeck, Richard and PopStefanija, Aleksandar and Porth, Oliver and P\"otzl, Felix M. and Prather, Ben and Preciado-L\'opez, Jorge A. and Psaltis, Dimitrios and Pu, Hung-Yi and Ramakrishnan, Venkatessh and Rao, Ramprasad and Rawlings, Mark G. and Raymond, Alexander W. and Ricarte, Angelo and Ripperda, Bart and Roelofs, Freek and Rogers, Alan and Ros, Eduardo and Rose, Mel and Roshanineshat, Arash and Rottmann, Helge and Roy, Alan L. and Ruszczyk, Chet and Rygl, Kazi L. J. and S\'anchez, Salvador and S\'anchez-Arguelles, David and Sasada, Mahito and Savolainen, Tuomas and Schloerb, F. Peter and Schuster, Karl-Friedrich and Shao, Lijing and Shen, Zhiqiang and Small, Des and Sohn, Bong Won and SooHoo, Jason and Sun, He and Tazaki, Fumie and Tetarenko, Alexandra J. and Tiede, Paul and Tilanus, Remo P. J. and Titus, Michael and Toma, Kenji and Torne, Pablo and Trent, Tyler and Traianou, Efthalia and Trippe, Sascha and van Bemmel, Ilse and van Langevelde, Huib Jan and van Rossum, Daniel R. and Wagner, Jan and Ward-Thompson, Derek and Wardle, John and Weintroub, Jonathan and Wex, Norbert and Wharton, Robert and Wielgus, Maciek and Wong, George N. and Wu, Qingwen and Yoon, Doosoo and Young, Andr\'e and Young, Ken and Yuan, Feng and Yuan, Ye-Fei and Zensus, J. Anton and Zhao, Guang-Yao and Zhao, Shan-Shan},
  collaboration = {EHT Collaboration},
  journal = {Phys. Rev. D},
  volume = {103},
  issue = {10},
  pages = {104047},
  numpages = {18},
  year = {2021},
  month = {May},
  publisher = {American Physical Society},
  doi = {10.1103/PhysRevD.103.104047},
  url = {https://link.aps.org/doi/10.1103/PhysRevD.103.104047}
}

@misc{elhashash2025waveformmodelsgravitationalwavememory,
      title={Waveform models for the gravitational-wave memory effect: II. Time-domain and frequency-domain models for nonspinning binaries}, 
      author={Arwa Elhashash and David A. Nichols},
      year={2025},
      eprint={2504.18635},
      archivePrefix={arXiv},
      primaryClass={gr-qc},
      url={https://arxiv.org/abs/2504.18635}, 
}

@misc{agazie2025nanograv15yeardataset,
      title={The NANOGrav 15-year Data Set: Search for Gravitational Wave Memory}, 
      author={Gabriella Agazie and Akash Anumarlapudi and Anne M. Archibald and Zaven Arzoumanian and Jeremy G. Baier and Paul T. Baker and Bence Becsy and Laura Blecha and Adam Brazier and Paul R. Brook and Sarah Burke-Spolaor and Rand Burnette and J. Andrew Casey-Clyde and Maria Charisi and Shami Chatterjee and Tyler Cohen and James M. Cordes and Neil J. Cornish and Fronefield Crawford and H. Thankful Cromartie and Kathryn Crowter and Megan E. DeCesar and Paul B. Demorest and Heling Deng and Lankeswar Dey and Timothy Dolch and Elizabeth C. Ferrara and William Fiore and Emmanuel Fonseca and Gabriel E. Freedman and Emiko C. Gardiner and Nate Garver-Daniels and Peter A. Gentile and Kyle A. Gersbach and Joseph Glaser and Deborah C. Good and Kayhan Gultekin and Jeffrey S. Hazboun and Ross J. Jennings and Aaron D. Johnson and Megan L. Jones and David L. Kaplan and Luke Zoltan Kelley and Matthew Kerr and Joey S. Key and Nima Laal and Michael T. Lam and William G. Lamb and Bjorn Larsen and T. Joseph W. Lazio and Natalia Lewandowska and Tingting Liu and Duncan R. Lorimer and Jing Luo and Ryan S. Lynch and Chung-Pei Ma and Dustin R. Madison and Alexander McEwen and James W. McKee and Maura A. McLaughlin and Natasha McMann and Bradley W. Meyers and Patrick M. Meyers and Chiara M. F. Mingarelli and Andrea Mitridate and Priyamvada Natarajan and Cherry Ng and David J. Nice and Stella Koch Ocker and Ken D. Olum and Timothy T. Pennucci and Benetge B. P. Perera and Polina Petrov and Nihan S. Pol and Henri A. Radovan and Scott M. Ransom and Paul S. Ray and Jessie C. Runnoe and Alexander Saffer and Shashwat C. Sardesai and Ann Schmiedekamp and Carl Schmiedekamp and Kai Schmitz and Brent J. Shapiro-Albert and Xavier Siemens and Joseph Simon and Magdalena S. Siwek and Sophia V. Sosa Fiscella and Ingrid H. Stairs and Daniel R. Stinebring and Kevin Stovall and Jerry P. Sun and Abhimanyu Susobhanan and Joseph K. Swiggum and Jacob Taylor and Stephen R. Taylor and Jacob E. Turner and Caner Unal and Michele Vallisneri and Rutger van Haasteren and Sarah J. Vigeland and Haley M. Wahl and Caitlin A. Witt and David Wright and Olivia Young},
      year={2025},
      eprint={2502.18599},
      archivePrefix={arXiv},
      primaryClass={gr-qc},
      url={https://arxiv.org/abs/2502.18599}, 
}

@misc{inchauspé2024measuringgravitationalwavememory,
      title={Measuring gravitational wave memory with LISA}, 
      author={Henri Inchauspé and Silvia Gasparotto and Diego Blas and Lavinia Heisenberg and Jann Zosso and Shubhanshu Tiwari},
      year={2024},
      eprint={2406.09228},
      archivePrefix={arXiv},
      primaryClass={gr-qc},
      url={https://arxiv.org/abs/2406.09228}, 
}

@article{PhysRevD.101.023011,
  title = {Measuring gravitational-wave memory in the first LIGO/Virgo gravitational-wave transient catalog},
  author = {H\"ubner, Moritz and Talbot, Colm and Lasky, Paul D. and Thrane, Eric},
  journal = {Phys. Rev. D},
  volume = {101},
  issue = {2},
  pages = {023011},
  numpages = {6},
  year = {2020},
  month = {Jan},
  publisher = {American Physical Society},
  doi = {10.1103/PhysRevD.101.023011},
  url = {https://link.aps.org/doi/10.1103/PhysRevD.101.023011}
}

@misc{bhattacharya2025displacementmemorybmemorygeneralised,
      title={Displacement memory and B-memory in generalised Ellis-Bronnikov wormholes}, 
      author={Soumya Bhattacharya and Suman Ghosh},
      year={2025},
      eprint={2502.03007},
      archivePrefix={arXiv},
      primaryClass={gr-qc},
      url={https://arxiv.org/abs/2502.03007}, 
}

@article{Datta_2024,
   title={Memory effect of gravitational wave pulses in PP-wave spacetimes},
   volume={99},
   ISSN={1402-4896},
   url={http://dx.doi.org/10.1088/1402-4896/ad5389},
   DOI={10.1088/1402-4896/ad5389},
   number={7},
   journal={Physica Scripta},
   publisher={IOP Publishing},
   author={Datta, Sucheta and Guha, Sarbari},
   year={2024},
   month=jun, pages={075023} }

@ARTICLE{1962RSPSA.269...21B,
       author = {{Bondi}, H. and {van der Burg}, M.~G.~J. and {Metzner}, A.~W.~K.},
        title = "{Gravitational Waves in General Relativity. VII. Waves from Axi-Symmetric Isolated Systems}",
      journal = {Proceedings of the Royal Society of London Series A},
         year = 1962,
        month = aug,
       volume = {269},
       number = {1336},
        pages = {21-52},
          doi = {10.1098/rspa.1962.0161},
       adsurl = {https://ui.adsabs.harvard.edu/abs/1962RSPSA.269...21B}
}

@article{Zaslavskii_2010,
   title={Regular black holes and energy conditions},
   volume={688},
   ISSN={0370-2693},
   url={http://dx.doi.org/10.1016/j.physletb.2010.04.031},
   DOI={10.1016/j.physletb.2010.04.031},
   number={4–5},
   journal={Physics Letters B},
   publisher={Elsevier BV},
   author={Zaslavskii, O.B.},
   year={2010},
   month=may, pages={278–280} }

@misc{theligoscientificcollaboration2025gw230814investigationloudgravitationalwave,
      title={GW230814: investigation of a loud gravitational-wave signal observed with a single detector}, 
      author={The LIGO Scientific Collaboration and The Virgo Collaboration and The Kagra Collaboration and Others},
      year={2025},
      eprint={2509.07348},
      archivePrefix={arXiv},
      primaryClass={gr-qc},
      url={https://arxiv.org/abs/2509.07348}, 
}

@article{PhysRevLett.119.161101,
  title = {GW170817: Observation of Gravitational Waves from a Binary Neutron Star Inspiral},
  author = {Abbott, B. P. and Abbott, R. and Abbott, T. D. and Acernese, F. and Ackley, K. and Adams, C. and Adams, T. and Addesso, P. and Adhikari, R. X. and Adya, V. B. and Affeldt, C. and Afrough, M. and Agarwal, B. and Agathos, M. and Agatsuma, K. and Aggarwal, N. and Aguiar, O. D. and Aiello, L. and Ain, A. and Ajith, P. and Allen, B. and Allen, G. and Allocca, A. and Altin, P. A. and Amato, A. and Ananyeva, A. and Anderson, S. B. and Anderson, W. G. and Angelova, S. V. and Antier, S. and Appert, S. and Arai, K. and Araya, M. C. and Areeda, J. S. and Arnaud, N. and Arun, K. G. and Ascenzi, S. and Ashton, G. and Ast, M. and Aston, S. M. and Astone, P. and Atallah, D. V. and Aufmuth, P. and Aulbert, C. and AultONeal, K. and Austin, C. and Avila-Alvarez, A. and Babak, S. and Bacon, P. and Bader, M. K. M. and Bae, S. and Bailes, M. and Baker, P. T. and Baldaccini, F. and Ballardin, G. and Ballmer, S. W. and Banagiri, S. and Barayoga, J. C. and Barclay, S. E. and Barish, B. C. and Barker, D. and Barkett, K. and Barone, F. and Barr, B. and Barsotti, L. and Barsuglia, M. and Barta, D. and Barthelmy, S. D. and Bartlett, J. and Bartos, I. and Bassiri, R. and Basti, A. and Batch, J. C. and Bawaj, M. and Bayley, J. C. and Bazzan, M. and B\'ecsy, B. and Beer, C. and Bejger, M. and Belahcene, I. and Bell, A. S. and Berger, B. K. and Bergmann, G. and Bernuzzi, S. and Bero, J. J. and Berry, C. P. L. and Bersanetti, D. and Bertolini, A. and Betzwieser, J. and Bhagwat, S. and Bhandare, R. and Bilenko, I. A. and Billingsley, G. and Billman, C. R. and Birch, J. and Birney, I. A. and Birnholtz, O. and Biscans, S. and Biscoveanu, S. and Bisht, A. and Bitossi, M. and Biwer, C. and Bizouard, M. A. and Blackburn, J. K. and Blackman, J. and Blair, C. D. and Blair, D. G. and Blair, R. M. and Bloemen, S. and Bock, O. and Bode, N. and Boer, M. and Bogaert, G. and Bohe, A. and Bondu, F. and Bonilla, E. and Bonnand, R. and Boom, B. A. and Bork, R. and Boschi, V. and Bose, S. and Bossie, K. and Bouffanais, Y. and Bozzi, A. and Bradaschia, C. and Brady, P. R. and Branchesi, M. and Brau, J. E. and Briant, T. and Brillet, A. and Brinkmann, M. and Brisson, V. and Brockill, P. and Broida, J. E. and Brooks, A. F. and Brown, D. A. and Brown, D. D. and Brunett, S. and Buchanan, C. C. and Buikema, A. and Bulik, T. and Bulten, H. J. and Buonanno, A. and Buskulic, D. and Buy, C. and Byer, R. L. and Cabero, M. and Cadonati, L. and Cagnoli, G. and Cahillane, C. and Calder\'on Bustillo, J. and Callister, T. A. and Calloni, E. and Camp, J. B. and Canepa, M. and Canizares, P. and Cannon, K. C. and Cao, H. and Cao, J. and Capano, C. D. and Capocasa, E. and Carbognani, F. and Caride, S. and Carney, M. F. and Carullo, G. and Casanueva Diaz, J. and Casentini, C. and Caudill, S. and Cavagli\`a, M. and Cavalier, F. and Cavalieri, R. and Cella, G. and Cepeda, C. B. and Cerd\'a-Dur\'an, P. and Cerretani, G. and Cesarini, E. and Chamberlin, S. J. and Chan, M. and Chao, S. and Charlton, P. and Chase, E. and Chassande-Mottin, E. and Chatterjee, D. and Chatziioannou, K. and Cheeseboro, B. D. and Chen, H. Y. and Chen, X. and Chen, Y. and Cheng, H.-P. and Chia, H. and Chincarini, A. and Chiummo, A. and Chmiel, T. and Cho, H. S. and Cho, M. and Chow, J. H. and Christensen, N. and Chu, Q. and Chua, A. J. K. and Chua, S. and Chung, A. K. W. and Chung, S. and Ciani, G. and Ciolfi, R. and Cirelli, C. E. and Cirone, A. and Clara, F. and Clark, J. A. and Clearwater, P. and Cleva, F. and Cocchieri, C. and Coccia, E. and Cohadon, P.-F. and Cohen, D. and Colla, A. and Collette, C. G. and Cominsky, L. R. and Constancio, M. and Conti, L. and Cooper, S. J. and Corban, P. and Corbitt, T. R. and Cordero-Carri\'on, I. and Corley, K. R. and Cornish, N. and Corsi, A. and Cortese, S. and Costa, C. A. and Coughlin, M. W. and Coughlin, S. B. and Coulon, J.-P. and Countryman, S. T. and Couvares, P. and Covas, P. B. and Cowan, E. E. and Coward, D. M. and Cowart, M. J. and Coyne, D. C. and Coyne, R. and Creighton, J. D. E. and Creighton, T. D. and Cripe, J. and Crowder, S. G. and Cullen, T. J. and Cumming, A. and Cunningham, L. and Cuoco, E. and Dal Canton, T. and D\'alya, G. and Danilishin, S. L. and D'Antonio, S. and Danzmann, K. and Dasgupta, A. and Da Silva Costa, C. F. and Dattilo, V. and Dave, I. and Davier, M. and Davis, D. and Daw, E. J. and Day, B. and De, S. and DeBra, D. and Degallaix, J. and De Laurentis, M. and Del\'eglise, S. and Del Pozzo, W. and Demos, N. and Denker, T. and Dent, T. and De Pietri, R. and Dergachev, V. and De Rosa, R. and DeRosa, R. T. and De Rossi, C. and DeSalvo, R. and de Varona, O. and Devenson, J. and Dhurandhar, S. and D\'{\i}az, M. C. and Dietrich, T. and Di Fiore, L. and Di Giovanni, M. and Di Girolamo, T. and Di Lieto, A. and Di Pace, S. and Di Palma, I. and Di Renzo, F. and Doctor, Z. and Dolique, V. and Donovan, F. and Dooley, K. L. and Doravari, S. and Dorrington, I. and Douglas, R. and Dovale \'Alvarez, M. and Downes, T. P. and Drago, M. and Dreissigacker, C. and Driggers, J. C. and Du, Z. and Ducrot, M. and Dudi, R. and Dupej, P. and Dwyer, S. E. and Edo, T. B. and Edwards, M. C. and Effler, A. and Eggenstein, H.-B. and Ehrens, P. and Eichholz, J. and Eikenberry, S. S. and Eisenstein, R. A. and Essick, R. C. and Estevez, D. and Etienne, Z. B. and Etzel, T. and Evans, M. and Evans, T. M. and Factourovich, M. and Fafone, V. and Fair, H. and Fairhurst, S. and Fan, X. and Farinon, S. and Farr, B. and Farr, W. M. and Fauchon-Jones, E. J. and Favata, M. and Fays, M. and Fee, C. and Fehrmann, H. and Feicht, J. and Fejer, M. M. and Fernandez-Galiana, A. and Ferrante, I. and Ferreira, E. C. and Ferrini, F. and Fidecaro, F. and Finstad, D. and Fiori, I. and Fiorucci, D. and Fishbach, M. and Fisher, R. P. and Fitz-Axen, M. and Flaminio, R. and Fletcher, M. and Fong, H. and Font, J. A. and Forsyth, P. W. F. and Forsyth, S. S. and Fournier, J.-D. and Frasca, S. and Frasconi, F. and Frei, Z. and Freise, A. and Frey, R. and Frey, V. and Fries, E. M. and Fritschel, P. and Frolov, V. V. and Fulda, P. and Fyffe, M. and Gabbard, H. and Gadre, B. U. and Gaebel, S. M. and Gair, J. R. and Gammaitoni, L. and Ganija, M. R. and Gaonkar, S. G. and Garcia-Quiros, C. and Garufi, F. and Gateley, B. and Gaudio, S. and Gaur, G. and Gayathri, V. and Gehrels, N. and Gemme, G. and Genin, E. and Gennai, A. and George, D. and George, J. and Gergely, L. and Germain, V. and Ghonge, S. and Ghosh, Abhirup and Ghosh, Archisman and Ghosh, S. and Giaime, J. A. and Giardina, K. D. and Giazotto, A. and Gill, K. and Glover, L. and Goetz, E. and Goetz, R. and Gomes, S. and Goncharov, B. and Gonz\'alez, G. and Gonzalez Castro, J. M. and Gopakumar, A. and Gorodetsky, M. L. and Gossan, S. E. and Gosselin, M. and Gouaty, R. and Grado, A. and Graef, C. and Granata, M. and Grant, A. and Gras, S. and Gray, C. and Greco, G. and Green, A. C. and Gretarsson, E. M. and Groot, P. and Grote, H. and Grunewald, S. and Gruning, P. and Guidi, G. M. and Guo, X. and Gupta, A. and Gupta, M. K. and Gushwa, K. E. and Gustafson, E. K. and Gustafson, R. and Halim, O. and Hall, B. R. and Hall, E. D. and Hamilton, E. Z. and Hammond, G. and Haney, M. and Hanke, M. M. and Hanks, J. and Hanna, C. and Hannam, M. D. and Hannuksela, O. A. and Hanson, J. and Hardwick, T. and Harms, J. and Harry, G. M. and Harry, I. W. and Hart, M. J. and Haster, C.-J. and Haughian, K. and Healy, J. and Heidmann, A. and Heintze, M. C. and Heitmann, H. and Hello, P. and Hemming, G. and Hendry, M. and Heng, I. S. and Hennig, J. and Heptonstall, A. W. and Heurs, M. and Hild, S. and Hinderer, T. and Ho, W. C. G. and Hoak, D. and Hofman, D. and Holt, K. and Holz, D. E. and Hopkins, P. and Horst, C. and Hough, J. and Houston, E. A. and Howell, E. J. and Hreibi, A. and Hu, Y. M. and Huerta, E. A. and Huet, D. and Hughey, B. and Husa, S. and Huttner, S. H. and Huynh-Dinh, T. and Indik, N. and Inta, R. and Intini, G. and Isa, H. N. and Isac, J.-M. and Isi, M. and Iyer, B. R. and Izumi, K. and Jacqmin, T. and Jani, K. and Jaranowski, P. and Jawahar, S. and Jim\'enez-Forteza, F. and Johnson, W. W. and Johnson-McDaniel, N. K. and Jones, D. I. and Jones, R. and Jonker, R. J. G. and Ju, L. and Junker, J. and Kalaghatgi, C. V. and Kalogera, V. and Kamai, B. and Kandhasamy, S. and Kang, G. and Kanner, J. B. and Kapadia, S. J. and Karki, S. and Karvinen, K. S. and Kasprzack, M. and Kastaun, W. and Katolik, M. and Katsavounidis, E. and Katzman, W. and Kaufer, S. and Kawabe, K. and K\'ef\'elian, F. and Keitel, D. and Kemball, A. J. and Kennedy, R. and Kent, C. and Key, J. S. and Khalili, F. Y. and Khan, I. and Khan, S. and Khan, Z. and Khazanov, E. A. and Kijbunchoo, N. and Kim, Chunglee and Kim, J. C. and Kim, K. and Kim, W. and Kim, W. S. and Kim, Y.-M. and Kimbrell, S. J. and King, E. J. and King, P. J. and Kinley-Hanlon, M. and Kirchhoff, R. and Kissel, J. S. and Kleybolte, L. and Klimenko, S. and Knowles, T. D. and Koch, P. and Koehlenbeck, S. M. and Koley, S. and Kondrashov, V. and Kontos, A. and Korobko, M. and Korth, W. Z. and Kowalska, I. and Kozak, D. B. and Kr\"amer, C. and Kringel, V. and Krishnan, B. and Kr\'olak, A. and Kuehn, G. and Kumar, P. and Kumar, R. and Kumar, S. and Kuo, L. and Kutynia, A. and Kwang, S. and Lackey, B. D. and Lai, K. H. and Landry, M. and Lang, R. N. and Lange, J. and Lantz, B. and Lanza, R. K. and Larson, S. L. and Lartaux-Vollard, A. and Lasky, P. D. and Laxen, M. and Lazzarini, A. and Lazzaro, C. and Leaci, P. and Leavey, S. and Lee, C. H. and Lee, H. K. and Lee, H. M. and Lee, H. W. and Lee, K. and Lehmann, J. and Lenon, A. and Leon, E. and Leonardi, M. and Leroy, N. and Letendre, N. and Levin, Y. and Li, T. G. F. and Linker, S. D. and Littenberg, T. B. and Liu, J. and Liu, X. and Lo, R. K. L. and Lockerbie, N. A. and London, L. T. and Lord, J. E. and Lorenzini, M. and Loriette, V. and Lormand, M. and Losurdo, G. and Lough, J. D. and Lousto, C. O. and Lovelace, G. and L\"uck, H. and Lumaca, D. and Lundgren, A. P. and Lynch, R. and Ma, Y. and Macas, R. and Macfoy, S. and Machenschalk, B. and MacInnis, M. and Macleod, D. M. and Maga\~na Hernandez, I. and Maga\~na-Sandoval, F. and Maga\~na Zertuche, L. and Magee, R. M. and Majorana, E. and Maksimovic, I. and Man, N. and Mandic, V. and Mangano, V. and Mansell, G. L. and Manske, M. and Mantovani, M. and Marchesoni, F. and Marion, F. and M\'arka, S. and M\'arka, Z. and Markakis, C. and Markosyan, A. S. and Markowitz, A. and Maros, E. and Marquina, A. and Marsh, P. and Martelli, F. and Martellini, L. and Martin, I. W. and Martin, R. M. and Martynov, D. V. and Marx, J. N. and Mason, K. and Massera, E. and Masserot, A. and Massinger, T. J. and Masso-Reid, M. and Mastrogiovanni, S. and Matas, A. and Matichard, F. and Matone, L. and Mavalvala, N. and Mazumder, N. and McCarthy, R. and McClelland, D. E. and McCormick, S. and McCuller, L. and McGuire, S. C. and McIntyre, G. and McIver, J. and McManus, D. J. and McNeill, L. and McRae, T. and McWilliams, S. T. and Meacher, D. and Meadors, G. D. and Mehmet, M. and Meidam, J. and Mejuto-Villa, E. and Melatos, A. and Mendell, G. and Mercer, R. A. and Merilh, E. L. and Merzougui, M. and Meshkov, S. and Messenger, C. and Messick, C. and Metzdorff, R. and Meyers, P. M. and Miao, H. and Michel, C. and Middleton, H. and Mikhailov, E. E. and Milano, L. and Miller, A. L. and Miller, B. B. and Miller, J. and Millhouse, M. and Milovich-Goff, M. C. and Minazzoli, O. and Minenkov, Y. and Ming, J. and Mishra, C. and Mitra, S. and Mitrofanov, V. P. and Mitselmakher, G. and Mittleman, R. and Moffa, D. and Moggi, A. and Mogushi, K. and Mohan, M. and Mohapatra, S. R. P. and Molina, I. and Montani, M. and Moore, C. J. and Moraru, D. and Moreno, G. and Morisaki, S. and Morriss, S. R. and Mours, B. and Mow-Lowry, C. M. and Mueller, G. and Muir, A. W. and Mukherjee, Arunava and Mukherjee, D. and Mukherjee, S. and Mukund, N. and Mullavey, A. and Munch, J. and Mu\~niz, E. A. and Muratore, M. and Murray, P. G. and Nagar, A. and Napier, K. and Nardecchia, I. and Naticchioni, L. and Nayak, R. K. and Neilson, J. and Nelemans, G. and Nelson, T. J. N. and Nery, M. and Neunzert, A. and Nevin, L. and Newport, J. M. and Newton, G. and Ng, K. K. Y. and Nguyen, P. and Nguyen, T. T. and Nichols, D. and Nielsen, A. B. and Nissanke, S. and Nitz, A. and Noack, A. and Nocera, F. and Nolting, D. and North, C. and Nuttall, L. K. and Oberling, J. and O'Dea, G. D. and Ogin, G. H. and Oh, J. J. and Oh, S. H. and Ohme, F. and Okada, M. A. and Oliver, M. and Oppermann, P. and Oram, Richard J. and O'Reilly, B. and Ormiston, R. and Ortega, L. F. and O'Shaughnessy, R. and Ossokine, S. and Ottaway, D. J. and Overmier, H. and Owen, B. J. and Pace, A. E. and Page, J. and Page, M. A. and Pai, A. and Pai, S. A. and Palamos, J. R. and Palashov, O. and Palomba, C. and Pal-Singh, A. and Pan, Howard and Pan, Huang-Wei and Pang, B. and Pang, P. T. H. and Pankow, C. and Pannarale, F. and Pant, B. C. and Paoletti, F. and Paoli, A. and Papa, M. A. and Parida, A. and Parker, W. and Pascucci, D. and Pasqualetti, A. and Passaquieti, R. and Passuello, D. and Patil, M. and Patricelli, B. and Pearlstone, B. L. and Pedraza, M. and Pedurand, R. and Pekowsky, L. and Pele, A. and Penn, S. and Perez, C. J. and Perreca, A. and Perri, L. M. and Pfeiffer, H. P. and Phelps, M. and Piccinni, O. J. and Pichot, M. and Piergiovanni, F. and Pierro, V. and Pillant, G. and Pinard, L. and Pinto, I. M. and Pirello, M. and Pitkin, M. and Poe, M. and Poggiani, R. and Popolizio, P. and Porter, E. K. and Post, A. and Powell, J. and Prasad, J. and Pratt, J. W. W. and Pratten, G. and Predoi, V. and Prestegard, T. and Prijatelj, M. and Principe, M. and Privitera, S. and Prix, R. and Prodi, G. A. and Prokhorov, L. G. and Puncken, O. and Punturo, M. and Puppo, P. and P\"urrer, M. and Qi, H. and Quetschke, V. and Quintero, E. A. and Quitzow-James, R. and Raab, F. J. and Rabeling, D. S. and Radkins, H. and Raffai, P. and Raja, S. and Rajan, C. and Rajbhandari, B. and Rakhmanov, M. and Ramirez, K. E. and Ramos-Buades, A. and Rapagnani, P. and Raymond, V. and Razzano, M. and Read, J. and Regimbau, T. and Rei, L. and Reid, S. and Reitze, D. H. and Ren, W. and Reyes, S. D. and Ricci, F. and Ricker, P. M. and Rieger, S. and Riles, K. and Rizzo, M. and Robertson, N. A. and Robie, R. and Robinet, F. and Rocchi, A. and Rolland, L. and Rollins, J. G. and Roma, V. J. and Romano, J. D. and Romano, R. and Romel, C. L. and Romie, J. H. and Rosi\ifmmode \acute{n}\else \'{n}\fi{}ska, D. and Ross, M. P. and Rowan, S. and R\"udiger, A. and Ruggi, P. and Rutins, G. and Ryan, K. and Sachdev, S. and Sadecki, T. and Sadeghian, L. and Sakellariadou, M. and Salconi, L. and Saleem, M. and Salemi, F. and Samajdar, A. and Sammut, L. and Sampson, L. M. and Sanchez, E. J. and Sanchez, L. E. and Sanchis-Gual, N. and Sandberg, V. and Sanders, J. R. and Sassolas, B. and Sathyaprakash, B. S. and Saulson, P. R. and Sauter, O. and Savage, R. L. and Sawadsky, A. and Schale, P. and Scheel, M. and Scheuer, J. and Schmidt, J. and Schmidt, P. and Schnabel, R. and Schofield, R. M. S. and Sch\"onbeck, A. and Schreiber, E. and Schuette, D. and Schulte, B. W. and Schutz, B. F. and Schwalbe, S. G. and Scott, J. and Scott, S. M. and Seidel, E. and Sellers, D. and Sengupta, A. S. and Sentenac, D. and Sequino, V. and Sergeev, A. and Shaddock, D. A. and Shaffer, T. J. and Shah, A. A. and Shahriar, M. S. and Shaner, M. B. and Shao, L. and Shapiro, B. and Shawhan, P. and Sheperd, A. and Shoemaker, D. H. and Shoemaker, D. M. and Siellez, K. and Siemens, X. and Sieniawska, M. and Sigg, D. and Silva, A. D. and Singer, L. P. and Singh, A. and Singhal, A. and Sintes, A. M. and Slagmolen, B. J. J. and Smith, B. and Smith, J. R. and Smith, R. J. E. and Somala, S. and Son, E. J. and Sonnenberg, J. A. and Sorazu, B. and Sorrentino, F. and Souradeep, T. and Spencer, A. P. and Srivastava, A. K. and Staats, K. and Staley, A. and Steinke, M. and Steinlechner, J. and Steinlechner, S. and Steinmeyer, D. and Stevenson, S. P. and Stone, R. and Stops, D. J. and Strain, K. A. and Stratta, G. and Strigin, S. E. and Strunk, A. and Sturani, R. and Stuver, A. L. and Summerscales, T. Z. and Sun, L. and Sunil, S. and Suresh, J. and Sutton, P. J. and Swinkels, B. L. and Szczepa\ifmmode \acute{n}\else \'{n}\fi{}czyk, M. J. and Tacca, M. and Tait, S. C. and Talbot, C. and Talukder, D. and Tanner, D. B. and T\'apai, M. and Taracchini, A. and Tasson, J. D. and Taylor, J. A. and Taylor, R. and Tewari, S. V. and Theeg, T. and Thies, F. and Thomas, E. G. and Thomas, M. and Thomas, P. and Thorne, K. A. and Thorne, K. S. and Thrane, E. and Tiwari, S. and Tiwari, V. and Tokmakov, K. V. and Toland, K. and Tonelli, M. and Tornasi, Z. and Torres-Forn\'e, A. and Torrie, C. I. and T\"oyr\"a, D. and Travasso, F. and Traylor, G. and Trinastic, J. and Tringali, M. C. and Trozzo, L. and Tsang, K. W. and Tse, M. and Tso, R. and Tsukada, L. and Tsuna, D. and Tuyenbayev, D. and Ueno, K. and Ugolini, D. and Unnikrishnan, C. S. and Urban, A. L. and Usman, S. A. and Vahlbruch, H. and Vajente, G. and Valdes, G. and Vallisneri, M. and van Bakel, N. and van Beuzekom, M. and van den Brand, J. F. J. and Van Den Broeck, C. and Vander-Hyde, D. C. and van der Schaaf, L. and van Heijningen, J. V. and van Veggel, A. A. and Vardaro, M. and Varma, V. and Vass, S. and Vas\'uth, M. and Vecchio, A. and Vedovato, G. and Veitch, J. and Veitch, P. J. and Venkateswara, K. and Venugopalan, G. and Verkindt, D. and Vetrano, F. and Vicer\'e, A. and Viets, A. D. and Vinciguerra, S. and Vine, D. J. and Vinet, J.-Y. and Vitale, S. and Vo, T. and Vocca, H. and Vorvick, C. and Vyatchanin, S. P. and Wade, A. R. and Wade, L. E. and Wade, M. and Walet, R. and Walker, M. and Wallace, L. and Walsh, S. and Wang, G. and Wang, H. and Wang, J. Z. and Wang, W. H. and Wang, Y. F. and Ward, R. L. and Warner, J. and Was, M. and Watchi, J. and Weaver, B. and Wei, L.-W. and Weinert, M. and Weinstein, A. J. and Weiss, R. and Wen, L. and Wessel, E. K. and We\ss{}els, P. and Westerweck, J. and Westphal, T. and Wette, K. and Whelan, J. T. and Whitcomb, S. E. and Whiting, B. F. and Whittle, C. and Wilken, D. and Williams, D. and Williams, R. D. and Williamson, A. R. and Willis, J. L. and Willke, B. and Wimmer, M. H. and Winkler, W. and Wipf, C. C. and Wittel, H. and Woan, G. and Woehler, J. and Wofford, J. and Wong, K. W. K. and Worden, J. and Wright, J. L. and Wu, D. S. and Wysocki, D. M. and Xiao, S. and Yamamoto, H. and Yancey, C. C. and Yang, L. and Yap, M. J. and Yazback, M. and Yu, Hang and Yu, Haocun and Yvert, M. and Zadro\ifmmode \dot{z}\else \.{z}\fi{}ny, A. and Zanolin, M. and Zelenova, T. and Zendri, J.-P. and Zevin, M. and Zhang, L. and Zhang, M. and Zhang, T. and Zhang, Y.-H. and Zhao, C. and Zhou, M. and Zhou, Z. and Zhu, S. J. and Zhu, X. J. and Zimmerman, A. B. and Zucker, M. E. and Zweizig, J.},
  collaboration = {LIGO Scientific Collaboration and Virgo Collaboration},
  journal = {Phys. Rev. Lett.},
  volume = {119},
  issue = {16},
  pages = {161101},
  numpages = {18},
  year = {2017},
  month = {Oct},
  publisher = {American Physical Society},
  doi = {10.1103/PhysRevLett.119.161101},
  url = {https://link.aps.org/doi/10.1103/PhysRevLett.119.161101}
}

@article{PhysRevD.103.124047,
  title = {Probing modified gravitational-wave propagation through tidal measurements of binary neutron star mergers},
  author = {Jiang, Nan and Yagi, Kent},
  journal = {Phys. Rev. D},
  volume = {103},
  issue = {12},
  pages = {124047},
  numpages = {14},
  year = {2021},
  month = {Jun},
  publisher = {American Physical Society},
  doi = {10.1103/PhysRevD.103.124047},
  url = {https://link.aps.org/doi/10.1103/PhysRevD.103.124047}
}

@article{PhysRevD.84.124013,
  title = {The gravitational-wave memory from eccentric binaries},
  author = {Favata, Marc},
  journal = {Phys. Rev. D},
  volume = {84},
  issue = {12},
  pages = {124013},
  numpages = {24},
  year = {2011},
  month = {Dec},
  publisher = {American Physical Society},
  doi = {10.1103/PhysRevD.84.124013},
  url = {https://link.aps.org/doi/10.1103/PhysRevD.84.124013}
}

@article{PhysRevD.80.024002,
  title = {Post-Newtonian corrections to the gravitational-wave memory for quasicircular, inspiralling compact binaries},
  author = {Favata, Marc},
  journal = {Phys. Rev. D},
  volume = {80},
  issue = {2},
  pages = {024002},
  numpages = {28},
  year = {2009},
  month = {Jul},
  publisher = {American Physical Society},
  doi = {10.1103/PhysRevD.80.024002},
  url = {https://link.aps.org/doi/10.1103/PhysRevD.80.024002}
}

@article{Favata_2009,
doi = {10.1088/1742-6596/154/1/012043},
url = {https://dx.doi.org/10.1088/1742-6596/154/1/012043},
year = {2009},
month = {mar},
publisher = {},
volume = {154},
number = {1},
pages = {012043},
author = {Marc Favata},
title = {Gravitational-wave memory revisited: Memory from the merger and recoil of binary black holes},
journal = {Journal of Physics: Conference Series}
}

@article{PhysRevD.111.044045,
  title = {Direct current memory effects in effective-one-body waveform models},
  author = {Grilli, Elisa and Placidi, Andrea and Albanesi, Simone and Grignani, Gianluca and Orselli, Marta},
  journal = {Phys. Rev. D},
  volume = {111},
  issue = {4},
  pages = {044045},
  numpages = {17},
  year = {2025},
  month = {Feb},
  publisher = {American Physical Society},
  doi = {10.1103/PhysRevD.111.044045},
  url = {https://link.aps.org/doi/10.1103/PhysRevD.111.044045}
}

@misc{chakraborty2025prospectscosmologicalconstraintsusing,
      title={Prospects for cosmological constraints using gravitational wave memory}, 
      author={Indranil Chakraborty and Susmita Jana and S. Shankaranarayanan},
      year={2025},
      eprint={2402.18083},
      archivePrefix={arXiv},
      primaryClass={gr-qc},
      url={https://arxiv.org/abs/2402.18083}, 
}

@article{PhysRevD.111.044044,
  title = {Measuring gravitational wave memory with LISA},
  author = {Inchausp\'e, Henri and Gasparotto, Silvia and Blas, Diego and Heisenberg, Lavinia and Zosso, Jann and Tiwari, Shubhanshu},
  journal = {Phys. Rev. D},
  volume = {111},
  issue = {4},
  pages = {044044},
  numpages = {19},
  year = {2025},
  month = {Feb},
  publisher = {American Physical Society},
  doi = {10.1103/PhysRevD.111.044044},
  url = {https://link.aps.org/doi/10.1103/PhysRevD.111.044044}
}

@article{Goncharov:2023woe,
    author = "Goncharov, Boris and Donnay, Laura and Harms, Jan",
    title = "{Inferring Fundamental Spacetime Symmetries with Gravitational-Wave Memory: From LISA to the Einstein Telescope}",
    eprint = "2310.10718",
    archivePrefix = "arXiv",
    primaryClass = "gr-qc",
    doi = "10.1103/PhysRevLett.132.241401",
    journal = "Phys. Rev. Lett.",
    volume = "132",
    number = "24",
    pages = "241401",
    year = "2024"
}

@article{yyv5-3y1c,
  title = {Gravitational-wave tails and memory effect for mergers in astrophysical environments},
  author = {Alnasheet, Qassim and Cardoso, Vitor and Duque, Francisco and Macedo, Rodrigo Panosso},
  journal = {Phys. Rev. D},
  volume = {112},
  issue = {4},
  pages = {044066},
  numpages = {8},
  year = {2025},
  month = {Aug},
  publisher = {American Physical Society},
  doi = {10.1103/yyv5-3y1c},
  url = {https://link.aps.org/doi/10.1103/yyv5-3y1c}
}

@article{article,
author = {Tolish, Alexander and Wald, Robert},
year = {2016},
month = {06},
pages = {},
title = {The Cosmological Memory Effect},
volume = {94},
journal = {Physical Review D},
doi = {10.1103/PhysRevD.94.044009}
}

@article{ZHANG2017743,
title = {The memory effect for plane gravitational waves},
journal = {Physics Letters B},
volume = {772},
pages = {743-746},
year = {2017},
issn = {0370-2693},
doi = {https://doi.org/10.1016/j.physletb.2017.07.050},
url = {https://www.sciencedirect.com/science/article/pii/S0370269317306068},
author = {P.-M. Zhang and C. Duval and G.W. Gibbons and P.A. Horvathy}
}

@misc{singh2025gravitationalmemorymeetsastrophysical,
      title={Gravitational memory meets astrophysical environments: exploring a new frontier through osculations}, 
      author={Rishabh Kumar Singh and Shailesh Kumar and Abhishek Chowdhuri and Arpan Bhattacharyya},
      year={2025},
      eprint={2509.01676},
      archivePrefix={arXiv},
      primaryClass={gr-qc},
      url={https://arxiv.org/abs/2509.01676}, 
}

@article{PhysRevLett.117.061102,
  title = {Detecting Gravitational-Wave Memory with LIGO: Implications of GW150914},
  author = {Lasky, Paul D. and Thrane, Eric and Levin, Yuri and Blackman, Jonathan and Chen, Yanbei},
  journal = {Phys. Rev. Lett.},
  volume = {117},
  issue = {6},
  pages = {061102},
  numpages = {5},
  year = {2016},
  month = {Aug},
  publisher = {American Physical Society},
  doi = {10.1103/PhysRevLett.117.061102},
  url = {https://link.aps.org/doi/10.1103/PhysRevLett.117.061102}
}

@article{Talbot:2018sgr,
    author = "Talbot, Colm and Thrane, Eric and Lasky, Paul D. and Lin, Fuhui",
    title = "{Gravitational-wave memory: waveforms and phenomenology}",
    eprint = "1807.00990",
    archivePrefix = "arXiv",
    primaryClass = "astro-ph.HE",
    doi = "10.1103/PhysRevD.98.064031",
    journal = "Phys. Rev. D",
    volume = "98",
    number = "6",
    pages = "064031",
    year = "2018"
}

@article{Islo:2019qht,
    author = "Islo, Kristina and Simon, Joseph and Burke-Spolaor, Sarah and Siemens, Xavier",
    title = "{Prospects for Memory Detection with Low-Frequency Gravitational Wave Detectors}",
    eprint = "1906.11936",
    archivePrefix = "arXiv",
    primaryClass = "astro-ph.HE",
    month = "6",
    year = "2019"
}

@article{Bieri:2024ios,
    author = "Bieri, Lydia and Polnarev, Alexander",
    title = "{Gravitational wave displacement and velocity memory effects}",
    eprint = "2402.02594",
    archivePrefix = "arXiv",
    primaryClass = "gr-qc",
    doi = "10.1088/1361-6382/ad4dfe",
    journal = "Class. Quant. Grav.",
    volume = "41",
    number = "13",
    pages = "135012",
    year = "2024"
}

@article{Pasterski:2015tva,
    author = "Pasterski, Sabrina and Strominger, Andrew and Zhiboedov, Alexander",
    title = "{New Gravitational Memories}",
    eprint = "1502.06120",
    archivePrefix = "arXiv",
    primaryClass = "hep-th",
    doi = "10.1007/JHEP12(2016)053",
    journal = "JHEP",
    volume = "12",
    pages = "053",
    year = "2016"
}

@article{Nichols:2017rqr,
    author = "Nichols, David A.",
    title = "{Spin memory effect for compact binaries in the post-Newtonian approximation}",
    eprint = "1702.03300",
    archivePrefix = "arXiv",
    primaryClass = "gr-qc",
    doi = "10.1103/PhysRevD.95.084048",
    journal = "Phys. Rev. D",
    volume = "95",
    number = "8",
    pages = "084048",
    year = "2017"
}

@article{Strominger:2014pwa,
    author = "Strominger, Andrew and Zhiboedov, Alexander",
    title = "{Gravitational Memory, BMS Supertranslations and Soft Theorems}",
    eprint = "1411.5745",
    archivePrefix = "arXiv",
    primaryClass = "hep-th",
    doi = "10.1007/JHEP01(2016)086",
    journal = "JHEP",
    volume = "01",
    pages = "086",
    year = "2016"
}

@article{Solanki:2023wmv,
    author = "Solanki, Divyesh N. and Bhattacharjee, Srijit",
    title = "{Soft theorems and memory effects at finite temperatures}",
    eprint = "2308.02445",
    archivePrefix = "arXiv",
    primaryClass = "hep-th",
    doi = "10.1140/epjc/s10052-023-12335-8",
    journal = "Eur. Phys. J. C",
    volume = "83",
    number = "12",
    pages = "1167",
    year = "2023"
}

@article{Mitman:2020pbt,
    author = "Mitman, Keefe and Moxon, Jordan and Scheel, Mark A. and Teukolsky, Saul A. and Boyle, Michael and Deppe, Nils and Kidder, Lawrence E. and Throwe, William",
    title = "{Computation of displacement and spin gravitational memory in numerical relativity}",
    eprint = "2007.11562",
    archivePrefix = "arXiv",
    primaryClass = "gr-qc",
    doi = "10.1103/PhysRevD.102.104007",
    journal = "Phys. Rev. D",
    volume = "102",
    number = "10",
    pages = "104007",
    year = "2020"
}

@article{Mitman:2021xkq,
    author = "Mitman, Keefe and others",
    title = "{Fixing the BMS frame of numerical relativity waveforms}",
    eprint = "2105.02300",
    archivePrefix = "arXiv",
    primaryClass = "gr-qc",
    doi = "10.1103/PhysRevD.104.024051",
    journal = "Phys. Rev. D",
    volume = "104",
    number = "2",
    pages = "024051",
    year = "2021"
}

@article{Mitman:2022kwt,
    author = "Mitman, Keefe and others",
    title = "{Fixing the BMS frame of numerical relativity waveforms with BMS charges}",
    eprint = "2208.04356",
    archivePrefix = "arXiv",
    primaryClass = "gr-qc",
    doi = "10.1103/PhysRevD.106.084029",
    journal = "Phys. Rev. D",
    volume = "106",
    number = "8",
    pages = "084029",
    year = "2022"
}

@article{Pollney:2010hs,
    author = "Pollney, Denis and Reisswig, Christian",
    title = "{Gravitational memory in binary black hole mergers}",
    eprint = "1004.4209",
    archivePrefix = "arXiv",
    primaryClass = "gr-qc",
    doi = "10.1088/2041-8205/732/1/L13",
    journal = "Astrophys. J. Lett.",
    volume = "732",
    pages = "L13",
    year = "2011"
}

@misc{akyüz2025potentialsciencegw250114,
      title={Potential science with GW250114 -- the loudest binary black hole merger detected to date}, 
      author={Aleyna Akyüz and Alex Correia and Jada Garofalo and Keisi Kacanja and Labani Roy and Kanchan Soni and Hung Tan and Vikas Jadhav Y and Alexander H. Nitz and Collin D. Capano},
      year={2025},
      eprint={2507.08789},
      archivePrefix={arXiv},
      primaryClass={gr-qc},
      url={https://arxiv.org/abs/2507.08789}, 
}

@article{Hinderer_2008,
doi = {10.1086/533487},
url = {https://doi.org/10.1086/533487},
year = {2008},
month = {apr},
publisher = {},
volume = {677},
number = {2},
pages = {1216},
author = {Hinderer, Tanja},
title = {Tidal Love Numbers of Neutron Stars},
journal = {The Astrophysical Journal}
}

@article{PhysRevLett.67.1486,
  title = {Nonlinear nature of gravitation and gravitational-wave experiments},
  author = {Christodoulou, Demetrios},
  journal = {Phys. Rev. Lett.},
  volume = {67},
  issue = {12},
  pages = {1486--1489},
  numpages = {0},
  year = {1991},
  month = {Sep},
  publisher = {American Physical Society},
  doi = {10.1103/PhysRevLett.67.1486},
  url = {https://link.aps.org/doi/10.1103/PhysRevLett.67.1486}
}

@article{PhysRevD.45.520,
  title = {Gravitational-wave bursts with memory: The Christodoulou effect},
  author = {Thorne, Kip S.},
  journal = {Phys. Rev. D},
  volume = {45},
  issue = {2},
  pages = {520--524},
  numpages = {0},
  year = {1992},
  month = {Jan},
  publisher = {American Physical Society},
  doi = {10.1103/PhysRevD.45.520},
  url = {https://link.aps.org/doi/10.1103/PhysRevD.45.520}
}

@article{Shankaranarayanan_2022,
   title={Modified theories of gravity: Why, how and what?},
   volume={54},
   ISSN={1572-9532},
   url={http://dx.doi.org/10.1007/s10714-022-02927-2},
   DOI={10.1007/s10714-022-02927-2},
   number={5},
   journal={General Relativity and Gravitation},
   publisher={Springer Science and Business Media LLC},
   author={Shankaranarayanan, S. and Johnson, Joseph P.},
   year={2022},
   month=may }

@article{PhysRevD.99.024031,
  title = {Geodesic congruences, impulsive gravitational waves, and gravitational memory},
  author = {O'Loughlin, Martin and Demirchian, Hovhannes},
  journal = {Phys. Rev. D},
  volume = {99},
  issue = {2},
  pages = {024031},
  numpages = {8},
  year = {2019},
  month = {Jan},
  publisher = {American Physical Society},
  doi = {10.1103/PhysRevD.99.024031},
  url = {https://link.aps.org/doi/10.1103/PhysRevD.99.024031}
}

@article{PhysRevD.89.084039,
  title = {Perturbative and gauge invariant treatment of gravitational wave memory},
  author = {Bieri, Lydia and Garfinkle, David},
  journal = {Phys. Rev. D},
  volume = {89},
  issue = {8},
  pages = {084039},
  numpages = {9},
  year = {2014},
  month = {Apr},
  publisher = {American Physical Society},
  doi = {10.1103/PhysRevD.89.084039},
  url = {https://link.aps.org/doi/10.1103/PhysRevD.89.084039}
}

@article{PhysRevD.90.044060,
  title = {Examination of a simple example of gravitational wave memory},
  author = {Tolish, Alexander and Bieri, Lydia and Garfinkle, David and Wald, Robert M.},
  journal = {Phys. Rev. D},
  volume = {90},
  issue = {4},
  pages = {044060},
  numpages = {9},
  year = {2014},
  month = {Aug},
  publisher = {American Physical Society},
  doi = {10.1103/PhysRevD.90.044060},
  url = {https://link.aps.org/doi/10.1103/PhysRevD.90.044060}
}

@article{PhysRevD.104.084026,
  title = {Note on electromagnetic memories},
  author = {Mao, Pujian},
  journal = {Phys. Rev. D},
  volume = {104},
  issue = {8},
  pages = {084026},
  numpages = {6},
  year = {2021},
  month = {Oct},
  publisher = {American Physical Society},
  doi = {10.1103/PhysRevD.104.084026},
  url = {https://link.aps.org/doi/10.1103/PhysRevD.104.084026}
}

@misc{caldwell2025persistencenonlineargravitationalwave,
      title={The Persistence of Nonlinear Gravitational Wave Memory}, 
      author={Robert R. Caldwell},
      year={2025},
      eprint={2506.20751},
      archivePrefix={arXiv},
      primaryClass={gr-qc},
      url={https://arxiv.org/abs/2506.20751}, 
}

@article{Winicour_2014,
   title={Global aspects of radiation memory},
   volume={31},
   ISSN={1361-6382},
   url={http://dx.doi.org/10.1088/0264-9381/31/20/205003},
   DOI={10.1088/0264-9381/31/20/205003},
   number={20},
   journal={Classical and Quantum Gravity},
   publisher={IOP Publishing},
   author={Winicour, J},
   year={2014},
   month=oct, pages={205003} }

@article{flanagan2019persistent,
  title={Persistent gravitational wave observables: general framework},
  author={Flanagan, {\'E}anna {\'E} and Grant, Alexander M and Harte, Abraham I and Nichols, David A},
  journal={Physical Review D},
  volume={99},
  number={8},
  pages={084044},
  year={2019},
  publisher={APS}
}

@article{PhysRevD.102.044009,
  title = {Probing modified gravity theories and cosmology using gravitational-waves and associated electromagnetic counterparts},
  author = {Mastrogiovanni, S. and Steer, D. A. and Barsuglia, M.},
  journal = {Phys. Rev. D},
  volume = {102},
  issue = {4},
  pages = {044009},
  numpages = {19},
  year = {2020},
  month = {Aug},
  publisher = {American Physical Society},
  doi = {10.1103/PhysRevD.102.044009},
  url = {https://link.aps.org/doi/10.1103/PhysRevD.102.044009}
}

@article{Wang_2020,
doi = {10.3847/1538-4365/aba2f3},
url = {https://doi.org/10.3847/1538-4365/aba2f3},
year = {2020},
month = {aug},
publisher = {The American Astronomical Society},
volume = {250},
number = {1},
pages = {6},
author = {Wang, Bo and Zhu, Zhenyu and Li, Ang and Zhao, Wen},
title = {Comprehensive Analysis of the Tidal Effect in Gravitational Waves and Implication for Cosmology},
journal = {The Astrophysical Journal Supplement Series}
}

@article{PhysRevD.44.R2945,
  title = {Christodoulou's nonlinear gravitational-wave memory: Evaluation in the quadrupole approximation},
  author = {Wiseman, Alan G. and Will, Clifford M.},
  journal = {Phys. Rev. D},
  volume = {44},
  issue = {10},
  pages = {R2945--R2949},
  numpages = {0},
  year = {1991},
  month = {Nov},
  publisher = {American Physical Society},
  doi = {10.1103/PhysRevD.44.R2945},
  url = {https://link.aps.org/doi/10.1103/PhysRevD.44.R2945}
}

@article{Cunningham_2025_1,
doi = {10.1088/1361-6382/ae09e8},
url = {https://doi.org/10.1088/1361-6382/ae09e8},
year = {2025},
month = {oct},
publisher = {IOP Publishing},
volume = {42},
number = {19},
pages = {199401},
author = {Cunningham, Kevin and Kavanagh, Chris and Pound, Adam and Trestini, David and Warburton, Niels and Neef, Jakob},
title = {Addendum: Gravitational memory: new results from post-Newtonian and self-force theory (2025 Class. Quantum Grav. 42 135009)},
journal = {Classical and Quantum Gravity}
}

@article{Grishchuk:1989qa,
    author = "Grishchuk, L. P. and Polnarev, A. G.",
    title = "{Gravitational wave pulses with 'velocity coded memory.'}",
    journal = "Sov. Phys. JETP",
    volume = "69",
    pages = "653--657",
    year = "1989"
}

@article{PhysRevD.107.064056,
  title = {Outlook for detecting the gravitational-wave displacement and spin memory effects with current and future gravitational-wave detectors},
  author = {Grant, Alexander M. and Nichols, David A.},
  journal = {Phys. Rev. D},
  volume = {107},
  issue = {6},
  pages = {064056},
  numpages = {24},
  year = {2023},
  month = {Mar},
  publisher = {American Physical Society},
  doi = {10.1103/PhysRevD.107.064056},
  url = {https://link.aps.org/doi/10.1103/PhysRevD.107.064056}
}

@article{Zhang_2018,
   title={Velocity Memory Effect for polarized gravitational waves},
   volume={2018},
   ISSN={1475-7516},
   url={http://dx.doi.org/10.1088/1475-7516/2018/05/030},
   DOI={10.1088/1475-7516/2018/05/030},
   number={05},
   journal={Journal of Cosmology and Astroparticle Physics},
   publisher={IOP Publishing},
   author={Zhang, P.-M. and Duval, C. and Gibbons, G.W. and Horvathy, P.A.},
   year={2018},
   month=may, pages={030–030} }

@article{Bieri_2012,
   title={The electromagnetic Christodoulou memory effect and its application to neutron star binary mergers},
   volume={29},
   ISSN={1361-6382},
   url={http://dx.doi.org/10.1088/0264-9381/29/21/215003},
   DOI={10.1088/0264-9381/29/21/215003},
   number={21},
   journal={Classical and Quantum Gravity},
   publisher={IOP Publishing},
   author={Bieri, Lydia and Chen, PoNing and Yau, Shing-Tung},
   year={2012},
   month=sep, pages={215003} }

@article{Blanchet_2023,
   title={Multipole expansion of gravitational waves: memory effects and Bondi aspects},
   volume={2023},
   ISSN={1029-8479},
   url={http://dx.doi.org/10.1007/JHEP07(2023)123},
   DOI={10.1007/jhep07(2023)123},
   number={7},
   journal={Journal of High Energy Physics},
   publisher={Springer Science and Business Media LLC},
   author={Blanchet, Luc and Compère, Geoffrey and Faye, Guillaume and Oliveri, Roberto and Seraj, Ali},
   year={2023},
   month=jul }

@article{Zeldovich:1974gvh,
    author = "Zel'dovich, Y. B. and Polnarev, A. G.",
    title = "{Radiation of gravitational waves by a cluster of superdense stars}",
    journal = "Sov. Astron.",
    volume = "18",
    pages = "17",
    year = "1974"
}

@article{Flanagan:2019ezo,
    author = "Flanagan, {\'E}anna {\'E}. and Grant, Alexander M. and Harte, Abraham I. and Nichols, David A.",
    title = "{Persistent gravitational wave observables: Nonlinear plane wave spacetimes}",
    eprint = "1912.13449",
    archivePrefix = "arXiv",
    primaryClass = "gr-qc",
    doi = "10.1103/PhysRevD.101.104033",
    journal = "Phys. Rev. D",
    volume = "101",
    number = "10",
    pages = "104033",
    year = "2020"
}

@article{Heisenberg_2023,
   title={Gravitational wave memory beyond general relativity},
   volume={108},
   ISSN={2470-0029},
   url={http://dx.doi.org/10.1103/PhysRevD.108.024010},
   DOI={10.1103/physrevd.108.024010},
   number={2},
   journal={Physical Review D},
   publisher={American Physical Society (APS)},
   author={Heisenberg, Lavinia and Yunes, Nicolás and Zosso, Jann},
   year={2023},
   month=jul }

\appendix

\section{Riemann Tensors for the Deviation analysis}

The non-zero components of $\omega$ corresponding to 
Eq.~\eqref{general metric with pulse1}, in the tetrad basis 
defined in Eq.~\eqref{tetrad basis}, are given by:
\begin{align*}
    \omega_1^0 &= \frac{f^\prime (r)}{2} du & \omega_0^2 &=  A (r, u)d\theta  & \omega_1^2 &= B(r, u) d\theta \\
    \omega_0^3 &= C (r, u) \sin \theta d\phi & \omega_1^3 &=  D (r, u)\sin \theta  d\phi & \omega_2^3 &= E(r, u) \cos \theta  d\phi
\end{align*}
Where, 
\begin{align*}
    A(r,u) &= \frac{rH^\prime (u)}{2 \sqrt{f(r)(r^2+ rH(u))}} & B(r, u) &= \frac{2rf(r)+ f(r) H(u) -rH^\prime (u)}{2 \sqrt{f(r)(r^2 + rH(u))}} & C(r,u) &=  \frac{-rH^\prime (u) }{2 \sqrt{f(r)(r^2 - rH(u))}}\\
    D(r, u)&= \frac{\left[2rf(r)- f(r) H(u) +rH^\prime (u)\right] }{2 \sqrt{f(r)(r^2 - rH(u))}} & E(r, u)& =  \sqrt{\frac{r^2 - rH(u)}{r^2 + r H(u)}}
\end{align*}
Here our notations are as follows: $\{u, r, \theta, \phi\} = \{0, 1, 2, 3\}$. In that tetrad basis, non-zero Riemann tensors are: 
\begin{align*}
    R^0_{110} &= \frac{f^{\prime \prime}(r)}{2} & R^2_{002} &= \frac{1}{\sqrt{f(r)[r^2 + r H(u)]}} \left[\frac{\partial A}{\partial u} - \frac{B f^\prime (r) }{2}\right] \\
    R^2_{012} &= \sqrt{\frac{f(r)}{r^2 + r H(u)}} \frac{\partial A}{\partial r} - R^2_{002} & R^2_{112} &= \sqrt{\frac{f(r)}{r^2 + r H(u)}} \frac{\partial B}{\partial r} - R^2_{102} \\
     R^3_{003} &= \frac{1}{\sqrt{f(r)[r^2 - r H(u)]}} \left[\frac{\partial C}{\partial u} - \frac{D f^\prime (r) }{2}\right] & R^3_{013} &=
     \sqrt{\frac{f(r)}{r^2 - r H(u)}} \frac{\partial C}{\partial r} - R^3_{003}\\
     R^3_{023} &= \frac{C-EA}{\sqrt{r^4-r^2 H^2 (u)}} \cot \theta & R^3_{113} &= \sqrt{\frac{f(r)}{r^2 - r H(u)}} \frac{\partial D}{\partial r} - R^3_{103}\\
     R^3_{123} &= \frac{D-EB}{\sqrt{r^4-r^2 H^2 (u)}} \cot \theta & R^3_{223} &= \frac{BD -E -AC}{\sqrt{r^4 - r^2 H^2 (u)}}
\end{align*}
\end{document}